\documentclass[a4paper,fleqn,usenatbib]{mnras}
\pdfoutput=1

\usepackage{mathptmx}
\usepackage[T1]{fontenc}
\usepackage{ae,aecompl}

\usepackage{graphicx}	% Including figure files
\usepackage{amsmath}	% Advanced maths commands
\usepackage{amssymb}	% Extra maths symbols
\newcommand{\lya}{Ly$\alpha$}
\newcommand{\unitcgssb}{erg s$^{-1}$ cm$^{-2}$ arcsec$^{-2}$}

\newcommand{\civ}{\ion{C}{iv}\ }
\newcommand{\heii}{\ion{He}{ii}\ }

\def \cgssb {{\rm\,erg\,s^{-1}\,cm^{-2}\,arcsec^{-2}}}

\title[QSO MUSEUM]{QSO MUSEUM I: A sample of 61 extended Ly$\alpha$-emission nebulae surrounding $z\sim3$ quasars}

\author[F. Arrigoni Battaia et al.]{
Fabrizio Arrigoni Battaia$^{1,2}$\thanks{E-mail: farrigon@eso.org}, % \newauthor
Joseph F. Hennawi$^{3,2}$, J. Xavier Prochaska$^{4,5}$, \and 
Jose O{\~{n}}orbe$^{6}$, Emanuele P. Farina$^{3}$, Sebastiano Cantalupo$^{7}$, and Elisabeta Lusso$^{8}$ 
\\
$^{1}$European Southern Observatory, Karl-Schwarzschild-Str. 2, D-85748 Garching bei M\"unchen, Germany\\
$^{2}$Max-Planck-Institut f\"ur Astronomie, K\"onigstuhl 17, D-69117 Heidelberg, Germany\\
$^{3}$Department of Physics, Broida Hall, University of California, Santa Barbara, CA 93106-9530, USA\\
$^{4}$Department of Astronomy and Astrophysics, University of California, 1156 High Street, Santa Cruz, California 95064, USA\\
$^{5}$University of California Observatories, Lick Observatory, 1156 High Street, Santa Cruz, California 95064, USA\\
$^{6}$Institute for Astronomy, University of Edinburgh, Royal Observatory Blackford Hill, EH9 3HJ, Edinburgh, United Kingdom\\
$^{7}$Department of Physics, ETH Zurich, CH-8093 Zurich, Switzerland\\
$^{8}$Centre for Extragalactic Astronomy, Department of Physics, Durham University, South Road, Durham, DH1 3LE, UK
}

\pubyear{2018}

\begin{document}
\label{firstpage}
\pagerange{\pageref{firstpage}--\pageref{lastpage}}
\maketitle

\begin{abstract}
Motivated by the recent discovery of rare Enormous Lyman-Alpha Nebulae (ELAN) around $z\sim2$ quasars, 
we have initiated a long-term observational campaign with the MUSE/VLT instrument 
to directly uncover the astrophysics of the gas around quasars.
We present here the first 61 targets of our effort under the acronym QSO MUSEUM ({\it Q}uasar {\it S}napshot {\it O}bservations with {\it MU}se: 
{\it S}earch for {\it E}xtended {\it U}ltraviolet e{\it M}ission).
These quasars are characterized by a median redshfit of $z=3.17$ ($3.03<z<3.46$), 
absolute $i$ magnitude in the range $-29.67\leq M_i(z=2)\leq-27.03$, and different levels of radio-loudness. 
This sample unveils diverse specimens of \lya\ nebulosities 
extending for tens of kiloparsecs around these quasars (on average out to a maximum projected distance of 80~kpc) above a surface brightness SB$>8.8\times10^{-19}$~\unitcgssb ($2\sigma$). 
Irrespective of the radio-loudness of the targets, 
the bulk of the extended \lya\ emission is within $R< 50$~kpc, and is characterized by relatively quiescent
kinematics, with average velocity dispersions of $\langle \sigma_{\rm Ly\alpha}\rangle < 400$~km~s$^{-1}$.
Therefore, the motions within all these \lya\ nebulosities have amplitudes consistent with gravitational motions expected
in dark matter halos hosting quasars at these redshifts, possibly reflecting the complexity in propagating a fast wind on large scales. 
Our current data suggest a combination of photoionization and resonant scattering as powering mechanisms of the 
\lya\ emission.
We discover the first $z\sim3$ ELAN, which confirms a very low probability ($\sim1\%$) 
of occurrence of such extreme systems at these cosmic epochs. 
Finally, we discuss the redshift evolution currently seen in extended \lya\ emission around radio-quiet quasars from $z\sim3$ to $z\sim2$, concluding that it is possibly linked to
a decrease of cool gas mass within the quasars' CGM from $z\sim3$ to $z\sim2$, and thus to the balance of cool vs hot media.
Overall, QSO MUSEUM opens the path to statistical and homogeneous surveys targeting the gas phases in 
quasars' halos along cosmic times.
\end{abstract}


%%%%%%%%%%%%%%%%%%%%%%%%%%%%%%%%%%%%%%%%%%%%%%%%%%

%%%%%%%%%%%%%%%%% BODY OF PAPER %%%%%%%%%%%%%%%%%%

\section{Introduction}
\label{sec:intro}

In the current paradigm of structure formation, most of the baryons at high redshift 
($z\gtrsim1.5$; \citealt{Meiksin2009} and references therein) are distributed in a web of diffuse filamentary structures in which 
galaxies form and evolve. The complex interplay between this rich reservoir of gas and the galaxies 
themselves is still a matter of investigation,  e.g. importance and strength of feedback from active galactic nuclei (AGN), existence of 
a cold mode of accretion onto galaxies, astrophysics of galactic outflows, build up of super-massive black holes in short timescales, angular momentum evolution 
(e.g., \citealt{Dekel2009,Brooks2009,DiMatteo2012,Shen2013,Dubois2013,Woods2014,Feng2014,AnglesAlcazar2014,Nelson2016, Stewart2016, Obreja2018}). 
Intergalactic and circumgalactic large-scale structures thus encode fundamental information to test
our current galaxy evolution theories.

So far, whether one focuses on large scales, i.e. on the intergalactic medium (IGM), or on smaller scales, 
ie. on the hundreds of kiloparsecs close to galaxies often referred to as the circumgalactic medium (CGM), 
the strongest constraints on the physical properties of these diffuse gas phases are obtained 
by analyzing absorption features along background sightlines (e.g., \citealt{Croft2002, Bergeron2002, Hennawi2006, QPQ2, Tumlinson2011, Farina2013b, 
Rudie2013, Turner2014, Farina2014, QPQ5, Lee2014}).
Direct imaging of the same gas phases greatly complement these absorption studies, allowing 
a spatial, morphological, physical, and kinematical characterisation, which is simply not possible 
with sparse one-dimensional information inherent to the absorption technique.       
Yet, predicted to be diffuse, the IGM and CGM is expected to be hard to detect in emission 
(SB$_{\rm Ly\alpha}\sim10^{-20}$~erg~s$^{-1}$~cm$^{-2}$~arcsec$^{-2}$; e.g., \citealt{Lowenthal1990, GW96, Bunker1998, Rauch2008}). 

Nevertheless, recently, the deployment of new advanced integral field spectrographs on 10-m class telescopes, 
i.e. the Multi-Unit Spectroscopic Explorer (MUSE; \citealt{Bacon2010}) 
on the ESO/VLT and the Keck Cosmic Web Imager (KCWI; \citealt{Morrissey2012}), opened up the possibility of 
routinely performing such an experiment by targeting very low levels of surface brightness (SB). Indeed, the latest observations 
are able to directly 
studying in emission at least the CGM of galaxies with reasonable observational times ($\sim$tens of hours).
In particular, these observations usually show extended \lya\ emission on scales of tens of 
kiloparsecs around the targeted $z\gtrsim1.7$ galaxies (\citealt{Wisotzki2016, Leclercq2017}), 
opening up a new parameter space for the study of the CGM gas phases, and ultimately
of galaxy evolution.

Even before the advent of the current new instrumentations, 
it has been shown that such extended \lya\ emission on CGM scales is more easily detected 
around high-redshift quasars or active galactic nuclei (AGN).
In this case, the \lya\ emission was detected around most of the objects ($50-70$\%) 
on $R<50$~kpc (e.g., \citealt{HuCowie1987,heckman91a,heckman91b,Moller2000b,Weidinger05,Christensen2006,North2012,qpq4,fab+16}), 
extending up to $R>100$~kpc in few rare bright cases known as Enormous Lyman-Alpha Nebulae (ELAN; \citealt{cantalupo14, hennawi+15,Cai2016, fab+2018,Cai2018}). 
The ELAN are indeed structures characterized by high observed surface brightnesses (SB$_{\rm Ly\alpha}> 10^{-17}$~erg~s$^{-1}$~cm$^{-2}$~arcsec$^{-2}$)
spanning continuously out to hundreds of kiloparsecs, and thus resulting in luminosities $L_{\rm Ly\alpha}> 10^{44}$~erg~s$^{-1}$ (\citealt{Cai2016}). 
The comparison between all these pioneering observations however is hampered by (i) the heterogeneity of the technique used 
(longslit spectroscopy, narrow-band imaging, integral-field spectroscopy), 
(ii) by the lower and different sensitivities inherent to the previous observational instruments 
(SB$_{\rm limit}> {\rm few}\times 10^{-18}$~\unitcgssb for some works and SB$_{\rm limit}\gtrsim 10^{-17}$~\unitcgssb for others), 
(iii) by uncertainties in 
the redshift of the targeted quasars, and (iv) by the 
difficulties in achieving a clean removal of the unresolved emission from the quasar, which
can easily outshine the faint large-scale emission (e.g., \citealt{Moller2000a}).

The advent of the aforementioned new sensitive IFU intruments together with the discoveries of 
the ELAN have motivated intense research on quasar halos, and have resulted in frequent new clear detections 
of the cool CGM gas. 
Overall, mainly due to the new sensitivities achieved ($<10^{-18}$~\unitcgssb), 
recent studies now routinely show detections on
$R\sim50$~kpc around $z\gtrsim3$ quasars (e.g., \citealt{Husband2015, Borisova2016, Fumagalli2016, Ginolfi2018}), 
i.e. the redshift range for which \lya\ is visible with MUSE. 
In very rare cases these newly discovered \lya\ nebulosities at $z\sim3$ match the observed SB and extent of the ELAN previously unveiled 
at $z\sim2$ (\citealt{fab+2018}).

In the presence of a quasar, the detected \lya\ emission on halo scales has been usually explained
as (i) recombination radiation produced after quasar photoionization ({\it a.k.a.} fluorescence; e.g.,\citealt{Weidinger04, Weidinger05, qpq4, cantalupo14, fab+15b, Borisova2016}), 
and/or (ii) \lya\ photons resonantly scattered within the gas distribution surrounding the quasar (e.g., \citealt{qpq4, cantalupo14, Borisova2016}), 
and/or (iii) recombination radiation produced after photoionization by several sources, e.g. quasar and active companions 
(e.g., \citealt{Husband2015, fab+2018, Husemann2018}). 
A clear determination of the contribution from the different mechanisms cannot be easily achieved by using only the information
enclosed in the \lya\ emission. Indeed \lya\ photons within the CGM gas are likely affected by resonant scattering (e.g., \citealt{Dijkstra2017,Gronke2017} and references therein),
which could lead to strong modification of the \lya\ spectral shape (e.g., \citealt{Dijkstra2017} and references therein), 
and the relative strength of the \lya\ line with respect to other diagnostics (e.g., \citealt{Neufeld_1990}). 
Notwithstanding these challenges, observations of the extended \lya\ emission -- together with the aforementioned absorption studies (e.g., \citealt{qpq3,QPQ5,QPQ7,qpq9}) -- 
currently paint a scenario in which quasar's halos are hosting a large reservoir of cool ($T\sim10^4$~K) gas. 
This gas is possibly tracing a complex set of astrophysical processes: gas/substructures infalling onto the central quasar (e.g., \citealt{Hu1991,Weidinger04,fab+2018}); 
strong turbulences or outflows (e.g., \citealt{Ginolfi2018}); interactions between substructures (e.g., \citealt{Hu1991,Husband2015,fab+2018,Husemann2018}); large-scale filaments (\citealt{cantalupo14}). 

Further, in the case of ELANe there are evidences that the \lya\ emitting gas on hundreds of kpc is composed by a population of 
cool and dense (volume density $n_{\rm H}> 1$~cm$^{-3}$) clumps (\citealt{cantalupo14}).
Indeed, if one assumes an ELAN to be powered by the radiation from the associated brightest quasar, the high  levels of \lya\ emission
together with the current stringent limits on the \ion{He}{ii}/\lya\ ratio can be matched by photoionization models only if 
very high densities ($n_{\rm H}\gtrsim3$~cm$^{-3}$; thus low ionization parameters log$U\lesssim-2$), and low column densities ($N_{\rm H}\lesssim10^{20}$~cm$^{-2}$) 
are used (\citealt{fab+15b,fab+2018}).
The same framework thus requires the clumps to have compact sizes $R \equiv N_{\rm H}/n_{\rm H}\lesssim20$~pc (\citealt{fab+15b,hennawi+15,fab+2018}). 
Current simulations are not able to achieve the resolutions needed to resolve such clumps in the CGM of galaxies, predicting low densities 
($n_{\rm H}\sim10^{-2}-10^{-3}$~cm$^{-3}$) for such a medium (see discussions in \citealt{cantalupo14,hennawi+15}).
This tension between observations and simulations motivated new research on hydrodinamical instabilities (e.g., \citealt{Mandelker2016}). 
In particular, very high resolution idealised hydrodinamical (\citealt{McCourt2018}) and magneto-hydrodinamical (\citealt{Ji2018}) simulations have 
shown that a mist of cool gas clouds could form and survive in the halo of galaxies, possibly 
explaining the high densities required by the observed levels of \lya\ emission around quasars.

Also, the effects of the quasar activity (e.g. radiation, outflows) on the diffuse gas phases and on the production of 
\lya\ photons is not fully 
understood, and could lead to effects on the morphology and physical properties of the surrounding gas distribution.
Extreme examples of what a powerful AGN can do are the high-redshift radio 
galaxies (HzRGs), whose  
powerfull radio jet
and UV radiation clearly alter the surrounding gas (e.g., \citealt{Nesvadba2017,Silva2018}). 
These objects show extended \lya\ emission with 
active kinematics (FWHM$>1000$~km~s$^{-1}$) associated with the radio emission, but a more 
extended (100 kpc scales) quiescent (FWHM$<700$~km~s$^{-1}$) gas phase consistent 
with gravitational motions (e.g., \citealt{vanOjik1997, VillarMartin2002, Humphrey2007, VillarM2007}), and/or 
large-scale structures (\citealt{Vernet2017}). 
This quiescent large-scale gas phase is similar to what is seen around quasars (\citealt{Husband2015, Borisova2016, fab+2018}).
Understanding the effects of AGN disruption mechanisms is fundamental given that these 
are invoked in cosmological simulations to modify the gas properties 
around and within massive galaxies to match observational constraints, 
e.g. star formation, halo mass versus stellar mass relation
(e.g., \citealt{Silk1998, Sijacki2007, Booth2009, Richardson2016}).

In this context, we have started to survey the $z\sim3$ quasar population 
with the main aim to characterise (i) the physical properties of the CGM/IGM in emission associated with such expected massive dark-matter halos 
(M$_{\rm DM}\sim10^{12.5}$~M$_{\odot}$, \citealt{white12,Trainor2012}), and (ii) the
frequency of detection of ELAN [${\rm SB}_{\rm Ly\alpha}\gtrsim 10^{-17}$\unitcgssb\ out to 100~kpc].
In this paper we focus on presenting the observations at the \lya\ transition for the first 61 targeted quasars under the acronym QSO MUSEUM 
({\it Q}uasar {\it S}napshot {\it O}bservations with {\it MU}se: 
{\it S}earch for {\it E}xtended {\it U}ltraviolet e{\it M}ission).
This work is part of an on-going multi-technique and multi-wavelength effort to unveil any dependence on the nature of each system (e.g., geometry, environment,
radio activity, luminosity) in the detection and properties of extended gaseous structures.

This work is structured as follows. In Section~\ref{sec:obs}, we describe our observations and data reduction.
In Section~\ref{sec:analysis} we explain our analysis procedures, and the current uncertainties on the
systemic redshifts of the quasars in our sample. 
In Section~\ref{sec:results}, we present the observational results. In particular, 
we reveal the discovery of extended \lya\ emisssion around our targeted quasars, and 
show their diverse morphologies, individual radial profiles, stacked profiles, 
average covering factor for the \lya\ emission, compact line emitters associated 
with the targeted systems, kinematics, and spectral shape of the \lya\ emission. 
In Section~\ref{sec:disc} we discuss our results in light 
of the current statistics for extended \lya\ emission around quasars, and the usually invoked
powering mechanisms. Finally, Section~\ref{sec:summ} summarises our findings.  

Throughout this paper, we adopt the cosmological parameters $H_0=70$~km~s$^{-1}$~Mpc$^{-1}$, $\Omega_M =0.3$ 
and $\Omega_{\Lambda}=0.7$. In this cosmology, 1\arcsec\ corresponds to about 7.6 physical kpc at $z=3.17$ (median redshift for our sample).
All magnitudes are in the AB system (\citealt{Oke1974}), and all distances are proper, unless otherwise specified.

\section{Observations and data reduction}
\label{sec:obs}

The survey QSO MUSEUM ({\it Q}uasar {\it S}napshot 
{\it O}bservations with {\it MU}se: {\it S}earch for {\it E}xtended {\it U}ltraviolet e{\it M}ission)
has been designed to target the population of $z\sim3$ quasars with the aim of 
(i) uncovering additional ELANe, similar to \citet{cantalupo14} and \citet{hennawi+15},  
(ii)  conducting a statistical census to determine the frequency of the ELAN phenomenon, (iii) studying the size, luminosity, 
covering factor of the extended \lya\ emission, and any relationship with quasar properties (e.g., luminosity, radio activity), 
(iv) looking for any evolutionary trend by comparing this sample with the $z\sim2$ quasar population (e.g., \citealt{fab+16}).
We have selected the targets from the last edition (13th) of the catalogue by \citet{VeronCetty2010}, and SDSS-DR12 Quasar catalogue (\citealt{Paris2017}), 
starting from the brightest quasars with $z<3.4$
not targeted by the MUSE Guaranteed Time Observation (GTO) team (i.e. \citealt{Borisova2016}). 
In addition, to facilitate detection of the expected low surface brightness features, 
we require our targeted fields (i) to have low extinction, i.e. $A_{V}<0.2$~mag, computed from the maps of \citet{sfd98}\footnote{We are aware of 
new extinction maps obtained by \citet{Schlafly2011}. However, here we decided to rely on \citet{sfd98} as their values are on average $14\%$ higher for our sample, 
thus making our selection more conservative.}, and (ii) 
to be devoid of stars brighter than 13 mag. 
Also, we gave higher priority for observations to the less crowded fields. For this purpose, we have visually inspected the density of sources 
using the optical catalogues available at the time of selection, in particular, SDSS-DR12 (\citealt{sdssdr12}), 
DSS2 (\citealt{Lasker1996}) and USNO-A2/B1 (\citealt{Monet1998, Monet2003}) catalogs.

To conduct the survey QSO MUSEUM, we have been awarded so far a total of 111 hours, of which $68\%$ have been executed.
The current resulting sample comprises 61 quasars\footnote{Additional four fields have been observed during our programmes, i.e. Q~1346+001, Q~0106-4137, Q~0029-3857, Q~0153-3951, but these sources happen to be
stars and not quasars as listed in the last edition (13th) of the catalogue by \citet{VeronCetty2010}.} 
with $i$-band magnitude in the range $17.4 < i < 19.0$ (median 18.02), or absolute $i$-band mangnitude normalized at $z=2$ (\citealt{Ross2013}) 
in the range $ -29.67 \le M_i(z=2) \le -27.03$ (or absolute magnitude at rest-frame 1450~\AA\ in the range $-28.29 \le M_{1450} \le -25.65$), 
and the redshift spanning $3.03 < z < 3.46$ (median 3.17).
Our survey thus expands on the work by \citet{Borisova2016}, both in number of targeted sources (19 vs 61) and by encompassing 
fainter sources\footnote{The faintest target in \citet{Borisova2016} has $M_i(z=2)= -28.23$.}. In Table~\ref{tab:sample} we summarize the 
information for our obtained sample. Of the 61 quasars targeted, 15 (or 25\%) are detected in radio, and fullfill 
the most used radio-loudness criteria $R=f_{\nu, 5 {\rm GHz}}/f_{\nu, 4400\AA}>10$ (\citealt{Kellermann1989})\footnote{The fraction
of radio-loud objects in our sample is thus larger than the value for the overall quasar population ($\sim10-20\%$; \citealt{Kellermann1989,Ivezic2002}). 
We initially aimed at building a characteristic sample of the quasar population, but our project has not been fully completed.}.
Some of these radio-loud objects have been covered by the Faint Images of the Radio Sky at Twenty-centimeters (FIRST) survey (\citealt{Becker1994})
with a resolution of 5\arcsec. This resolution is sufficient to start comparing the location/morphology of extended \lya\ emission and 
radio emission (see Section~\ref{sec:powering} and Appendix~\ref{app:FIRST}).

The observations for QSO MUSEUM have been acquired in service-mode under the ESO programmes 
094.A-0585(A), 095.A-0615(A/B), and 096.A-0937(B) with 
the Multi Unit Spectroscopic Explorer (MUSE; \citealt{Bacon2010}) on the VLT 8.2m telescope YEPUN (UT4) during ``dark time''.
All the observations consisted of exposures of 900~s each, rotated by 90 degrees with respect to each other.
The survey has been designed to get a total exposure time on source per object of 45 minutes.
However, SDSS~J0817+1053 and SDSS~J1342+1702 have only two usable exposures. In addition, we integrated longer on 
UM~683 (180 min) and UM~672 (90 minutes) motivated
by the extended \lya\ emission already clearly detected in the preliminary reduction using the v1.0 MUSE pipeline (\citealt{Weilbacher2014}) 
at the time of first data acquisition. For the same reason, the quasar PKS~1017+109 has been 
observed for additional 4.5 hours as presented in \citet{fab+2018}.

The observations were carried out in weather conditions classified by ESO as clear (CL; $45\%$ of the objects), photometric (PH; $27\%$),  
clear with high-wind (CL-WI; $18\%$), photometric with high humidity (PH-WI; $5\%$), photometric with high-wind (PH-WI; $3\%$), 
and thin (TN; $2\%$). The average (median) seeing of these observations is $1.07\arcsec$ ($0.98\arcsec$) (FWHM of a Moffat profile 
computed from the white-light image obtained by collapsing the final MUSE datacube for each target), which is great
given the any-seeing nature of our programmes. 
In Table~\ref{tab:ObsLog} we summarize the observation log for our survey. 

We have reduced the data using the MUSE pipeline recipes 
v1.4 (\citealt{Weilbacher2014}). Specifically, we have performed bias subtraction, flat fielding, twilight and illumination correction, 
and wavelength calibration using this software. In addition, the pipeline flux-calibrate each exposure using a spectrophotometric
standard star observed during the same night of each individual observing block. 
To improve the flat-fielding and to enable the detection of very low surface-brightness signals, 
we have performed a flat-fielding correction and subtract the sky with the procedures {\tt CubeFix} and {\tt CubeSharp} 
within the \textsc{cubExtractor} package (Cantalupo in prep., \citealt{Borisova2016}). 
We combined the individual exposures using an average $3\sigma$-clipping algorithm. To improve the 
removal of self-calibration effects, we applied a second time {\tt CubeFix} and {\tt CubeSharp}.
In this way, we are left with a final science datacube and a variance datacube. The latter has been
obtained by taking into account the propagation of errors for the MUSE pipeline and during the
combination of the different exposures. The variance is then rescaled by a constant factor to take into account the correlated noise
as done in \citet{Borisova2016}.
The final MUSE datacubes result in an average $2\sigma$ surface 
brightness limit of SB$_{\rm Ly\alpha}=8.8\times10^{-19}\cgssb$ (in 1 arcsec$^2$ aperture) 
in a single channel (1.25\AA) at the wavelength of the \lya\ line for the extended emission around each quasar.
The deepest narrow-band (NB) studies in the literature usually use NB filters with FWHM$\sim30$\AA.
For comparison, our datacubes have an average $2\sigma$ surface 
brightness limit of SB$_{\rm Ly\alpha}=4.2\times10^{-18}\cgssb$ (in 1 arcsec$^2$ aperture) in 
NB images of $30$\AA, centered at the \lya\ line for the extended emission around each quasar.
In Table~\ref{tab:ObsLog} we list these SB limits for each datacube.

\begin{table*}
\scriptsize
\caption{QSO MUSEUM: the sample.}
\centering
\begin{tabular}{lccccccccc}
\hline
\hline
ID  & Quasar	& $z_{\rm systemic}^{\rm a}$	& $z_{\rm peak\, QSO\, Ly\alpha}^{\rm b}$ & $z_{\rm peak\, Ly\alpha}^{\rm c}$  & $i^{\rm d}$ & $M_i(z=2)^{\rm e}$ & $M_{1450}$ & Radio Flux$^{\rm f}$ & $A_V^{\rm g}$ \\	
    &	        &                               &                                   &   & (mag)	& (mag) & (mag) & (mJy) & (mag) \\
\hline
1   & SDSS~J2319-1040    &  3.166  & 3.166 &  3.172   & 18.17 & -28.61 & -27.22 &<0.396 	 & 0.09 \\   
2   & UM~24	        &  3.133  &  3.161 & 3.163   & 17.08  & -29.67 & -28.29 &<0.396 	 & 0.10 \\   
3   & J~0525-233         &  3.110  & 3.114 &  3.123   & 17.71 & -29.04 & -27.64 &398.0* 	 & 0.10 \\   
4   & Q-0347-383         &  3.219  & 3.227 &  3.230   & 17.63 & -29.18 & -27.80 &<0.210*	 & 0.03 \\   
5   & SDSS~J0817+1053    &  3.320  & 3.323 &  3.336   & 18.08 & -28.82 & -27.43 &<0.381 	 & 0.08 \\   
6   & SDSS~J0947+1421    &  3.029  & 3.039 &  3.073   & 17.04 & -29.65 & -28.25 &<0.423 	 & 0.09 \\   
7   & SDSS~J1209+1138    &  3.117  & 3.117 &  3.126   & 17.63 & -29.12 & -27.73 &<0.432 	 & 0.08 \\   
8   & UM683	        &  3.132  &  3.125 & 3.132   & 18.59  & -28.17 & -26.78 &<0.450*	 & 0.09 \\   
9   & Q-0956+1217        &  3.301  & 3.306 &  3.316   & 17.48 & -29.40 & -28.01 &<0.414 	 & 0.11 \\   
10  & SDSS~J1025+0452    &  3.227  & 3.244 &  3.243   & 17.92 & -28.89 & -27.51 &<0.435 	 & 0.07 \\   
11  & Q-N1097.1          &  3.078  & 3.101 &  3.099   & 18.81 & -27.92 & -26.52 &<0.180*	 & 0.07 \\   
12  & SDSS~J1019+0254    &  3.376  & 3.384 &  3.395   & 18.26 & -28.66 & -27.29 &<0.432 	 & 0.12 \\   
13  & PKS-1017+109       &  3.164  & 3.166 &  3.167   & 18.13 & -28.65 & -27.26 &<0.438 	 & 0.10 \\   
14  & SDSS~J2100-0641    &  3.126  & 3.138 &  3.136   & 18.29 & -28.46 & -27.07 &<0.471 	 & 0.10 \\   
15  & SDSS~J1550+0537    &  3.141  & 3.143 &  3.147   & 18.08 & -28.68 & -27.29 &<1.458 	 & 0.19 \\   
16  & SDSS~J2348-1041    &  3.142  & 3.186 &  3.190   & 18.17 & -28.58 & -27.20 &<0.426 	 & 0.16 \\   
17  & SDSS~J0001-0956    &  3.340  & 3.349 &  3.348   & 18.56 & -28.35 & -26.97 &<0.447$^{\rm l}$ & 0.10 \\   
18  & SDSS~J1557+1540    &  3.265  & 3.276 &  3.288   & 18.46 & -28.38 & -27.00 &<0.429 	 & 0.13 \\   
19  & SDSS~J1307+1230    &  3.188  & 3.213 &  3.229   & 17.58 & -29.20 & -27.83 &<0.408 	 & 0.12 \\   
20  & SDSS~J1429-0145    &  3.395  & 3.419 &  3.425   & 17.90 & -29.02 & -27.66 &<0.786 	 & 0.08 \\   
21  & CT-669	         &  3.219  & 3.219 &  3.218   & 18.13 & -28.68 & -27.30 &<0.990*	 & 0.16 \\   
22  & Q-2139-4434        &  3.176  & 3.229 &  3.229   & 17.95 & -28.82 & -27.45 &---		 & 0.11 \\   
23  & Q-2138-4427        &  3.061  & 3.143 &  3.142   & 18.13 & -28.57 & -27.18 &---		 & 0.06 \\   
24  & SDSS~J1342+1702    &  3.062  & 3.055 &  3.053   & 18.55 & -28.17 & -26.76 &<0.414 	 & 0.06 \\   
25  & SDSS~J1337+0218    &  3.307  & 3.342 &  3.344   & 18.37 & -28.51 & -27.13 &<0.432 	 & 0.05 \\   
26  & Q-2204-408         &  3.181  & 3.178 &  3.179   & 17.17 & -29.62 & -28.23 &---		 & 0.07 \\   
27  & Q-2348-4025        &  3.318  & 3.332 &  3.334   & 17.98 & -28.91 & -27.53 &---		 & 0.04 \\   
28  & Q-0042-269         &  3.344  & 3.358 &  3.357   & 19.10 & -27.81 & -26.43 &<0.060*	 & 0.08 \\   
29  & Q-0115-30 	 &  3.180  & 3.221 &  3.221   & 17.95 & -28.83 & -27.45 &<0.540*	 & 0.04 \\   
30  & SDSS~J1427-0029	 &  3.357  & 3.359 &  3.354   & 18.27 & -28.67 & -27.27 &<0.513 	 & 0.05 \\   
31  & UM670		 &  3.131  & 3.204 &  3.203   & 17.64 & -29.10 & -27.73 &<0.450 	 & 0.14 \\   
32  & Q-0058-292	 &  3.069  & 3.098 &  3.101   & 18.53 & -28.19 & -26.79 &<3.570*	 & 0.12 \\   
33  & Q-0140-306	 &  3.125  & 3.133 &  3.132   & 18.33 & -28.43 & -27.03 &3.7*$^{\rm m}$  & 0.08 \\   
34  & Q-0057-3948	 &  3.223  & 3.242 &  3.251   & 19.49 & -27.32 & -25.94 &<0.150*	 & 0.05 \\   
35  & CTS-C22.31	 &  3.247  & 3.244 &  3.246   & 19.50 & -27.33 & -25.95 &---		 & 0.05 \\   
36  & Q-0052-3901A	 &  3.197  & 3.190 &  3.203   & 19.10 & -27.69 & -26.31 &27.3*$^{\rm n}$ & 0.06 \\   
37  & UM672		 &  3.130  & 3.128 &  3.127   & 18.52 & -28.24 & -26.85 &46.0*  	 & 0.04 \\   
38  & SDSS~J0125-1027	 &  3.348  & 3.351 &  3.319   & 18.24 & -28.68 & -27.29 &<0.438 	 & 0.04 \\   
39  & SDSS~J0100+2105	 &  3.100  & 3.096 &  3.097   & 18.15 & -28.59 & -27.19 &<2.22* 	 & 0.09 \\   
40  & SDSS~J0250-0757	 &  3.342  & 3.337 &  3.336   & 18.19 & -28.72 & -27.34 &<0.471 	 & 0.09 \\   
41  & SDSS~J0154-0730	 &  3.321  & 3.334 &  3.337   & 18.66 & -28.24 & -26.85 &<0.399 	 & 0.13 \\   
42  & SDSS~J0219-0215	 &  3.042  & 3.034 &  3.036   & 18.29 & -28.42 & -27.01 &<0.492 	 & 0.06 \\   
43  & CTSH22.05 	 &  3.087  & 3.124 &  3.127   & 18.66 & -28.06 & -26.67 &---		 & 0.08 \\   
44  & SDSS~J2321+1558	 &  3.197  & 3.233 &  3.241   & 18.29 & -28.50 & -27.12 &<1.770*	 & 0.08 \\   
45  & FBQS~J2334-0908	 &  3.312  & 3.359 &  3.361   & 17.96 & -28.93 & -27.54 &27.65  	 & 0.10 \\   
46  & Q2355+0108	 &  3.385  & 3.389 &  3.395   & 17.61 & -29.32 & -27.94 &<0.423 	 & 0.09 \\   
47  & 6dF~J0032-0414	 &  3.156  & 3.154 &  3.162   & 18.88 & -27.89 & -26.50 &34.55  	 & 0.08 \\   
48  & UM679		 &  3.197  & 3.204 &  3.215   & 18.77 & -28.02 & -26.64 &<1.020*	 & 0.14 \\   
49  & PKS0537-286	 &  3.109  & 3.139 &  3.141   & 18.93 & -27.81 & -26.42 &862.2* 	 & 0.09 \\   
50  & SDSS~J0819+0823	 &  3.197  & 3.213 &  3.205   & 18.33 & -28.46 & -27.08 &<0.423 	 & 0.08 \\   
51  & SDSS~J0814+1950	 &  3.138  & 3.133 &  3.137   & 18.74 & -28.02 & -26.63 &4.45		 & 0.08 \\   
52  & SDSS~J0827+0300	 &  3.144  & 3.137 &  3.137   & 18.09 & -28.68 & -27.29 &7.64		 & 0.12 \\   
53  & SDSS~J0905+0410	 &  3.149  & 3.152 &  3.165   & 19.73 & -27.03 & -25.65 &194.64 	 & 0.16 \\   
54  & S31013+20 	 &  3.109  & 3.108 &  3.108   & 19.19 & -27.56 & -26.16 &727.01 	 & 0.12 \\   
55  & SDSS~J1032+1206	 &  3.188  & 3.191 &  3.195   & 18.66 & -28.13 & -26.75 &2.29		 & 0.07 \\   
56  & TEX1033+137	 &  3.089  & 3.090 &  3.097   & 18.09 & -28.64 & -27.24 &102.52$^{\rm o}$& 0.10 \\   
57  & SDSS~J1057-0139	 &  3.406  & 3.453 &  3.452   & 18.40 & -28.51 & -27.16 &21.13  	 & 0.11 \\   
58  & Q1205-30  	 &  3.037  & 3.048 &  3.047   & 17.95 & -28.75 & -27.34 &<0.660*	 & 0.14 \\   
59  & LBQS1244+1129	 &  3.118  & 3.155 &  3.157   & 18.19 & -28.55 & -27.17 &<0.408 	 & 0.19 \\   
60  & SDSS~J1243+0720	 &  3.183  & 3.178 &  3.178   & 19.16 & -27.63 & -26.24 &2.29		 & 0.11 \\   
61  & LBQS1209+1524	 &  3.061  & 3.059 &  3.075   & 18.02 & -28.69 & -27.29 &<0.450 	 & 0.07 \\   
%Q-1346+001	  &  13:49:17.800	&  -00:07:03.00   &   &         & & 0.88 &0.1077 CL	 &  &  095.A-0615(A) \\ It's a star!!
\hline
\end{tabular}
\label{tab:sample}
\vspace{-0.2cm}
\flushleft{$^{\rm a}$ Quasar systemic redshift from peak of \civ line, correcting for the expected shift (\citealt{Shen2016}). The
intrinsic uncertainty on this correction is $\sim415$~km~s$^{-1}$ and dominates the error budget ($\Delta z\approx0.007$).}
\vspace{-0.2cm}
\flushleft{$^{\rm b}$ Redshift corresponding to the peak of the \lya\ emission in the observed spectrum of each quasar.}
\vspace{-0.2cm}
\flushleft{$^{\rm c}$ Redshift of the nebulosities estimated as flux-weighted centroid of the Ly$\alpha$ emission for a 
circular aperture with a diameter of $3\arcsec$, at the peak of the SB map of
each nebula. As we masked the $1\arcsec\times 1\arcsec$ region used to normalised the PSF, the peaks are always at $>1\arcsec$
projected distance from the quasar position.} 
\vspace{-0.2cm}
\flushleft{$^{\rm d}$ Extracted from our data using the SDSS filter transmission curve and a circular aperture with a diameter of $3\arcsec$. 
The average uncertainty is $\sim5\%$.}
\vspace{-0.2cm}
\flushleft{$^{\rm e}$ Absolute $i$-band magnitude normalized at $z=2$ following \citet{Ross2013}.}
\vspace{-0.2cm}
\flushleft{$^{\rm f}$ Integrated flux at 1.4~GHz from the FIRST survey for northern sources or peak flux at 1.4~GHz from the NRAO VLA Sky
Survey (NVSS; \citealt{Condon1998}) for the southern sources (indicated by *). The 6 sources with Dec$<-40$ are currently not covered by any radio survey. 
For the non-detections in each survey, and thus currently considered radio-quiet objects, 
we report the $3\times$ rms in mJy/beam from each catalogue. All the detected sources are radio-loud following the most used definition by \citealt{Kellermann1989}, 
i.e. $R=f_{\nu, 5 {\rm GHz}}/f_{\nu, 4400\AA}>10.$} 
\vspace{-0.2cm}
\flushleft{$^{\rm g}$ Extinction from \citet{sfd98}.} %FAB: beta parameter for the moffat is between 2.15 and 3.45 (see my paper for discussion on this...)
\vspace{-0.2cm}
\flushleft{$^{\rm l}$ This quasar has a radio source of 50.25~mJy (integrated flux) at $\approx13.5\arcsec$, but likely not physically associated 
(Appendix \ref{app:FIRST}).}
\vspace{-0.2cm}
\flushleft{$^{\rm m}$ The radio source is at $\approx8\arcsec$ from the quasar.} % (NVSS; \citealt{Condon1998}).}
\vspace{-0.2cm}
\flushleft{$^{\rm n}$ The radio source is at $\approx2.3\arcsec$ from the quasar.} % (NVSS; \citealt{Condon1998}).}
\vspace{-0.2cm}
\flushleft{$^{\rm o}$ This source has two lobes separated by $\approx8.5\arcsec$ (Appendix \ref{app:FIRST}). 
We report the sum of the integrated flux of the two lobes.} 
\end{table*}

%Removed because it's the same as Q2355+0108	
%28  & SDSS~J2358+0125    &  3.375  &  3.395  & 17.41 & -29.51 & <0.423          & 0.04 \\   

\begin{table*}
\scriptsize
\caption{Observation Log}
\centering
\setlength\tabcolsep{4pt}
\begin{tabular}{lcccccccccc}
\hline
\hline
ID  & Quasar           & RA	           & Dec             &  Seeing$^{\rm a}$ & Weather$^{\rm b}$ & Exp. T.$^{\rm c}$  & SB limit$_{\rm layer}^{\rm d}$ & SB limit$_{\rm 30\AA\, NB}^{\rm e}$ & UT date & ESO Prog. \\ 
    &	               & (J2000)           & (J2000)         & (arcsec)	& 	& (min) & ($10^{-19}$~cgs)& ($10^{-18}$~cgs)& (dd/mm/yyyy) & \\
\hline
1   & SDSS~J2319-1040  &  23:19:34.800	   &  -10:40:36.00   & 1.40 & CL-WI    & 45.	     &  8.2 & 4.1  & 14/10/2014 &094.A-0585(A) \\ %FAB: DONE! REDO PSF SUB - larger radius?
2   & UM~24	       &  00:15:27.400	   &  +06:40:12.00   & 1.62 & CL       & 45.	     &  8.8 & 4.2  & 14/10/2014 &094.A-0585(A) \\ %FAB: DONE! REDO PSF SUB - larger radius?
3   & J~0525-233       &  05:25:06.500	   &  -23:38:10.00   & 0.86 & CL       & 45.	     &  8.5 & 3.5  & 17/01/2015 &094.A-0585(A) \\
4   & Q-0347-383       &  03:49:43.700	   &  -38:10:31.00   & 1.40 & CL       & 45.	     &  7.6 & 3.7  & 25/01/2015 &094.A-0585(A) \\
5   & SDSS~J0817+1053  &  08:17:52.099	   &  +10:53:29.68   & 1.47 & CL       & 30.	     &  8.0 & 3.9  & 14/02/2015 &094.A-0585(A) \\ %there are cosmic rays as expected...remove them from NB image
6   & SDSS~J0947+1421  &  09:47:34.200	   &  +14:21:17.00   & 1.21 & CL       & 45.	     & 10.4 & 4.6  & 14/02/2015 &094.A-0585(A) \\ %big star nearby in NB: mask it! 
7   & SDSS~J1209+1138  &  12:09:18.000	   &  +11:38:31.00   & 1.22 & CL       & 45.	     &  8.5 & 4.2  & 14/02/2015 &094.A-0585(A) \\ %there are 2 nebulae...CHECK again if the second is real...
8   & UM~683	       &  03:36:26.900	   &  -20:19:39.00   & 1.05 & CL-PH    & 45.         &  5.0 & 2.0  & 18/02/2015 &094.A-0585(A) \\
    &     	       &         	   &                 &      & PH-WI    & 135.        &      &      & 14/08-21/09/2015 &095.A-0615(A) \\
9   & Q-0956+1217      &  09:58:52.200	   &  +12:02:45.00   & 1.12 & PH       & 45.	     &  8.4 & 4.2  & 18/02/2015 &094.A-0585(A) \\
10  & SDSS~J1025+0452  &  10:25:09.600	   &  +04:52:46.00   & 1.10 & PH       & 45.	     &  8.5 & 3.7  & 18/02/2015 &094.A-0585(A) \\
11  & Q-N1097.1	       &  02:46:34.200	   &  -30:04:55.00   & 1.18 & PH-HU    & 45.	     & 10.6 & 4.2  & 19/02/2015 &094.A-0585(A) \\
12  & SDSS~J1019+0254  &  10:19:08.255	   &  +02:54:31.94   & 1.22 & PH-HU    & 45.	     &  7.4 & 3.9  & 19/02/2015 &094.A-0585(A) \\
13  & PKS-1017+109     &  10:20:10.000	   &  +10:40:02.00   & 0.91 & PH-HU    & 45.	     &  8.2 & 4.2  & 19/02/2015 &094.A-0585(A) \\
14  & SDSS~J2100-0641  &  21:00:25.030     &  -06:41:45.00   & 0.85 & CL       & 45.	     & 10.5 & 4.9  & 11/06/2015 &095.A-0615(B) \\ %FAB: SOLVED!! there might be problems in removing three bright nearby sources
15  & SDSS~J1550+0537  &  15:50:36.806     &  +05:37:50.07   & 1.01 & CL-WI    & 45.	     & 10.2 & 4.1  & 17/08/2015 &095.A-0615(B) \\ %comment on the presence of the close-by galaxy (there is no nebular emission at that location)
16  & SDSS~J2348-1041  &  23:48:56.488     &  -10:41:31.17   & 1.72 & CL-WI    & 45.	     &  8.0 & 3.9  & 17/08/2015 &095.A-0615(B) \\
17  & SDSS~J0001-0956  &  00:01:44.886     &  -09:56:30.83   & 1.81 & PH-WI    & 45.         &  7.8 & 3.7  & 17/08/2015 &095.A-0615(B) \\
18  & SDSS~J1557+1540  &  15:57:43.300     &  +15:40:20.00   & 0.81 & CL       & 45.	     &  7.4 & 4.5  & 22/05/2015 &095.A-0615(A) \\ % rms calculation wrong...checked-->there is a bright star (select region for better calc)
19  & SDSS~J1307+1230  &  13:07:10.200     &  +12:30:21.00   & 0.89 & CL       & 45.	     &  8.3 & 3.8  & 11/06/2015 &095.A-0615(A) \\
20  & SDSS~J1429-0145  &  14:29:03.033     &  -01:45:19.00   & 0.79 & CL       & 45.	     &  7.7 & 3.5  & 11/06/2015 &095.A-0615(A) \\
21  & CT-669	       &  20:34:26.300	   &  -35:37:27.00   & 0.74 & CL       & 45.	     & 11.9 & 3.8  & 11/06/2015 &095.A-0615(A) \\
22  & Q-2139-4434      &  21:42:25.900     &  -44:20:18.00   & 1.02 & CL       & 45.         &  8.9 & 4.4  & 11/06/2015 &095.A-0615(A) \\
23  & Q-2138-4427      &  21:41:59.500     &  -44:13:26.00   & 0.81 & CL       & 45.         &  7.7 & 3.9  & 17/06/2015 &095.A-0615(A) \\
24  & SDSS~J1342+1702  &  13:42:33.200     &  +17:02:46.00   & 1.03 & CL       & 30.	     & 10.3 & 4.4  & 08/07/2015 &095.A-0615(A) \\
25  & SDSS~J1337+0218  &  13:37:57.900     &  +02:18:21.00   & 0.75 & CL-PH    & 45.         & 14.2 & 3.5  & 09/07/2015 &095.A-0615(A) \\ %check this nebula...companions? bright emission at lower redshift
26  & Q-2204-408       &  22:07:34.300     &  -40:36:57.00   & 0.84 & TN       & 45.         & 25.3 & 15.8 & 13/07/2015 &095.A-0615(A) \\ %FAB: SOLVED (worse depth than others)!! check this cube...qso stronger than expected...FAB: worse depth: there is small nebula and companion nebula
27  & Q-2348-4025      &  23:51:16.100     &  -40:08:36.00   & 0.87 & CL       & 45.         &  7.2 & 3.4  & 16/07/2015 &095.A-0615(A) \\
28  & Q-0042-269       &  00:44:52.300     &  -26:40:09.00   & 0.80 & CL       & 45.	     &  6.8 & 3.4  & 25/07/2015 &095.A-0615(A) \\ %check this nebula...bright narrow emission...also velocity map has problems
29  & Q-0115-30        &  01:17:34.000     &  -29:46:29.00   & 0.91 & CL       & 45.	     &  7.1 & 3.6  & 25/07/2015 &095.A-0615(A) \\
30  & SDSS~J1427-0029  &  14:27:55.800     &  -00:29:51.00   & 0.95 & CL       & 45.	     & 17.1 & 12.4 & 28/07/2015 &095.A-0615(A) \\ %(not very good data) %check this cube...seems no nebula % one star to be removed from NB (to do profile...)
31  & UM~670	       &  01:17:23.300     &  -08:41:32.00   & 0.80 & CL       & 45.	     &  7.5 & 3.6  & 28/07/2015 &095.A-0615(A) \\
32  & Q-0058-292       &  01:01:04.700     &  -28:58:03.00   & 1.42 & CL-WI    & 45.	     &  7.2 & 3.6  & 11/08/2015 &095.A-0615(A) \\
33  & Q-0140-306       &  01:42:54.700     &  -30:23:45.00   & 1.39 & CL-WI    & 45.	     &  8.3 & 3.5  & 11/08/2015 &095.A-0615(A) \\
34  & Q-0057-3948      &  00:59:53.200     &  -39:31:58.00   & 1.09 & CL-WI    & 45.	     &  9.2 & 3.3  & 11/08/2015 &095.A-0615(A) \\
35  & CTS-C22.31       &  02:04:35.500     &  -45:59:23.00   & 1.18 & CL-WI    & 45.	     &  7.2 & 3.6  & 11/08/2015 &095.A-0615(A) \\
36  & Q-0052-3901A     &  00:54:45.400     &  -38:44:15.00   & 1.25 & CL-WI    & 45.	     &  7.2 & 3.4  & 11/08/2015 &095.A-0615(A) \\
37  & UM672	       &  01:34:38.600     &  -19:32:06.00   & 1.87 & CL       & 45.	     &  7.2 & 2.6  & 07/09/2015 & 095.A-0615(A) \\ %FAB: DONE! REDO PSF subtraction -- larger radius?
    &     	       &         	   &                 &      & CL-WI    & 45.         &      &      & 04/11/2015 & 096.A-0937(B) \\ %FAB: DONE! REDO PSF subtraction -- larger radius?
38  & SDSS~J0125-1027  &  01:25:30.900     &  -10:27:39.00   & 0.74 & CL       & 45.	     &  7.3 & 3.7  & 15/09/2015 &095.A-0615(A) \\ %check this cube...bright narrow emission
39  & SDSS~J0100+2105  &  01:00:27.661     &  +21:05:41.57   & 0.73 & PH       & 45.	     &  9.5 & 4.4  & 17/09/2015 &095.A-0615(A) \\
40  & SDSS~J0250-0757  &  02:50:21.800     &  -07:57:50.00   & 0.67 & PH       & 45.	     &  7.5 & 3.6  & 17/09/2015 &095.A-0615(A) \\
41  & SDSS~J0154-0730  &  01:54:40.328     &  -07:30:31.85   & 0.75 & PH       & 45.	     &  6.7 & 3.5  & 21/09/2015 &095.A-0615(A) \\
42  & SDSS~J0219-0215  &  02:19:38.732     &  -02:15:40.47   & 0.64 & PH       & 45.	     &  9.3 & 4.2  & 21/09/2015 &095.A-0615(A) \\
43  & CTSH22.05        &  01:48:18.130     &  -53:27:02.00   & 1.34 & CL-WI    & 45.	     &  7.5 & 3.8  & 02/11/2015 &096.A-0937(B) \\
44  & SDSS~J2321+1558  &  23:21:54.980     &  +15:58:34.24   & 1.61 & CL       & 45.	     &  7.7 & 3.8  & 04/11/2015 &096.A-0937(B) \\
45  & FBQS~J2334-0908  &  23:34:46.400     &  -09:08:12.24   & 1.18 & CL       & 45.	     &  7.3 & 3.7  & 04/11/2015 &096.A-0937(B) \\
46  & Q2355+0108       &  23:58:08.540     &  +01:25:07.20   & 1.23 & CL       & 45.	     &  7.4 & 3.4  & 04/11/2015 &096.A-0937(B) \\
47  & 6dF~J0032-0414   &  00:32:05.380     &  -04:14:16.21   & 1.59 & CL       & 45.	     &  9.6 & 3.9  & 04/11/2015 &096.A-0937(B) \\
48  & UM~679	       &  02:51:48.060     &  -18:14:29.00   & 0.96 & PH       & 45.	     &  7.0 & 3.6  & 18/11/2015 &096.A-0937(B) \\
49  & PKS0537-286      &  05:39:54.267     &  -28:39:56.00   & 0.74 & PH       & 45.	     &  6.7 & 3.5  & 18/11/2015 &096.A-0937(B) \\
50  & SDSS~J0819+0823  &  08:19:40.580     &  +08:23:57.98   & 0.73 & PH       & 45.	     &  7.7 & 3.5  & 15/12/2015 &096.A-0937(B) \\
51  & SDSS~J0814+1950  &  08:14:53.449     &  +19:50:18.62   & 0.75 & PH       & 45.	     & 10.2 & 4.2  & 12/01/2016 &096.A-0937(B) \\ %FAB: DONE!  REDO PSF subtraction!! too much subtracted!
52  & SDSS~J0827+0300  &  08:27:21.968     &  +03:00:54.74   & 0.96 & CL       & 45.	     &  7.8 & 3.3  & 14/01/2016 &096.A-0937(B) \\
53  & SDSS~J0905+0410  &  09:05:49.058     &  +04:10:10.15   & 0.93 & CL       & 45.	     &  7.1 & 3.2  & 14/01/2016 &096.A-0937(B) \\
54  & S31013+20        &  10:16:44.319     &  +20:37:47.29   & 0.92 & PH       & 45.	     &  7.9 & 3.7  & 14/01/2016 &096.A-0937(B) \\
55  & SDSS~J1032+1206  &  10:32:12.886     &  +12:06:12.83   & 0.89 & CL-WI    & 45.	     &  7.7 & 3.7  & 16/01/2016 &096.A-0937(B) \\
56  & TEX1033+137      &  10:36:26.886     &  +13:26:51.75   & 0.70 & PH       & 45.	     &  9.0 & 4.2  & 17/01/2016 &096.A-0937(B) \\
57  & SDSS~J1057-0139  &  10:57:13.250     &  -01:39:13.79   & 1.31 & PH       & 45.	     & 11.2 & 5.4  & 19/01/2016 &096.A-0937(B) \\ %check this nebula...might be that the nebula is at lower z than expected (do dlower cut?)
58  & Q1205-30         &  12:08:12.730     &  -30:31:07.00   & 0.63 & PH       & 45.	     &  9.2 & 4.3  & 09/02/2016 &096.A-0937(B) \\
59  & LBQS1244+1129    &  12:46:40.370     &  +11:13:02.92   & 1.45 & PH       & 45.	     &  9.8 & 4.6  & 04/03/2016 &096.A-0937(B) \\
60  & SDSS~J1243+0720  &  12:43:53.960     &  +07:20:15.47   & 1.29 & PH       & 45.	     &  8.5 & 3.9  & 08/03/2016 &096.A-0937(B) \\
61  & LBQS1209+1524    &  12:12:32.040     &  +15:07:25.63   & 1.42 & PH       & 45.	     &  8.9 & 4.1  & 17/03/2016 &096.A-0937(B) \\
%Q-1346+001	  &  13:49:17.800	&  -00:07:03.00   &   &         & & 0.88 & CL      &  &  095.A-0615(A) \\ It's a star!!
\hline
\end{tabular}
\label{tab:ObsLog}
\flushleft{$^{\rm a}$ Seeing computed from the ``white-light'' image of the final MUSE datacubes as FWHM of a Moffat profile.} %FAB: beta parameter for the moffat is between 2.15 and 3.45 (see my paper for discussion on this...)
\flushleft{$^{\rm b}$ Sky conditions during observations as from the ESO observations log: PH-photometric; CL-clear; WI-windy; TN-thin cirrus; HU-humid}
\flushleft{$^{\rm c}$ Exposure time on source (i.e. without overheads).}
\flushleft{$^{\rm d}$ $2\sigma$ SB limit in 1 arcsec$^{2}$ computed for the layer (1.25\AA) at the redshift of the Ly$\alpha$ emission for each target. The cgs units are erg~s$^{-1}$~cm$^{-2}$~arcsec$^{-2}$.}
\flushleft{$^{\rm e}$ $2\sigma$ SB limit in 1 arcsec$^{2}$ computed for the $30$\AA\ NB image centered at the redshift of the Ly$\alpha$ emission for each target. The cgs units are erg~s$^{-1}$~cm$^{-2}$~arcsec$^{-2}$.}
\end{table*}

%Removed because it's the same as Q2355+0108
%28  & SDSS~J2358+0125  &  23:58:08.000     &  +01:25:07.00   & 0.89 & CL       & 45.	     & 10.8 & 4.9  & 16/07/2015 &095.A-0615(A) \\

%\flushleft{$^{\rm b}$ Absolute $i$-band magnitude normalized at $z=2$ following \citet{Ross2013}.}
%$M_i(z=2)^{\rm b}	
% -28.61
% -29.67
% -29.04
% -29.18
% -28.82
% -29.65
% -29.12
% -28.17
% -29.40
% -28.89
% -27.92
% -28.66
% -28.65
% -28.46
% -28.68
% -28.58
% -28.35
% -28.38
% -29.20
% -29.02
% -28.68
% -28.82
% -28.57
% -28.17
% -28.51
% -29.62
% -28.91
% -29.51
% -27.81
% -28.83
% -28.67
% -29.10
% -28.19
% -28.43
% -27.32
% -27.33
% -27.69
% -28.24
% -28.68
% -28.59
% -28.72
% -28.24
% -28.42
% -28.06
% -28.50
% -28.93
% -29.32
% -27.89
% -28.02
% -27.81
% -28.46
% -28.02
% -28.68
% -27.03
% -27.56
% -28.13
% -28.64
% -28.51
% -28.75
% -28.55
% -27.63
% -28.69

\section{Detecting large-scale emission associated with a quasar}
\label{sec:analysis}

As many works in the literature have illustrated (e.g., \citealt{HuCowie1987,Moller2000a,heckman91a,bergeron99,Husemann2013,qpq4,fab+16}), 
the detection of extended emission around a quasar inevitably requires
that one removes the unresolved emission from the quasar itself (continuum and line emission) and the contribution
of any other continuum sources.
To perform these operations and to extract the extended emission from each of the MUSE datacubes, 
we rely on the same procedures described in \citet{Borisova2016}, 
which we briefly summarize in the following section, while leaving the presentation of the results to section~\ref{sec:results}. 
We use this approach to allow for a one-to-one comparison with that work, and to facilitate the creation of a 
homogeneous sample.

\subsection{Empirical point-spread-function subtraction, continuum source subtraction, and extraction of extended emission}
\label{sec:cubex}

The point spread function (PSF) of the quasar has been subtracted with the 
{\tt CubePSFSub} method within \textsc{CubExtractor} (Cantalupo in prep.).
The algorithm obtains the PSF of the quasar as pseudo-NB images of 
user-defined wavelength ranges within the MUSE datacube.
For each wavelength range, the empirical PSF is then rescaled to the flux
within $5\times5$ pixels ($1\arcsec \times 1\arcsec$) around the quasar position and
subtracted from a circular region with a radius of about five times the
seeing. Here we use the same wavelength ranges as in \citet{Borisova2016} to construct the pseudo-NB images, i.e. 150 
spectral pixels ($187.5$~\AA).
This method is unsuitable for the study of emission within the region used for the
PSF rescaling, as data show complex residuals on these scales ($\sim10$~kpc). 
In particular, this method assumes the quasar to greatly outshine the host galaxy,
whose contribution in the central region is thus considered to be negligible.

Subsequently, we have applied the {\tt CubeBKGSub} algorithm within \textsc{CubExtractor} 
to remove any continuum 
source in the MUSE datacube. This procedure is based on a fast median-filtering approach described in 
\citet{Borisova2016}, and allows us to avoid any contamination from continuum sources in the 
search for extended line emission associated with each quasar.

We then run \textsc{CubExtractor} on each of the PSF and continuum subtracted cubes to select, if any, 
extended \lya\ emission. Specifically, we only use the portion of the cubes at about $\pm11000$~km~s$^{-1}$ ($\pm150$ layers) 
around the quasar systemic redshift. 
As \citet{Borisova2016} found \lya\ nebulae with 
a minimum ``volume'' of $10000$ connected voxels (volume pixels) above a signal-to-noise ratio ${\rm S/N} > 2$
around all of their targets, 
we first use such selection criteria to identify the \lya\ extended emission around our sample. However, we find that 
some of our targets are not characterized by such large volumes. For this reason, we have then decreased the threshold 
on the number of connected voxels in steps of $1000$ to search for detections. All our targets have a detected \lya\ nebula 
greater than 1000 voxels. 
We list in Table~\ref{Tab:LyaNeb} the number 
of connected voxels for each of them. 
These numbers should be considered with caution as, by construction, the extraction process is strongly dependent 
on the correct determination of the noise in MUSE datacubes. 
In this respect, we rely on the variance cube produced by our pipeline. This cube is propagated during the reduction steps 
and then rescaled to take into account the correlated noise (see details in \citealt{Borisova2016}).

\textsc{CubExtractor} generates a three-dimensional (3D) mask that includes all the connected voxels 
satisfing the selection criteria, and which can be used to compute several diagnostic images. 
First, for each quasar, the 3D mask is used to construct 
an ``optimally extracted'' NB image for the \lya\ emission 
by summing the flux along the wavelength direction for only the voxels selected by this 3D mask. 
Following \citet{Borisova2016}, a layer of the MUSE cube corresponding to the peak of the \lya\ emission 
has been added as ``background'' layer to each ``optimally extracted'' NB image.
In other words, as the projected field-of-view described by the 3D mask does not cover the whole MUSE
field-of-view, we replaced the empty pixels with values from the respective pixels in the layer at the peak of the
\lya\ emission.

In addition, the 3D masks -- and thus only the voxels selected -- are also used to compute the first and second moments (in velocity space) of the flux distribution 
at each spatial location around each quasar. Finally, we have also constructed NB images of $30$~\AA\ by collapsing the MUSE cubes 
centered at the \lya\ peak of each nebulosity.   
Altough all this analysis approach can be used to extract and present any line-emission
covered by the MUSE datacubes for our sample (e.g., \ion{C}{iv}, \ion{He}{ii}), in this work we focus only
on the \lya\ emission and we leave to a future paper (Arrigoni Battaia et al., in prep.) the analysis of 
additional extended line diagnostics.

\subsection{The quasars' systemic redshift}
\label{sec:systemic_redshift}

Having an accurate systemic redshift for the targeted quasars is a first fundamental step towards the 
understanding of the geometry and kinematics of each system (e.g., \citealt{Weidinger05}). In particular, a precise determination of the 
systemic redshift would allow (i) to say if the detected extended line emission is physically related to the studied quasars, 
or if it is a structure at larger distances in the foreground or background; and (ii) 
to compare the kinematic informations available from observations to simulation studies.
However, such precise estimates of the systemic redshift of $z>2$ quasars are difficult to determine (e.g., \citealt{Gaskell1982,Tytler1992,Hewett2010}), 
and requires expensive campaigns of follow-up observations in the near-IR (e.g., \citealt{McIntosh1999}) or molecular tracers 
to bring the uncertainties down to the order of few tens of km~s$^{-1}$ (e.g., \citealt{Venemans2017}).

For our sample, at the moment we can only rely on the sytemic redshift estimated from the \civ line after correcting 
for the known luminosity-dependent blueshift of this line with respect to systemic (\citealt{Richards2002,Richards2011,Shen2016}). 
We have estimated the peak of the \civ line for each of the quasars from our MUSE sample, as specified in \citet{Shen2016}, and
added to the estimated redshift the luminosity dependent velocity shift determined following the relation in \citet{Shen2016}.
In Table~~\ref{tab:sample} we list such systemic redshifts for our targeted quasars. 
With an intrinsic uncertainty of $415$~km~s$^{-1}$ for the empirical calibration in 
\citet{Shen2016}, these estimates are not optimal for a detailed studied of our systems in which we expect 
velocities of the order of hundreds of km~s$^{-1}$ 
to be in place (e.g., \citealt{Goerdt2015,Borisova2016,fab+2018}).  
Future efforts are surely needed to better constrain 
this fundamental parameter. We have indeed initiated a survey in molecular tracers to obtain the systemic redshift of these quasars
down to few tens of km~s$^{-1}$ uncertainties.

\section{Results: diverse specimens of extended Ly$\alpha$ emission}
\label{sec:results}

In this section we present the results of our QSO MUSEUM survey concerning the detection of 
extended \lya\ emission around the targeted quasars. We stress once again that our data do cover
the wavelengths of additional emission lines and continuum diagnostics of which we will present the data in dedicated papers.

\subsection{The morphology of the extended Ly$\alpha$ emission}
\label{sec:morph}

\begin{figure*}
       \includegraphics[width=0.85\textwidth]{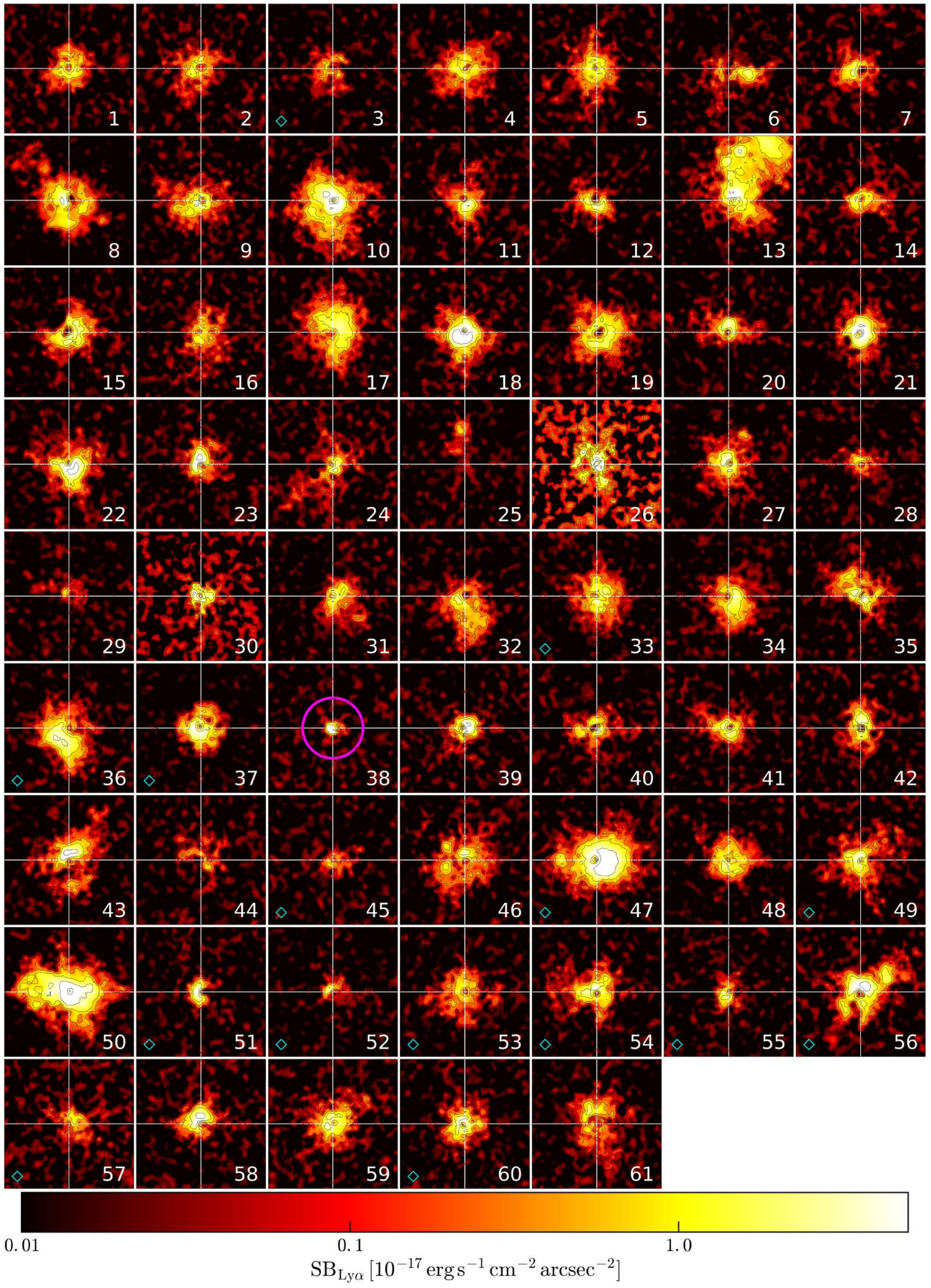}
    \caption{Atlas of the 61 ``optimally extracted'' NB images for the \lya\ emission around the quasars in the QSO MUSEUM sample. 
    Each image shows the SB maps of $30\arcsec\times30\arcsec$ (or about $230$~kpc~$\times\, 230$~kpc) centered on each quasar after PSF and continuum subtraction 
    (see Section~\ref{sec:cubex} for details). 
    Each of the system is numbered following Table~\ref{tab:sample}.
    In each image the white crosshair indicates the position of the quasar prior to PSF subtraction. For reference, we overlay a magenta circle with 
    a radius of 50~kpc on the image with ID~38. To highlight the significance of the detected emission, 
    for each SB map we indicate the contours for ${\rm S/N}=2,\, 4,\, 10,\, 20$, and $50$.
    A cyan diamond in the bottom-left corner indicates when the quasar is radio-loud.
    The detected extended \lya\ emission shows a diversity of extents, geometries and substructures. North is up, east is to the left.}
    \label{fig:SBmaps}
\end{figure*}

In Figure~\ref{fig:SBmaps} we show the atlas of the $61$ ``optimally extracted'' NB images for the 
\lya\ nebulae extracted from the PSF and continuum subtracted datacubes following the procedure mentioned in section~\ref{sec:cubex}.   
In this atlas, each image represents $30\arcsec\, \times\, 30\arcsec$ (or about $230$~kpc~$\times\, 230$~kpc) around each 
quasar, whose position before PSF subtraction is indicated by the white crosshair.
For simplicity, we identify each target by the index from Table~\ref{tab:sample}, and we report this number in the 
bottom-right corner of each image.

As it is not feasible to visually assess the noise level in these ``optimally extracted'' images, we overlay the 
S/N contours computed from the propagated variance and taking into account the number of spaxels contributing to each spatial pixel.
Specifically, we show the contours for ${\rm S/N}=2,\, 4,\, 10,\, 20$, and $50$.

Qualitatively, Figure~\ref{fig:SBmaps} shows extended \lya\ emission around each system however with different extents, 
geometries, and substructures. 
All the objects are characterized by \lya\ emission on radii $R\lesssim50$~kpc (magenta circle on ID~38 in Figure~\ref{fig:SBmaps}), 
which appears to be mostly symmetric on such scales. 
The average maximum projected distance at which we detect \lya\ emission from the targeted quasar down 
to the $2\sigma$ SB limit is 81~kpc or 10.7\arcsec (see Table~\ref{Tab:LyaNeb}).
Objects with asymmetric extended \lya\ emission appear to be very rare in our sample. Indeed, we discover only $1$ ELAN (around the quasar with ID~13 or
PKS~1017+109; \citealt{fab+2018}).

\begin{figure*}
       \includegraphics[width=0.95\textwidth]{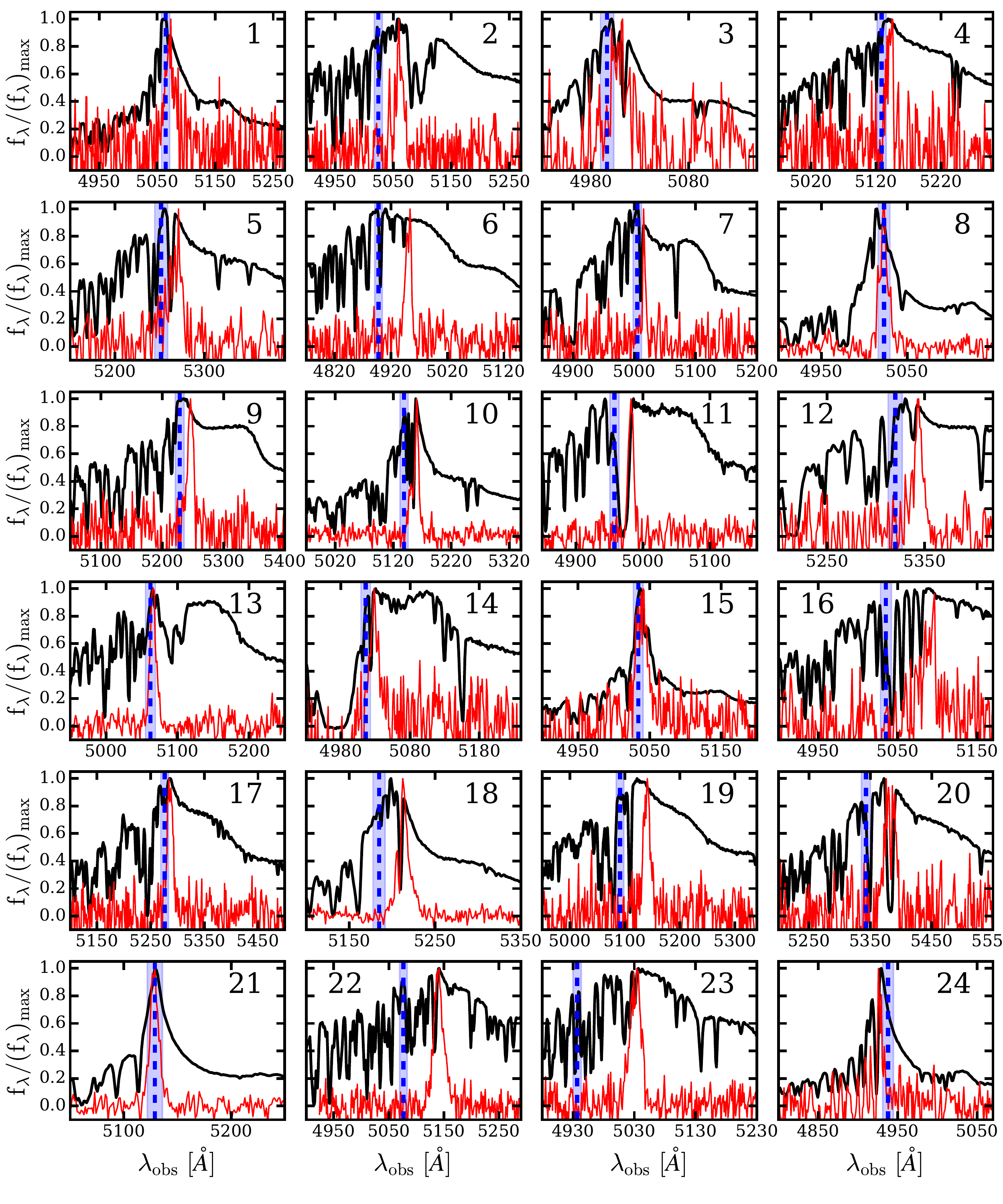}
    \caption{1D spectra for the quasars in our sample (black) and discovered nebulosities (red) at the location of the \lya\ emission, in the 
    observed frame. Each spectrum has been extracted in a circular aperture of $1.5$~arcsec radius, and it is normalized to its peak in this wavelength range. 
    The dashed blue lines indicate the systemic redshift of the quasars as listed in Table~\ref{tab:sample}, 
    and the blue shaded regions indicate the calibration uncertainty on this estimates ($415$~km~s$^{-1}$). Following Table~\ref{tab:sample}, we report the quasars' ID numbers 
    in each panel. The nebulosities show a much narrower line profile than quasars, and appear to be shifted from the 
    current estimate for the quasars' systemic redshift.}
    \label{fig:1DspecComparison}
\end{figure*}

In all our targets, we are confident that the extended \lya\ emission is not due to a systematic effect in underestimating the 
PSF of each quasar (see also appendix~\ref{app:PSFsubtraction}). Indeed, in each case, the extended \lya\ emission is characterized by a much narrower line profile
than the broad \lya\ emission of the quasar itself. We show these differences in Figure~\ref{fig:1DspecComparison}, 
where we compare the normalized \lya\ line features for each quasar (black) and surrounding nebulosity (red).
The quasar and nebulosity spectra reported in these figures have been extracted in circular apertures with $1.5$ arcsec radius,
and centered at the peak of the unresolved quasar or at the peak of the extended \lya\ emission, respectively. 

From Figure~\ref{fig:1DspecComparison},  
it is also evident that some of the 
extended nebulosities appear to have a shift in redshift with respect to the current uncertain estimate for the quasar systemic. 
Specifically, we have estimated the redshift for the nebulosities as the location of the flux-weighted centroid of the \lya\ emission in these 1D spectra.
In Figure~\ref{fig:histoDeltaV} we show the histogram\footnote{We use bins with the same size as the intrinsic uncertainty on the estimate of the 
quasars redshift, i.e. 415~km~s$^{-1}$} of such velocity shifts between the quasar systemics and the nebulosities, which 
cover the range $-2000\, {\rm km\, s^{-1}}\, \leq \Delta v \leq 5980$~ km~s$^{-1}$, 
with the quasars with ID~38 (or SDSSJ0125-1027) and ID~23 (or Q-2138-4427) having the 
bluest and the reddest shift, respectively. Overall, 75\% of the nebulosities have their peaks at positive velocity 
shifts with respect to the systemic of the quasar, 
making the median value of such shift $\Delta v_{\rm median} = 782$~km~s$^{-1}$. 

Additionally, if we instead compute the velocity shift between the nebulosities and 
the observed peak of the \lya\ emission of the quasars themselves ($z_{\rm peak\, QSO\, Ly\alpha}$), we find definitely smaller shifts (orange histogram in Figure~\ref{fig:histoDeltaV}). 
Such velocity displacements cover 
the range $-2205\, {\rm km\, s^{-1}}\, \leq \Delta v \leq 2524$~ km~s$^{-1}$
with the quasars with ID~38 (or SDSSJ0125-1027) and ID~6 (or SDSSJ0947+1421) being two outliers with the 
bluest and the reddest shift, respectively. The median value is $\Delta v_{\rm median} = 144$~km~s$^{-1}$. 
Thus, intriguingly, even though the peak of the \lya\ line of the quasars is highly affected by the presence 
of multiple absorption features, it appears close in wavelength to the \lya\ nebulosities. 
This occurrence highlight the important of a precise determination of the quasar systemic, and might be related to the
physical mechanism producing the \lya\ emission within the nebulosities, and/or to the overall kinematics of the observed gas 
(see discussion in section~\ref{sec:powering}).
We report all the estimates of the redshifts used for this plot in Table~\ref{tab:sample}\footnote{Our calculation of the quasar systemic redshifts gives 
consistent results with works in the literature (e.g., \citealt{Weidinger05}; see Appendix~\ref{app:individual_fields}).}. 

We compared the nebulae here discovered with the work by \citet{Borisova2016}, finding an overall agreement with our results.
\citet{Borisova2016} also found mainly spatially symmetric nebulosities and only two of their targets (ID 1 and 3 in their work) show asymmetric 
extended nebulosities.
Overall, this result is consistent with the expectations from previous works in the literature which showed that 50-70\% of the targeted quasars
have extended \lya\ emission in their surroundings on radii $<50$~kpc (see Section~\ref{sec:intro}). 
It is important to stress that previous works targeting $z\sim3$ quasars were achieving SB limits of the order of $\sim10^{-17}$~\unitcgssb\ evaluated 
for equivalent $60$~\AA\ NB filters 
(e.g. \citealt{heckman91a, Christensen2006}), 
which is much poorer (at least one order of magnitude) than any recent MUSE data, including our work.

Regarding the velocity shift between the \lya\ nebulosities and the quasars systemic, the values reported in \citet{Borisova2016} (their Table~2) 
show that all nebulosities have positive velocity shifts in the range $959\, {\rm km\, s^{-1}}\, \leq \Delta v \leq 4011$~ km~s$^{-1}$, and with a median value of
$\Delta v_{\rm median} = 1821$~km~s$^{-1}$ \footnote{The quasars' systemic redshift in \citet{Borisova2016} is obtained as in this work. Thus, it is similarly uncertain.}. 
However, while looking at their Figure~2 it appears that the peak of the \lya\ emission from the nebulosities is 
closer to the peak of the \lya\ emission of the quasars themselves, being sometimes also blueshifted from the quasar's systemic (e.g. their ID 14, 16). 
The closer association between the nebulosities and the peak of the \lya\ emission of the quasars themselves seems thus in agreement with our analysis.\\

\begin{table*}
\scriptsize
\caption{QSO MUSEUM: properties of the extended Ly$\alpha$ emission.}
\centering
\begin{tabular}{lccccccccc}
\hline
\hline
ID & Quasar	& Luminosity	& Area	& $\Delta \lambda^{\rm a}$	& Max. Distance$^{\rm b}$ & $\alpha^{\rm c}$ & $d_{\rm QSO-Neb}^{\rm d}$ & $\phi_{\rm NE}^{\rm e}$ & Voxels$^{\rm f}$ \\	
	& & ($10^{43}$~erg~s$^{-1}$) & (arcsec$^{2}$) & (\AA) & (arcsec / kpc) & & (arcsec / kpc) & (degree) & \\
\hline
1  & SDSS~J2319-1040   &  3.17    &  102.4    & 36.2   &  7.37 / 55.8  & 0.80  & 0.29 / 2.2 &  -38.8 & 12059 \\
2  & UM~24	       &  2.57    &  156.7    & 32.5   & 10.32 / 78.2  & 0.77  & 1.17 / 9.0 &  -30.4 & 10812 \\
3  & J~0525-233        &  1.17    &  79.5     & 30.0   &  9.21 / 70.1  & 0.84  & 1.11 / 8.4 &  -1.7  &  6406 \\
4  & Q-0347-383        &  4.11    &  177.5    & 27.5   & 12.12 / 91.2  & 0.80  & 0.40 / 3.0 &  -60.0 & 16760 \\
5  & SDSS~J0817+1053   &  4.99    &  147.3    & 36.2   & 12.98 / 96.6  & 0.80  & 1.43 / 10.7 & -22.9 & 17726  \\
6  & SDSS~J0947+1421   &  2.04    &  86.2     & 33.8   & 11.71 / 89.5  & 0.34  & 3.14 / 24.1 & -83.8 &  8936 \\
7  & SDSS~J1209+1138   &  2.32    &  93.0     & 32.5   &  7.97 / 60.6  & 0.90  & 1.71 / 13.0 & -27.8 &  8793 \\
8  & UM683	       &  7.92    &  274.0    & 43.8   & 19.09 / 145.1 & 0.64  & 3.83 / 29.1 & 23.9  & 44094  \\
9  & Q-0956+1217       &  5.34    &  158.4    & 50.0   & 12.50 / 93.2  & 0.73  & 1.78 / 13.3 & -80.0 & 16921  \\
10 & SDSS~J1025+0452   &  18.66   &  288.9    & 51.2   & 14.97 / 112.5 & 0.78  & 1.37 / 10.3 & 19.2  & 50282  \\
11 & Q-N1097.1         &  2.96    &  103.1    & 26.2   & 10.83 / 82.6  & 0.70  & 0.39 / 3.0 &  16.8  & 10313 \\
12 & SDSS~J1019+0254   &  3.09    &  76.7     & 33.8   & 10.35 / 76.6  & 0.45  & 1.12 / 8.3 &  63.4  &  9643\\
13 & PKS-1017+109      &  24.87   &  466.6    & 47.5   & 27.28 / 206.6 & 0.49  & 8.50 / 64.4 & -28.7 & 74865  \\
14 & SDSS~J2100-0641   &  2.14    &  63.0     & 48.8   &  8.63 / 65.6  & 0.60  & 0.84 / 6.4 &  -82.6 & 13645 \\
15 & SDSS~J1550+0537   &  5.63    &  133.6    & 35.0   & 10.35 / 78.5  & 0.61  & 0.61 / 4.7 &  -41.9 & 19547 \\
16 & SDSS~J2348-1041   &  2.28    &  115.6    & 32.5   &  8.53 / 64.5  & 0.71  & 0.99 / 7.5 &  -8.8  & 11316 \\
17 & SDSS~J0001-0956   &  10.80   &  243.1    & 40.0   & 13.11 / 97.5  & 0.76  & 1.52 / 11.3 & 19.6  & 34978  \\
18 & SDSS~J1557+1540   &  21.83   &  161.5    & 52.5   & 11.33 / 84.8  & 0.84  & 1.10 / 8.3 &  68.9  & 32057 \\
19 & SDSS~J1307+1230   &  6.98    &  189.4    & 41.2   & 12.15 / 91.5  & 0.87  & 1.12 / 8.4 &  -45.3 & 28534 \\
20 & SDSS~J1429-0145   &  3.55    &  90.6     & 41.2   &  8.61 / 63.5  & 0.65  & 1.60 / 11.8 & 84.1  & 13666  \\
21 & CT-669	       &  10.95   &  129.7    & 41.2   &  9.29 / 70.0  & 0.76  & 0.74 / 5.6 &  4.0   & 20077 \\
22 & Q-2139-4434       &  10.36   &  172.5    & 56.2   & 11.14 / 83.8  & 0.88  & 0.68 / 5.1 &  -50.3 & 26745 \\
23 & Q-2138-4427       &  4.65    &  91.3     & 38.8   &  8.47 / 64.3  & 0.82  & 1.52 / 11.6 & -9.1  & 13164  \\
24 & SDSS~J1342+1702   &  2.41    &  134.5    & 23.8   & 15.47 / 118.5 & 0.40  & 2.31 / 17.7 & -44.3 &  6820 \\
25 & SDSS~J1337+0218   &  0.52    &  36.7     & 27.5   & 12.41 / 92.3  & 0.37  & 6.91 / 51.6 & -11.8 &  3235 \\
26 & Q-2204-408        &  1.10    &  12.5     & 15.0   &  3.40 / 25.7  & 0.36  & 0.84 / 6.3 &  22.5  &  1192 \\
27 & Q-2348-4025       &  5.13    &  149.4    & 40.0   &  9.44 / 70.3  & 0.79  & 1.14 / 8.5 &  -23.6 & 16359 \\
28 & Q-0042-269        &  0.96    &  53.7     & 22.5   &  7.94 / 59.0  & 0.65  & 0.77 / 5.7 &  69.0  &  3925 \\
29 & Q-0115-30         &  0.33    &  28.9     & 22.5   &  5.73 / 43.1  & 0.65  & 1.65 / 12.5 & 79.5  &  2580 \\
30 & SDSS~J1427-0029   &  4.11    &  34.6     & 17.5   &  5.26 / 39.1  & 0.71  & 0.28 / 2.0 &  76.3  &  2676 \\
31 & UM670	       &  2.79    &  115.6    & 30.0   & 10.04 / 75.7  & 0.75  & 2.14 / 16.3 & -5.2  & 11133 \\
32 & Q-0058-292        &  3.77    &  161.6    & 40.0   & 11.82 / 90.2  & 0.60  & 3.11 / 23.8 & 31.2  & 19912  \\
33 & Q-0140-306        &  5.10    &  174.3    & 37.5   & 11.74 / 89.3  & 0.73  & 0.58 / 4.4 &  -3.8  & 19124 \\
34 & Q-0057-3948       &  5.06    &  160.2    & 42.5   & 10.84 / 81.4  & 0.69  & 1.56 / 11.7 & -0.3  & 19711  \\
35 & CTS-C22.31        &  5.01    &  141.7    & 33.8   & 11.71 / 88.0  & 0.53  & 1.83 / 13.7 & 54.9  & 18334  \\
36 & Q-0052-3901A      &  9.81    &  198.5    & 40.0   & 12.31 / 92.9  & 0.63  & 2.56 / 19.3 & 40.3  & 30792  \\
37 & UM672	       &  6.51    &  117.9    & 47.5   &  7.92 / 60.2  & 0.95  & 1.22 / 9.3 &  -82.6 & 23973  \\
38 & SDSS~J0125-1027   &  2.25    &  26.0     & 21.2   &  6.65 / 49.6  & 0.98  & 0.45 / 3.4 &  61.3  &  3000 \\
39 & SDSS~J0100+2105   &  5.02    &  59.8     & 38.8   &  6.42 / 49.0  & 0.79  & 0.38 / 2.9 &  -41.4 &  8969 \\
40 & SDSS~J0250-0757   &  2.88    &  103.9    & 38.8   & 11.58 / 86.2  & 0.68  & 0.57 / 4.3 &  -78.2 & 10602 \\
41 & SDSS~J0154-0730   &  3.27    &  97.2     & 42.5   &  9.89 / 73.6  & 0.62  & 0.92 / 6.9 &  68.5  & 13112 \\
42 & SDSS~J0219-0215   &  3.51    &  101.7    & 35.0   &  8.92 / 68.5  & 0.53  & 0.92 / 7.0 &  0.3   & 10631 \\
43 & CTSH22.05         &  7.36    &  205.2    & 47.5   & 15.60 / 118.6 & 0.75  & 1.51 / 11.6 & -53.0 & 26134  \\
44 & SDSS~J2321+1558   &  0.89    &  69.2     & 21.2   &  9.78 / 73.5  & 0.68  & 1.33 / 10.0 & 37.0  &  5248 \\
45 & FBQS~J2334-0908   &  0.67    &  39.1     & 25.0   &  8.67 / 64.4  & 0.68  & 1.11 / 8.3 &  -83.0 &  3382 \\
46 & Q2355+0108        &  5.37    &  208.4    & 42.5   & 12.63 / 93.5  & 0.75  & 0.85 / 6.3 &  -49.5 & 24127 \\
47 & 6dF~J0032-0414    &  35.73   &  305.0    & 47.5   & 15.28 / 115.8 & 0.90  & 1.09 / 8.3 &  -84.0 & 60203 \\
48 & UM679	       &  4.49    &  123.1    & 37.5   & 10.70 / 80.6  & 0.91  & 0.72 / 5.4 &  -40.4 & 18594 \\
49 & PKS0537-286       &  4.63    &  170.0    & 38.8   & 10.85 / 82.4  & 0.81  & 0.29 / 2.2 &  -26.1 & 20509 \\
50 & SDSS~J0819+0823   &  38.70   &  342.7    & 58.8   & 17.20 / 129.8 & 0.56  & 0.56 / 4.2 &  74.4  & 73149 \\
51 & SDSS~J0814+1950   &  2.58    &  33.8     & 21.2   &  6.24 / 47.4  & 0.48  & 0.52 / 3.9 &  -2.0  &  4995 \\
52 & SDSS~J0827+0300   &  0.84    &  43.5     & 21.2   &  5.53 / 42.0  & 0.72  & 0.85 / 6.5 &  81.6  &  2277 \\
53 & SDSS~J0905+0410   &  2.79    &  131.8    & 41.2   & 12.01 / 91.0  & 0.80  & 1.23 / 9.3 &  -12.7 & 15633 \\
54 & S31013+20         &  5.63    &  163.0    & 47.5   & 10.74 / 81.8  & 0.70  & 1.10 / 8.4 &  -75.1 & 21659 \\
55 & SDSS~J1032+1206   &  1.54    &  45.6     & 32.5   &  5.84 / 44.1  & 0.70  & 0.81 / 6.2 &  -1.3  &  5293 \\
56 & TEX1033+137       &  12.67   &  202.4    & 37.5   & 14.16 / 108.0 & 0.68  & 1.37 / 10.4 & -56.3 & 31894  \\
57 & SDSS~J1057-0139   &  1.70    &  61.8     & 33.8   &  8.92 / 65.7  & 0.64  & 1.28 / 9.5 &  34.1  &  6488 \\
58 & Q1205-30	       &  5.78    &  86.8     & 40.0   &  8.92 / 68.4  & 0.64  & 1.88 / 14.4 & -53.2 & 12046  \\
59 & LBQS1244+1129     &  5.21    &  138.1    & 48.8   & 10.62 / 80.6  & 0.73  & 0.83 / 6.3 &  -48.4 & 17614  \\
60 & SDSS~J1243+0720   &  5.07    &  107.4    & 45.0   &  8.66 / 65.5  & 0.86  & 0.23 / 1.7 &  -49.1 & 14714 \\
61 & LBQS1209+1524     &  3.21    &  156.9    & 43.8   & 13.67 / 104.5 & 0.60  & 0.26 / 2.0 &  8.7   & 15957  \\
\hline
\end{tabular}
\label{Tab:LyaNeb}
\flushleft{$^{\rm a}$ Maximum spectral width of the ``3-dimensional mask'' extracted above the $2\sigma$ threshold (see Section~\ref{sec:cubex})}
\flushleft{$^{\rm b}$ Maximum distance from the quasar spanned by the Ly$\alpha$ emission within the $2\sigma$ isophote.}
\flushleft{$^{\rm c}$ Ratio between the semiminor axis $b$ and semimajor axis $a$ obtained from the second order moments of the \lya\ emission within the $2\sigma$ isophote.}
\flushleft{$^{\rm d}$ Distance between the position of the quasar and the flux-weighted centroid of the \lya\ emission within the $2\sigma$ isophote.}
\flushleft{$^{\rm e}$ Position angle (East of North) of the major axis $a$ obtained from the second order moments of the \lya\ emission within the $2\sigma$ isophote.}
\flushleft{$^{\rm f}$ Number of connected voxels above the threshold of S/N=2 used for the extraction of the nebulosities.}
\end{table*}
	
%SDSS~J2358+0125	  &  6.64    &  159.9	 & 53.8   & 12.53 / 92.7  \\

\begin{figure*}
       \includegraphics[width=0.95\textwidth]{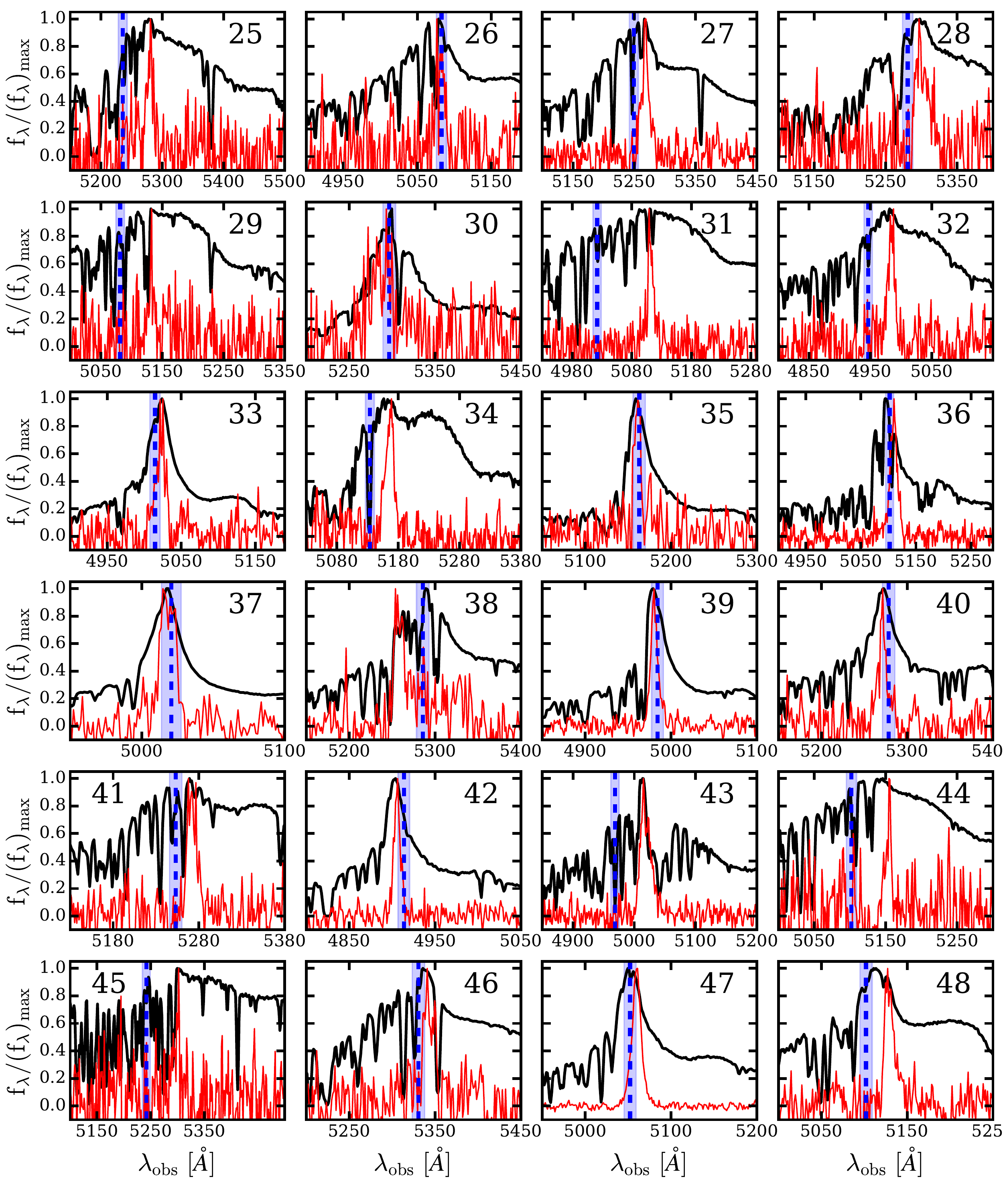}
    \smallskip
    \centerline{{\bf Figure 2} -- {\it continued}}
    \smallskip
\end{figure*}

\begin{figure*}
       \includegraphics[width=0.95\textwidth]{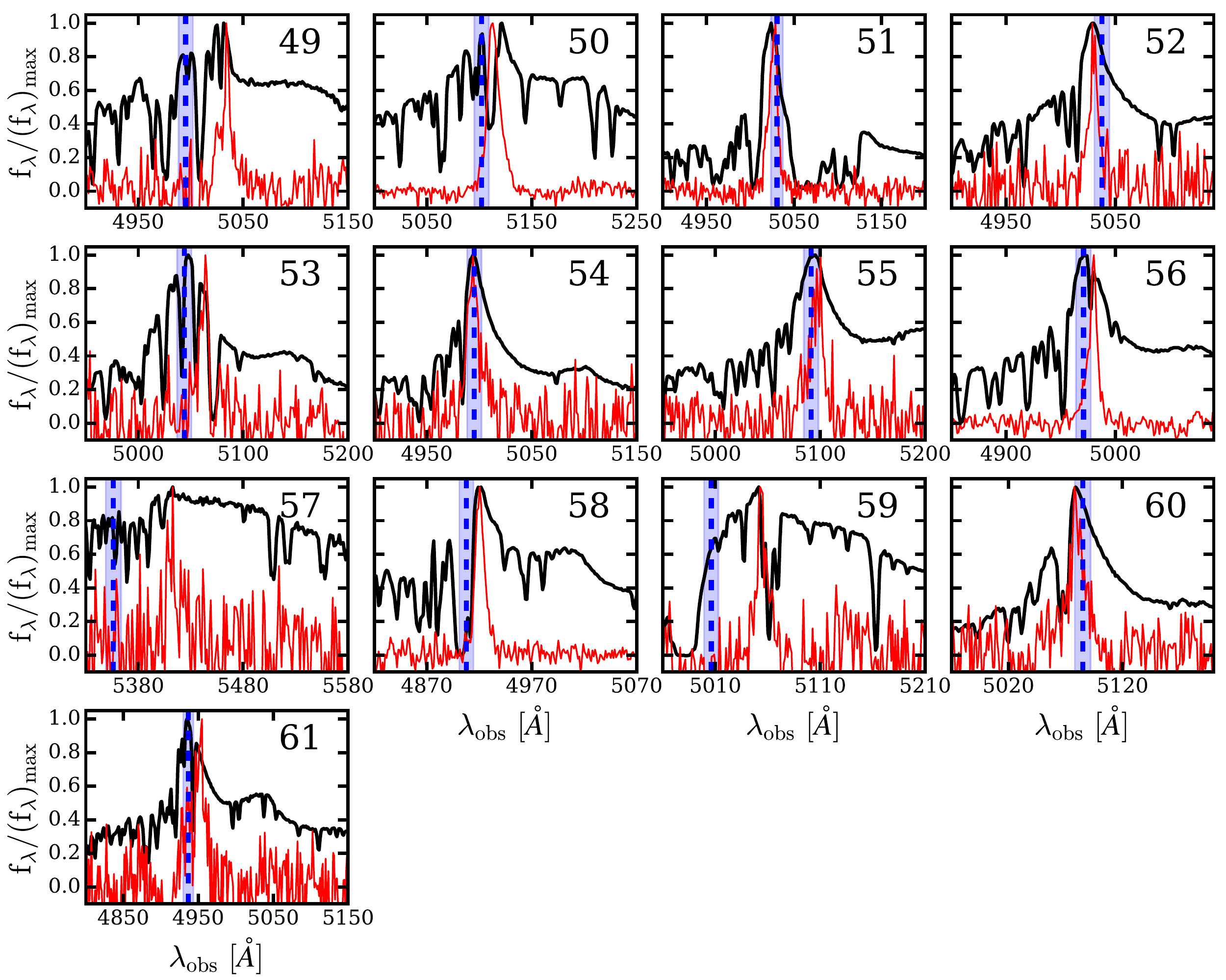}
    \smallskip
    \centerline{{\bf Figure 2} -- {\it continued}}
    \smallskip
\end{figure*}

\begin{figure}
       \includegraphics[width=0.95\columnwidth]{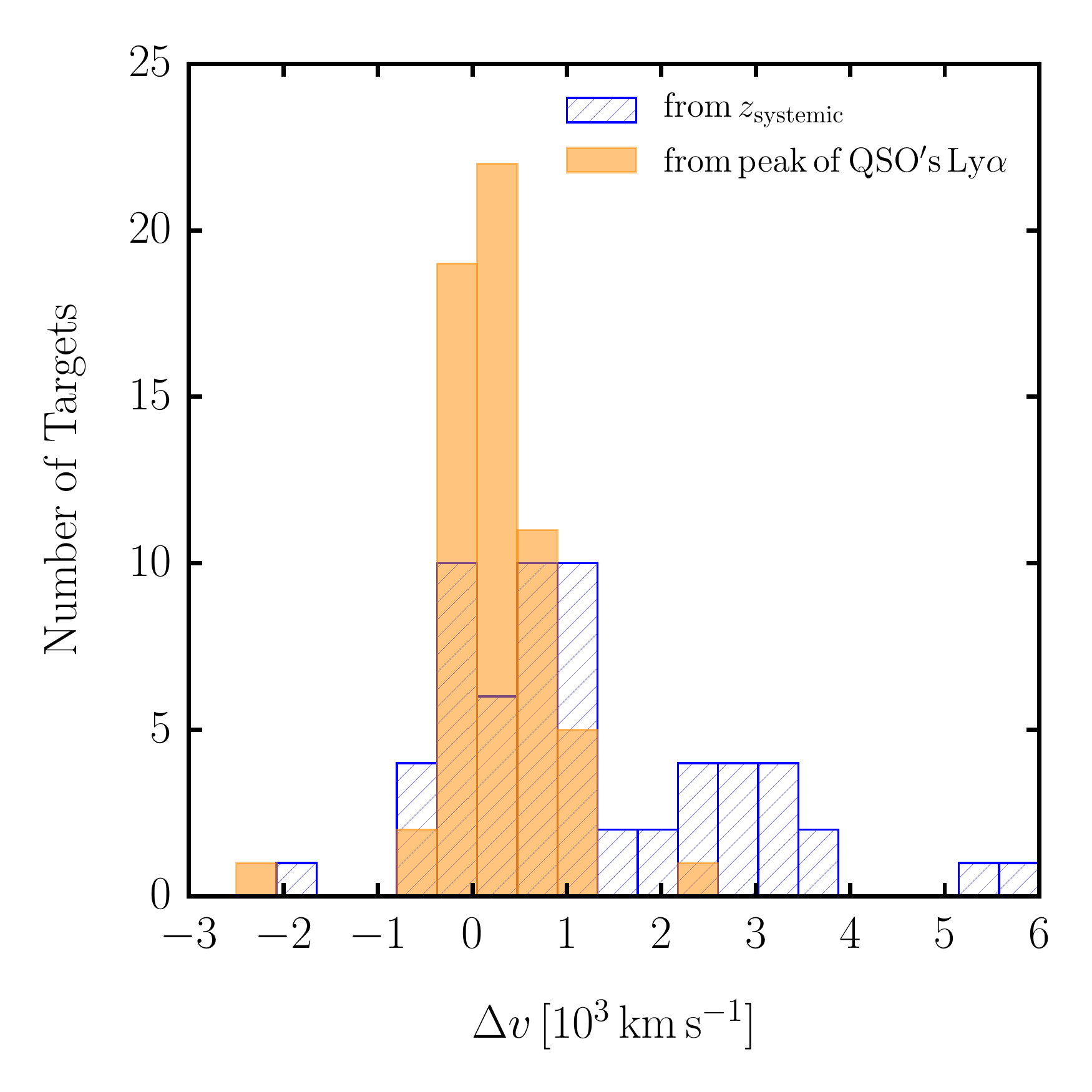}
    \caption{Histogram of the velocity shifts between the flux-weighted centroid of the \lya\ emission peak of the nebulosities $z_{\rm peak\, Ly\alpha}$ 
    and the systemic redshift of
    the quasars $z_{\rm systemic}$ (blue), both listed in Table~\ref{tab:sample}. We compare this histogram with the histogram of the 
    velocity shifts between $z_{\rm peak\, Ly\alpha}$ and 
    the \lya\ emission peak of the quasar itself $z_{\rm peak\, QSO\, Ly\alpha}$ (orange, see Table~\ref{tab:sample}). 
    The \lya\ nebulosities appear to be closer to $z_{\rm peak\, QSO\, Ly\alpha}$ than to the current (uncertain) $z_{\rm systemic}$ of the quasars.}
    \label{fig:histoDeltaV}
\end{figure}

\subsubsection{Quantifying the spatial asymmetry of the emission}
\label{sec:asymm}

A better understanding of the discovered \lya\ nebulae requires quantitative informations also on their morphology and on the level of asymmetry of the structures above our current
detection limits. Indeed, 
these informations should be used to infer the physical properties, geometry, and scales of the
\lya\ emitting gas both observationally and in comparison to current cosmological simulations post-processed 
with radiative transfer codes (e.g., \citealt{Weidinger05,cantalupo14}).
We have thus quantified the asymmetry of the \lya\ light distribution in the discovered nebulosities by estimating (i) the 
distance between the quasar and the flux-weighted centroid of the emission within the $2\sigma$ isophote of each nebulosity, $d_{\rm QSO-Neb}$; (ii) the asymmetry of 
the emission within the $2\sigma$ isophote using the second order moments of each nebulosity's light.

In particular, after estimating the flux-weighted centroid for each nebulosity $(x_{\rm Neb}, y_{\rm Neb})$ within the $2\sigma$ isophote
in the 2D image shown in Figure~\ref{fig:SBmaps}\footnote{Also for this analysis the $1\arcsec \times 1\arcsec$ region around each quasar, and used for the PSF rescaling, has been masked.}, 
we have measured the flux-weighted second order moments following for example \citet{stoughton02}.
These moments are defined as

\begin{eqnarray}
M_{xx}&\equiv&\langle\frac{(x-x_{\rm Neb})^2}{r^2}\rangle_{f}, \ \ \ M_{yy}\equiv\langle\frac{(y-y_{\rm Neb})^2}{r^2}\rangle_{f},\\ \nonumber
M_{xy}&\equiv&\langle\frac{(x-x_{\rm Neb})(y-y_{\rm Neb})}{r^2}\rangle_{f},
\end{eqnarray}
where $r$ is the distance of a point $(x,y)$ from the flux-weigthed centroid.
Using the second order moments one can define the so-called ``Stokes parameters''

\begin{equation}
Q\equiv M_{xx} - M_{yy}, \ \ \ U\equiv 2M_{xy},
\end{equation} 
and derive the asymmetry (or the ratio between the semiminor and semimajor axis $b$ and $a$), 
and the position angle  
describing the light distribution:

\begin{equation}
\centering
\alpha = b/a = \frac{(1-\sqrt{Q^2+U^2})}{1+\sqrt{Q^2+U^2}}, \ \ \ \phi = \arctan\left(\frac{U}{Q}\right).
\end{equation}

In particular, the angle $\phi$ is the angle between the semimajor axis and the closest $x$- or $y$-axis.
In Table~\ref{Tab:LyaNeb} we list the obtained parameters following this formalism: distance between each quasar and the center of the nebulosity $d_{\rm QSO-Neb}$, 
the asymmetry $\alpha$, and the angle $\phi$ after converting it to the more common angle East of North.

\begin{figure}
       \includegraphics[width=0.95\columnwidth]{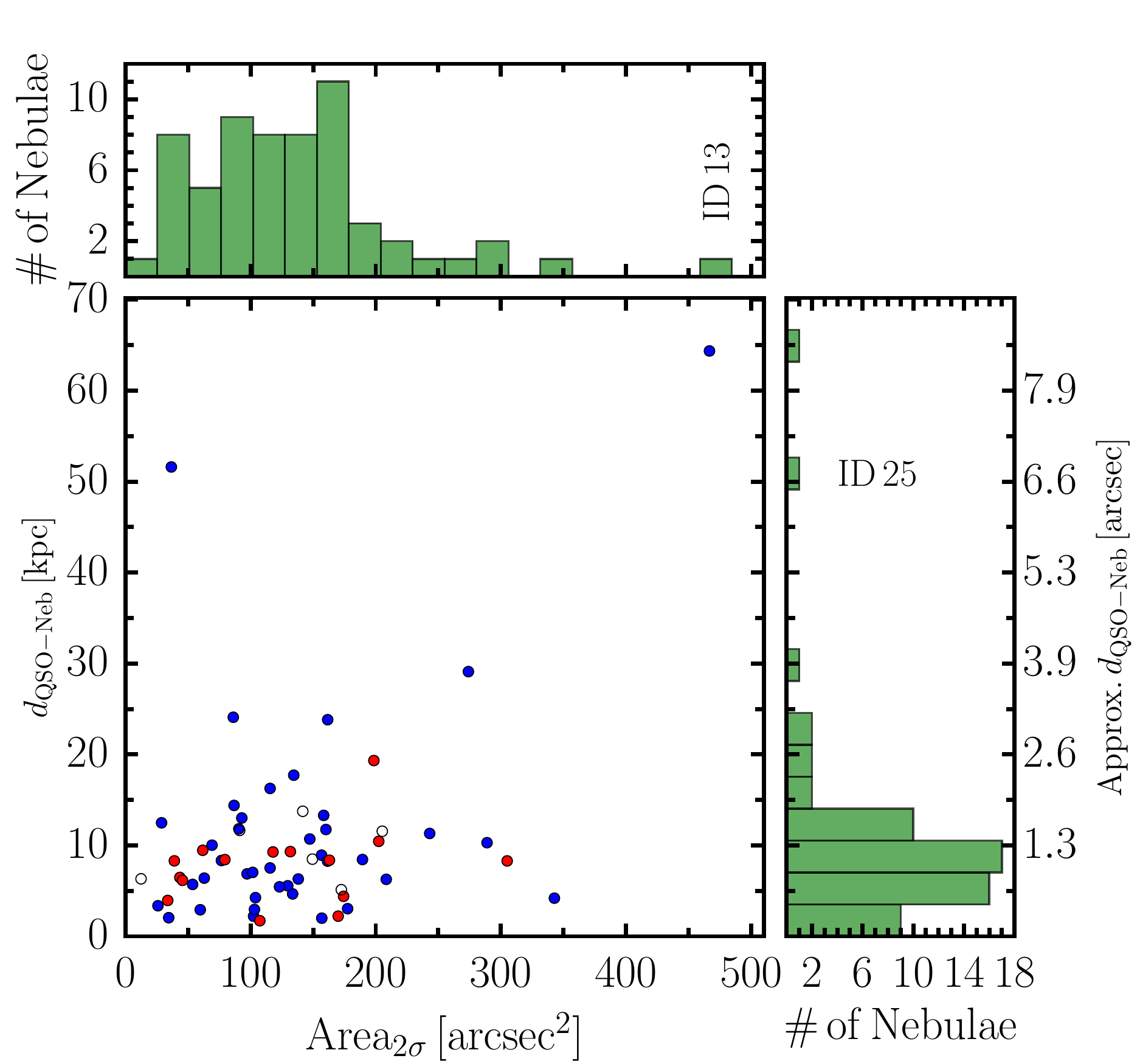}\\
       \includegraphics[width=0.95\columnwidth]{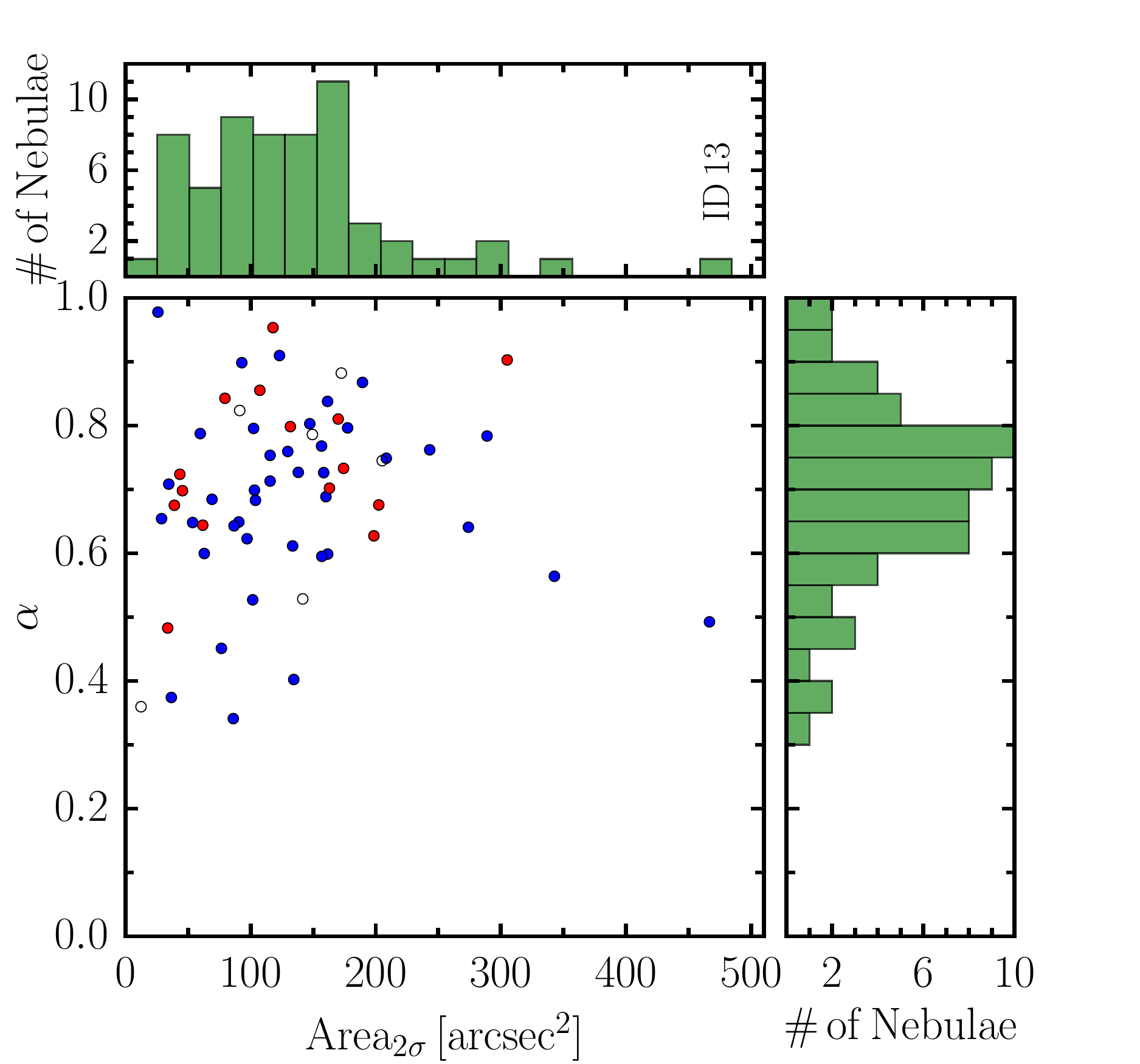}
    \caption{Top panel: plot of the distance between the quasar position and the flux-weighted centroid of the \lya\ nebulosity, $d_{\rm QSO-Neb}$, versus the area enclosed by the $2\sigma$
    isophote. Most of the nebulosities are centered at small separations from the quasar, with ID 13 (or PKS~1017+109) being a clear outlier. Lower panel: plot of the 
    asymmetry $\alpha$, i.e. the ratio between the semiminor axis $b$ and semimajor axis $a$, versus the area enclosed by the $2\sigma$
    isophote. ID 13 (or PKS~1017+109) appears to be a clear outlier with the small asymmetry driven by structures on large scales.
    In both panels, blue circles indicate the radio-quiet systems, red circles the radio-loud, and the open circles the six cases where no radio data are available.}
    \label{fig:asymmetry}
\end{figure}

In Figure~\ref{fig:asymmetry} we plot the distance $d_{\rm QSO-Neb}$, and the asymmetry $\alpha$ versus the area enclosed
by the $2\sigma$ isophote of each \lya\ nebulosity in the top panel and bottom panel, respectively.
To help the visualization of the data we also show the histogram for each quantity.
This figure clearly shows that most of the nebulae have their flux-weighted centroid in close proximity of the targeted quasar, $(d_{\rm QSO-Neb})_{\rm median}=8.3$~kpc (or $1.1$~arcsec), 
and that appear to have a more circular shape, $\alpha_{\rm median}=0.71$. On the contrary, there are a handful of cases for which the centroid of the nebulosity 
is clearly separated from the quasar ($>20$~kpc), with ID 13 (or PKS~1017+109) being the extreme case ($d_{\rm QSO-Neb}=64.4$~kpc).
Also, 7 nebulosities show an asymmetry $\alpha<0.5$. However, while most of these small asymmetries are driven by features on small scales, ID 13 (or PKS~1017+109)
appear to be a clear outlier with the small asymmetry driven by structures on large scales, 
i.e. encompassing an area $\geq 3.4\times$ that of the other 6 nebulosities with $\alpha<0.5$.  
These characteristics are probably evidences of  the powering mechanism for the \lya\ emission (see Section~\ref{sec:powering}), of the local environment of the 
system studied (e.g., \citealt{hennawi+15,fab+2018}), and possibly encode information on the different orientations of the systems with respect 
to our observational vantage point. 
In Figure~\ref{fig:asymmetry} we also indicate weather a quasar is radio-quiet (blue) or radio-loud (red), showing that the two populations 
do not differ on these plots.

\subsubsection{Individual circularly averaged surface brightness profiles}
\label{sec:ind_profiles}

In this section we compute the radial SB profiles for the discovered \lya\ nebulosities as commonly done in the literature.
Such profiles provide additional information on the diverse morphologies of the nebulosities, while can 
shed light on the mechanism powering the \lya\ line and on the cool gas distribution around quasars at this redshift.
In particular, we aim to obtain an average profile for our sample, and compare it to the average profile of \lya\ seen from the quasar 
CGM at different redshifts (e.g., \citealt{fab+16}).

First, to use the full information contained in the MUSE datacubes we decided not to use the 
``optimally extracted'' images. Indeed these images are obtained by cutting the data with a 
S/N threshold which would inevitably result in the loss of information at large distances from the quasar.  
From Table~\ref{Tab:LyaNeb} -- where we report the maximum projected distance reached by each \lya\ structure from the quasar position -- it is evident that 
32 nebulosities do not have emission above $2\sigma$ extending at distances $>80$~kpc from the quasar. In addition, 
the noise characterization in the ``optimally extracted'' images is not
trivial and would introduce an additional uncertainty in the estimate of the profiles (as explained in \citealt{Borisova2016}).

\begin{figure*}
       \centering
       \includegraphics[width=0.68\textwidth]{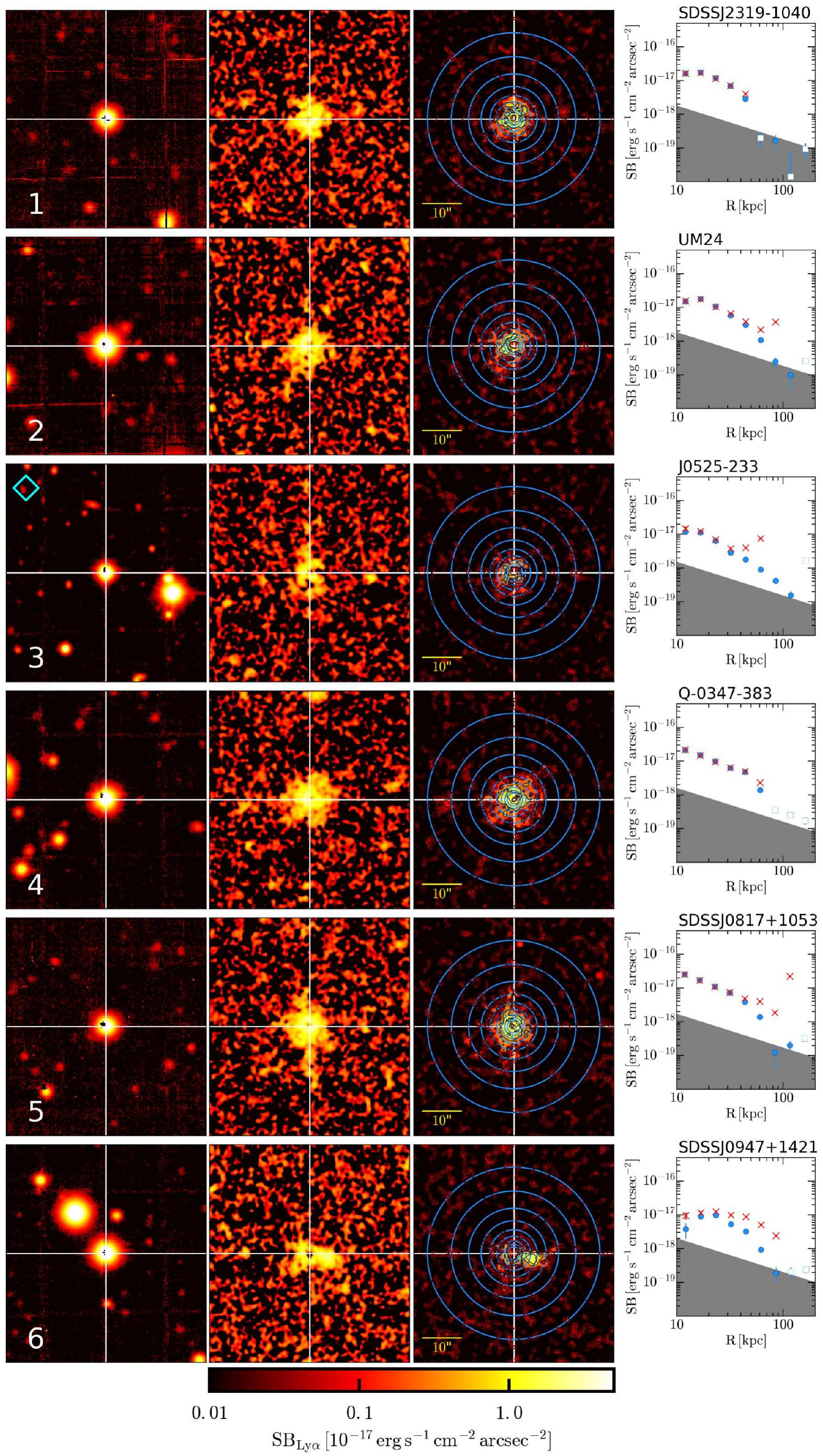}
    \caption{Each row shows $50\arcsec\times50\arcsec$ (or approx. $380$~kpc~$\times380$~kpc) images (white-light image, $30$\AA\ NB image, ``optimally extracted'' NB image) and 
    the circularly average SB profiles for the targets in our QSO
    MUSEUM survey. A cyan diamond in the top-left corner of the white-light image indicates that the quasar is radio-loud.
    For comparison, the NB images and ``optimally extracted'' NB images are 
    shown on the same color scheme and the position of the quasar prior to PSF subtraction is
    indicated by the white crosshair in all images. 
    We overlay on the ``optimally extracted'' images the circular apertures used for the extraction of the profiles.
    In the rightmost panel we show the circularly average SB profile extracted from the NB image (blue) and from the ``optimally extracted'' 
    NB image (red) (see section~\ref{sec:ind_profiles} for details). The gray shaded
    region represents the $2\sigma$ SB limit expected within each aperture in the case it is assumed a perfect sky and continuum subtraction.
    North is up, east is to the left.}
    \label{fig:Prof_one}
\end{figure*}

For these reasons we extracted the circularly averaged SB profiles from NB images with fixed width ($30$~\AA) 
obtained from the PSF and continuum subtracted datacubes, 
and centered at the \lya\ peak of each nebulosity. 
For each of these NB images, we have propagated the corresponding subsection of the variance cube to obtain a variance image.
As anticipated in Section~\ref{sec:obs}, in Table~\ref{tab:ObsLog} we list the $2\sigma$ SB limit for each of these NB images, 
which are roughly $5\times$ higher than for a single layer (1.25~\AA).
While such a fixed width does not enclose the whole \lya\ emission above the ${\rm S/N} = 2$ threshold used to extract the 3D mask, 
it is a good compromise to include most of the bright 
\lya\ emission while being still sensitive to emission at very low SB levels on large scales ($\sim100$~kpc). 
Also, this width is similar to the wavelength range spanned by the 
NB filter used in the literature to detect \lya\ emission, and thus enables a more direct comparison with those studies. 
Figures~\ref{fig:Prof_one} show all the NB images obtained from the PSF and continuum subtracted
datacubes together with the white-light images and the ``optimally extracted'' images.  
This figure clearly display the higher noise level in comparison to the single layer ($1.25$~\AA) used 
as background in the ``optimally extracted'' NB images (see section~\ref{sec:cubex}).

Before proceeding with the extraction of the profiles, we further prepared the NB images by masking artifacts and compact sources, which would otherwise contaminate the profiles. 
In particular, even though we have subtracted the
continuum using the {\tt CubeBKGSub} algorithm within \textsc{CubExtractor}, our images are affected by residuals (both positive or negative) 
expecially at the position of bright continuum sources, i.e. stars or galaxies. To avoid such contamination, we constructed a mask for the continuum sources using the white-light image 
of each field. This approach is very conservative as the subtraction of the continuum has been tested in previous works (e.g., \citealt{Borisova2016}).
In particular, in analogy to what was done in \citet{fab+16}, we run \textsc{CubExtractor} to identify all the continuum sources above 
${\rm S/N}=2$ and down to very small objects 
(5 pixels minimum) within the white-light images. 
\textsc{CubExtractor} produces a ``segmentation'' image that is then used to create a mask, 
after removing the region corresponding to the quasar that otherwise would 
hide the emission of interest here.

In addition, to avoid the detection of signal from compact line emitters around the quasars, i.e., 
objects which do not have a continuum detection,
we obtain a similar mask using the NB image itself. However, at this step we only identify compact sources, 
with area between 5 pixels and 20 pixels (roughly a seeing disc).
By combining the two individual masks, we then obtained a final mask which allow us to neglect in 
our analysis both pixels from artifacts and compact sources. Overall this masking procedure
reduces the fraction of the field-of-view that is usable around each quasar to an average of $88\%$.
This analysis and the final available field-of-view is similar to what has been used to extract radial 
SB profiles around quasars in NB images of similar depth (\citealt{fab+16}).

We then calculated the circularly averaged SB profile for each quasar in radial logarithmic bins centered at the quasar positions with an 
unweighted average of all the not-masked pixels
within each annulus. 
We then consistently propagated the errors from the variance images. To facilitate the visualization of the profiles and the comparison with the NB images, 
we overplot on the ``optimally extracted'' images in Figures~\ref{fig:Prof_one}
the annuli within which we calculate the radial SB profiles.
In the same Figures we show the SB profiles for each of the discovered \lya\ nebulosities (blue data points with errors), together with the $2\sigma$ error estimates expected 
within each aperture in the case one assumes a perfect sky and continuum subtraction (gray shaded area)\footnote{This is obtained using 
the formula SB$_{\rm limit}($area$)=$SB$_{\rm limit}^{1\, {\rm arcsec^2}}/\sqrt{\rm area}$
where SB$_{\rm limit}^{1\, {\rm arcsec^2}}$ 
is the SB limit in $1$~arcsec$^{2}$ reported in Table~\ref{tab:ObsLog}. 
This formula is true only if a perfect sky and continuum subtraction is performed. For this reason the shaded region has to be regarded as only indicative.}. 
In addition, to compare such profiles with the ``optimally extracted'' NB images, 
we compute the profiles within the same annuli, but only using the emission enclosed by the $2\sigma$ isophote (red data points). By construction, 
these profiles do not extend further than the contour enclosing the maximum distance defined by 
the $2\sigma$ isophote\footnote{The last data-point in some of the profiles extracted using the 3D mask (red crosses) is driven to high SB values because of the 
small portion of nebula within the last ring.}.
While extracting both radial profiles, we masked the central $1\arcsec \times 1\arcsec$ region around the quasar position used for the PSF rescaling. 
For this reason the radial profiles in Figures~\ref{fig:Prof_one} are shown for radii $R>10$~kpc.

Overall, this analysis once again shows the diversity of the morphologies and extents of the \lya\ emission around the targeted quasars, 
while emphasizing the differences
in SB between the different objects. In particular, consistently with the 2D images, the profiles extend out to the maximum 
annulus which include the $2\sigma$ isophote, 
and beyond this radius they appear to monotonically decrease. 
This decrease is not driven by the current depth of the data as otherwise we would have
had larger nebulae in the ``optimally extracted'' images than in the NB images. This can be clearly seen by comparing the data-points from the 
NB images (blue) and the data-points from the ``optimally extracted'' images (red). 
The decrease is visible in both symmetric and asymmetric nebulosities. Further, we do not see 
any particular difference between the individual profiles of the radio-quiet and radio-loud objects.

\begin{figure*}
       \centering
       \includegraphics[width=0.68\textwidth]{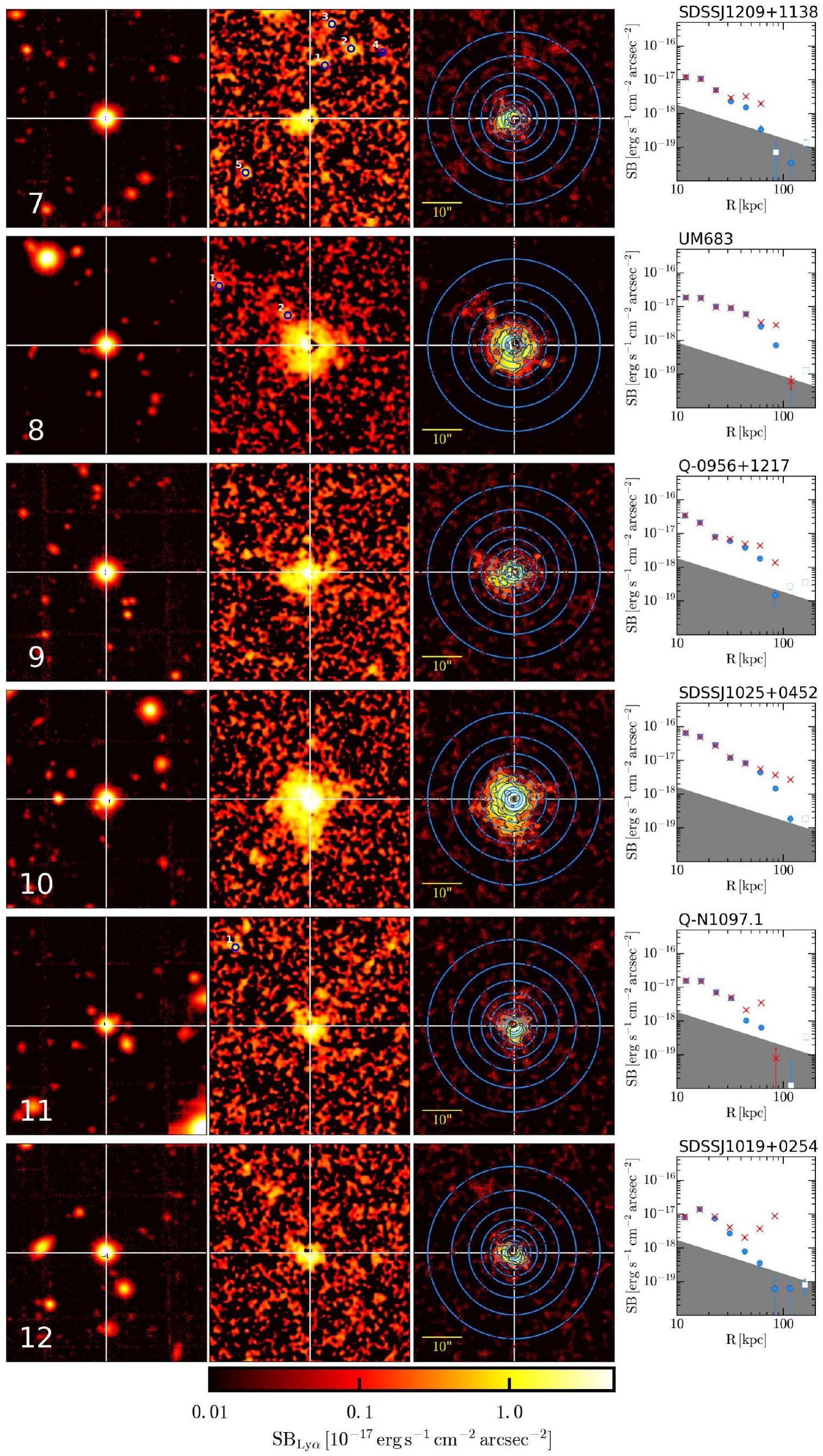}
    \smallskip
    \centerline{{\bf Figure 5} -- {\it continued}}
    \smallskip
\end{figure*}

\begin{figure*}
       \centering
       \includegraphics[width=0.68\textwidth]{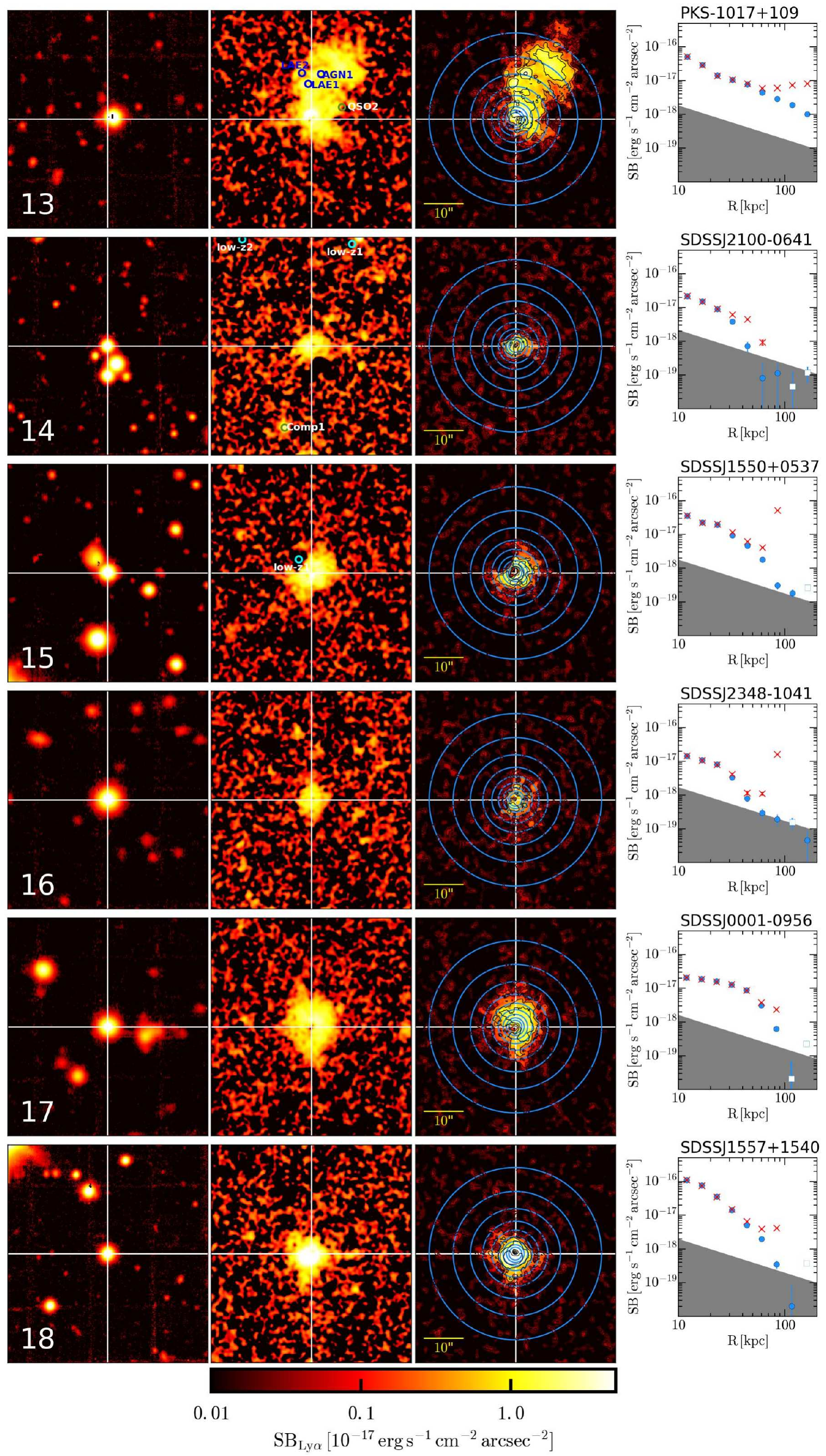}
    \smallskip
    \centerline{{\bf Figure 5} -- {\it continued}}
    \smallskip
\end{figure*}

\begin{figure*}
       \centering
       \includegraphics[width=0.68\textwidth]{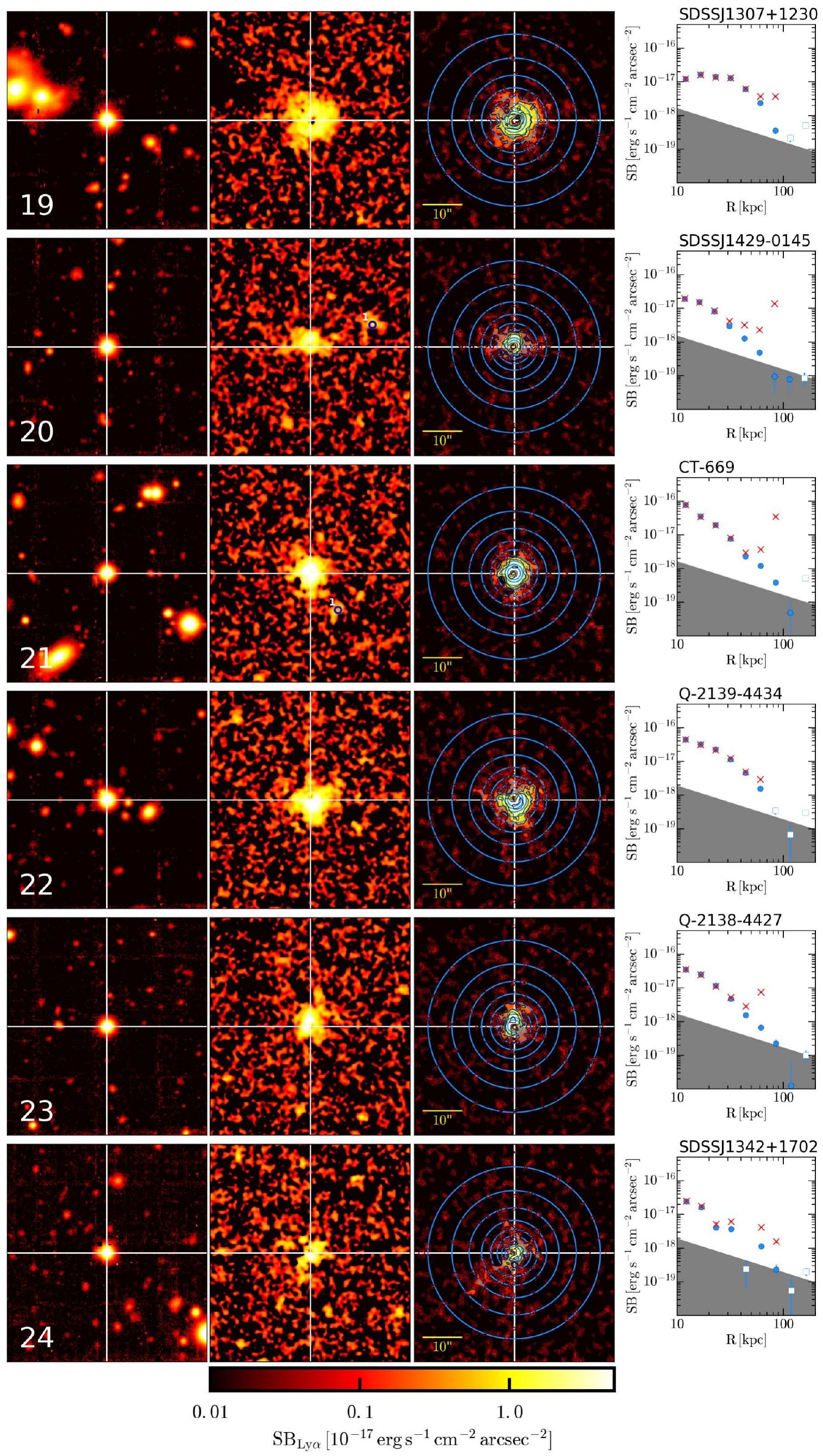}
    \smallskip
    \centerline{{\bf Figure 5} -- {\it continued}}
    \smallskip
\end{figure*}

\begin{figure*}
       \centering
       \includegraphics[width=0.68\textwidth]{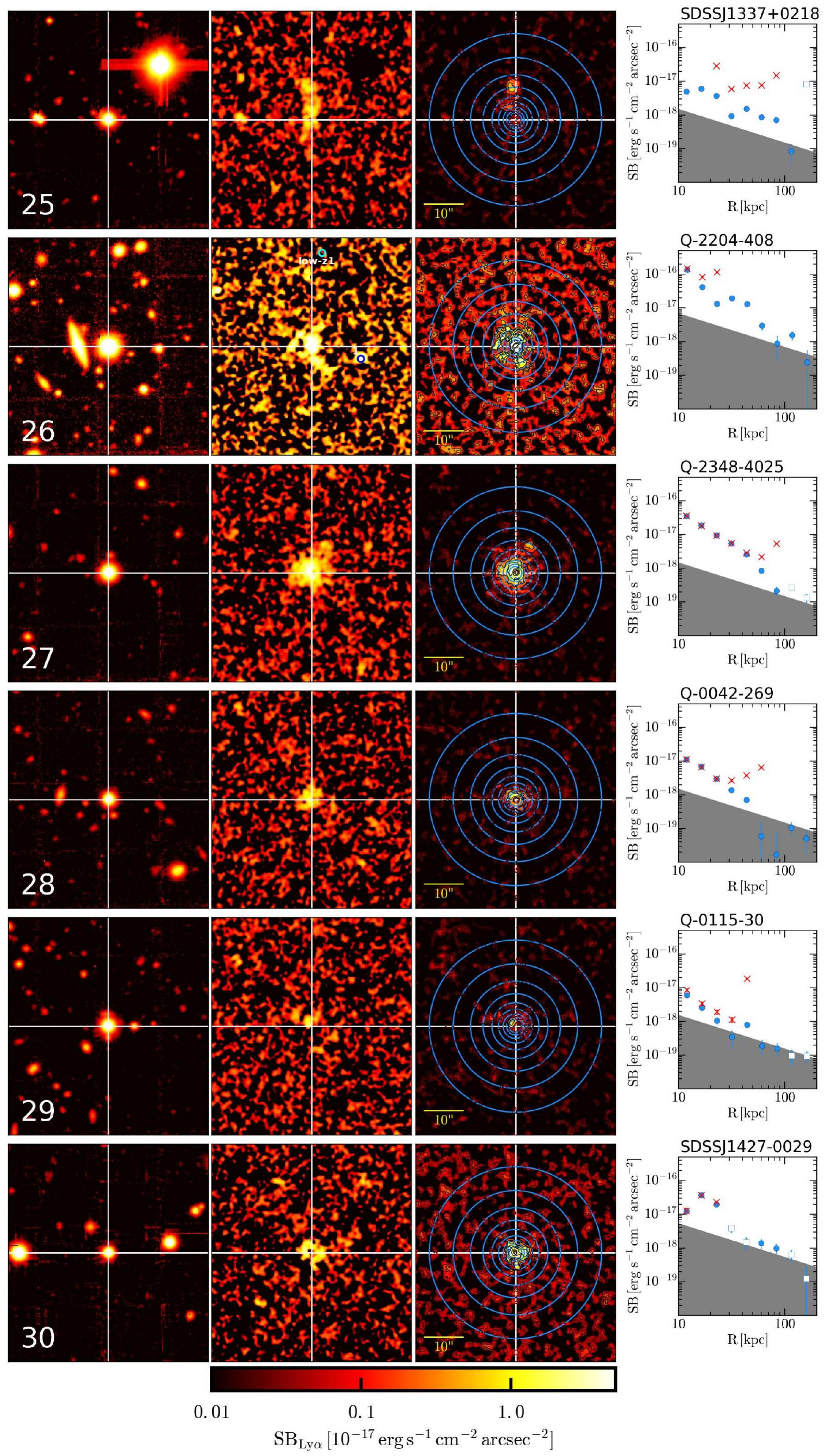}
    \smallskip
    \centerline{{\bf Figure 5} -- {\it continued}}
    \smallskip
\end{figure*}

\begin{figure*}
       \centering
       \includegraphics[width=0.68\textwidth]{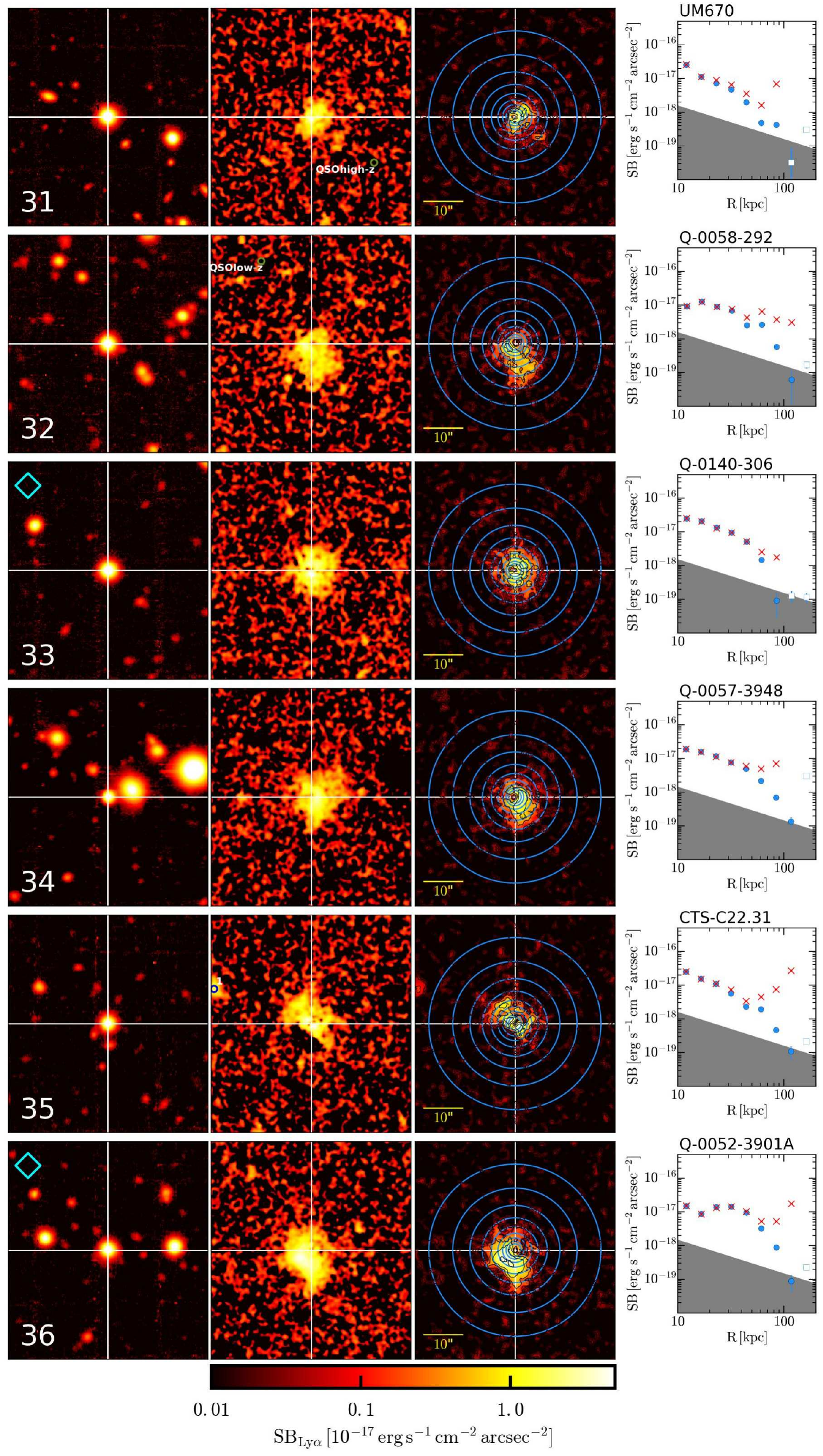}
    \smallskip
    \centerline{{\bf Figure 5} -- {\it continued}}
    \smallskip
\end{figure*}

\clearpage

\begin{figure*}
       \centering
       \includegraphics[width=0.68\textwidth]{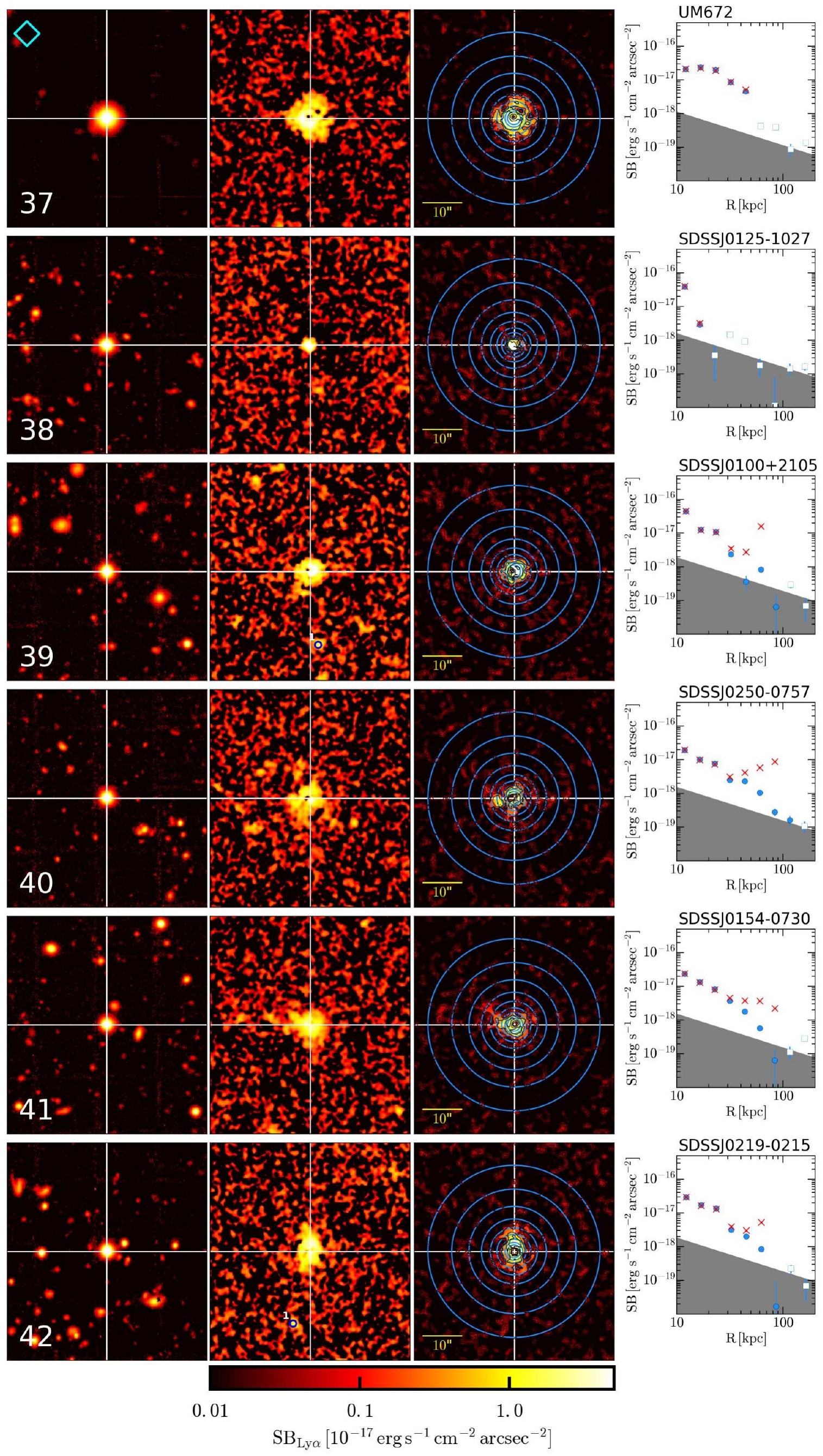}
    \smallskip
    \centerline{{\bf Figure 5} -- {\it continued}}
    \smallskip
\end{figure*}

\begin{figure*}
       \includegraphics[width=0.68\textwidth]{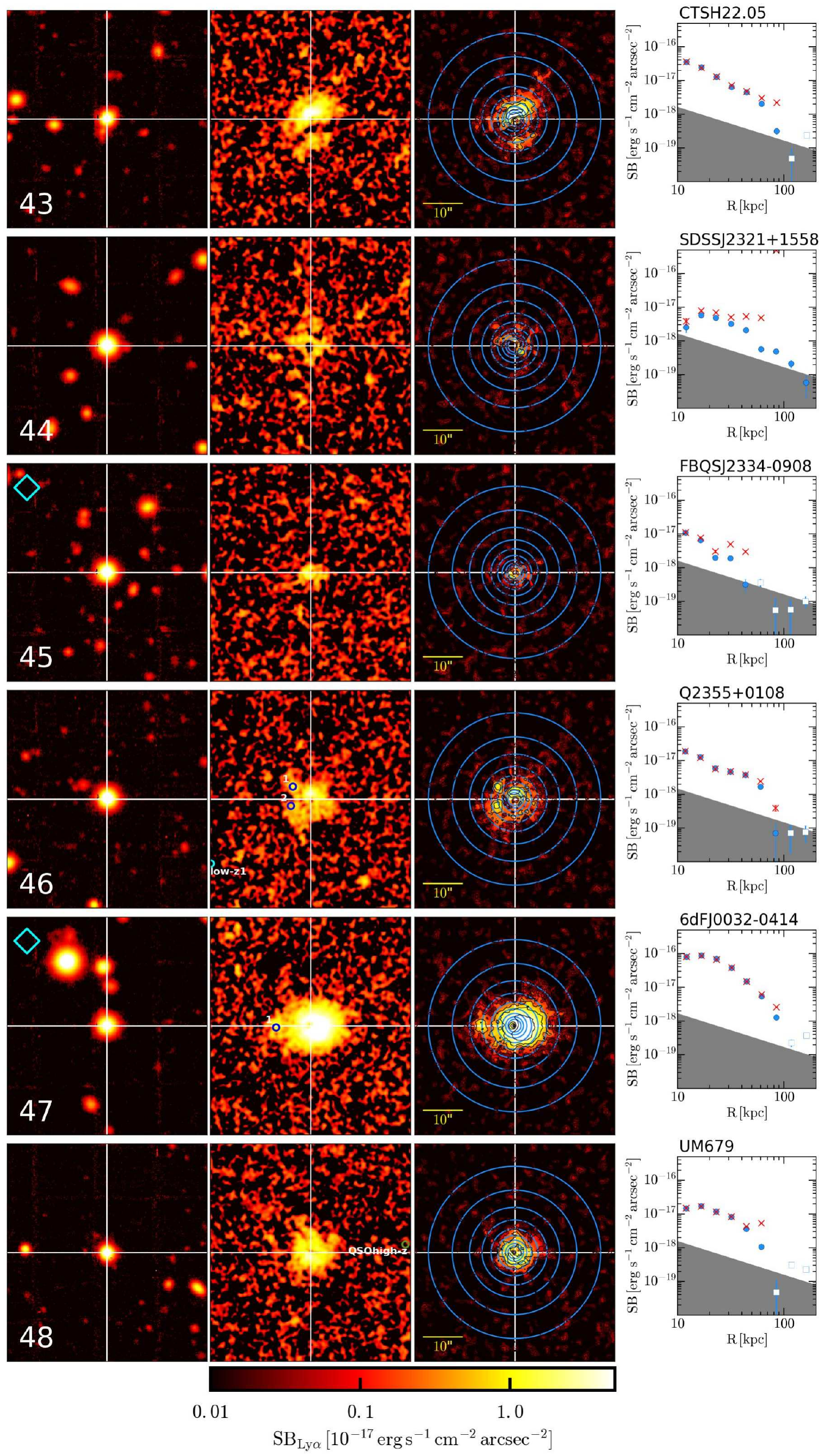}
    \smallskip
    \centerline{{\bf Figure 5} -- {\it continued}}
    \smallskip
\end{figure*}

\begin{figure*}
       \centering
       \includegraphics[width=0.68\textwidth]{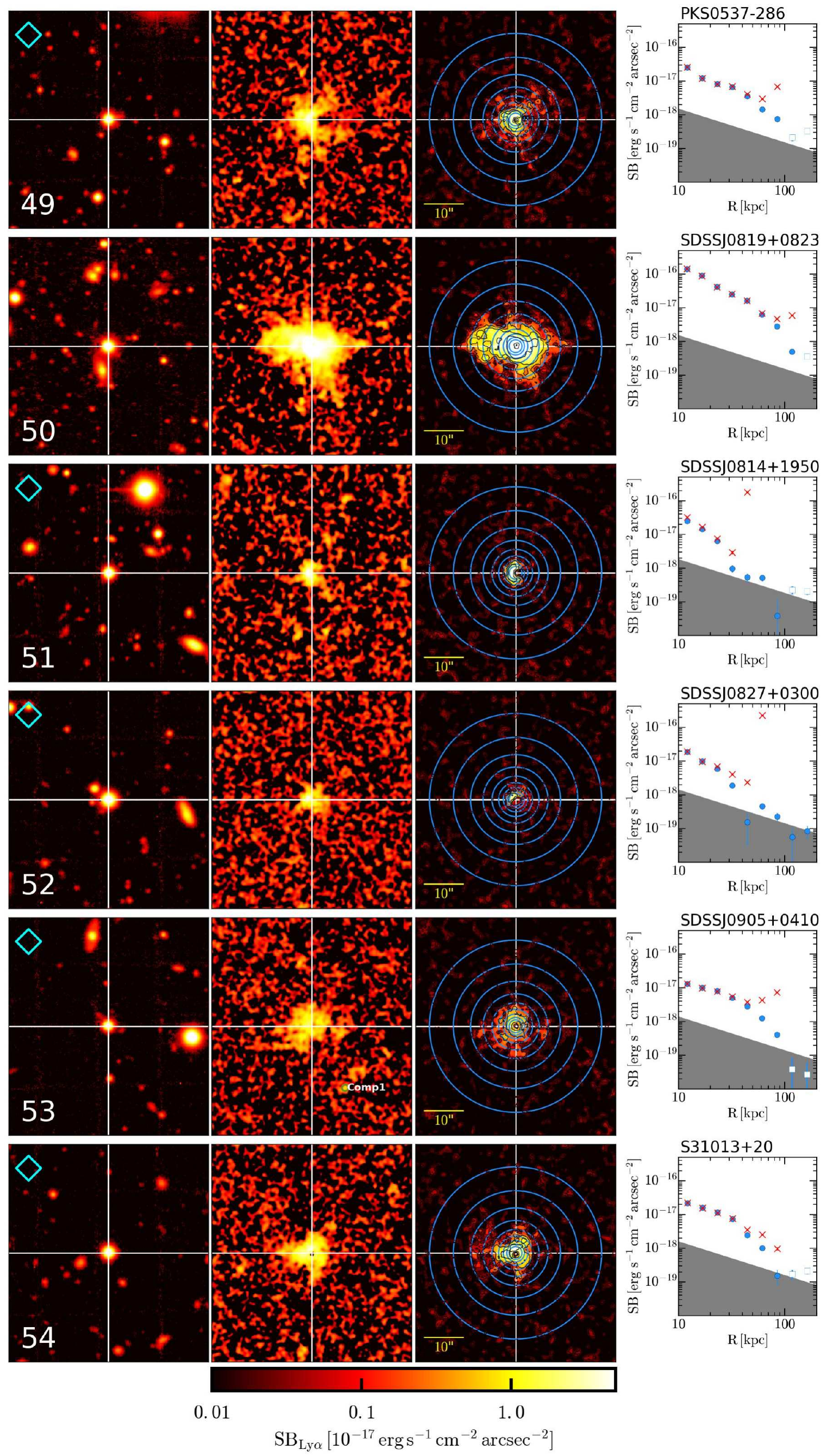}
    \smallskip
    \centerline{{\bf Figure 5} -- {\it continued}}
    \smallskip
\end{figure*}

\begin{figure*}
       \centering
       \includegraphics[width=0.68\textwidth]{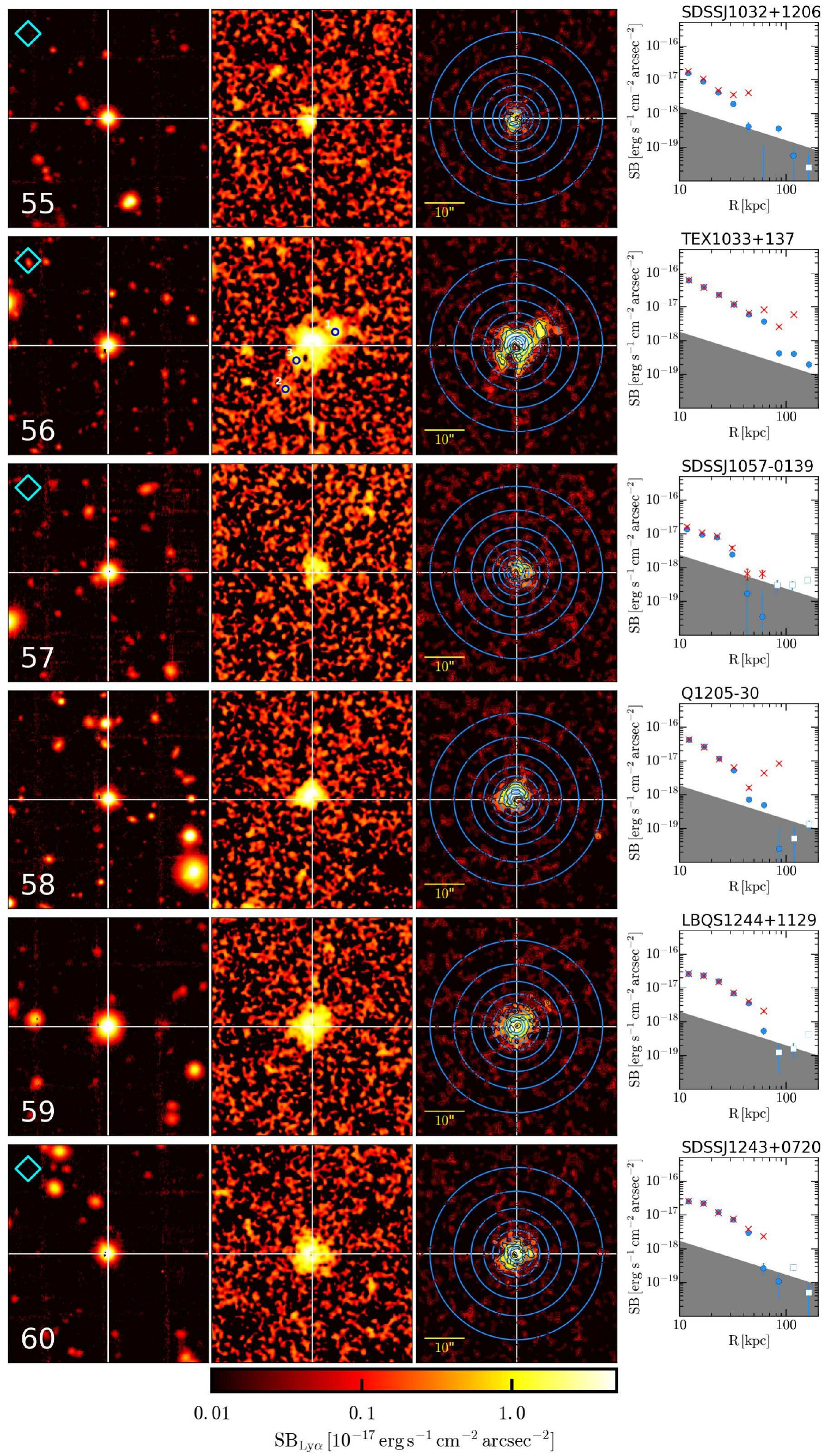}
    \smallskip
    \centerline{{\bf Figure 5} -- {\it continued}}
    \smallskip
\end{figure*}

\begin{figure*}
       \centering
       \includegraphics[width=0.68\textwidth]{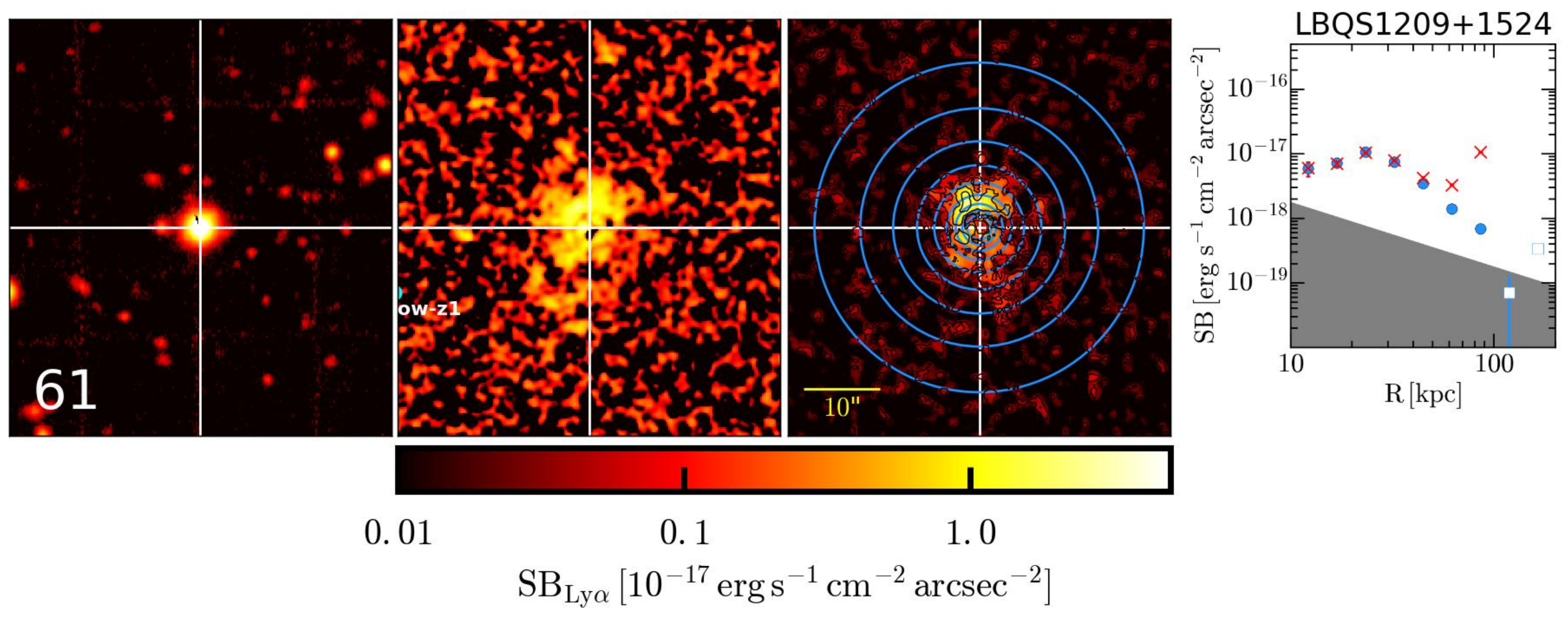}
    \smallskip
    \centerline{{\bf Figure 5} -- {\it continued}}
    \smallskip
\end{figure*}

\subsubsection{The average surface brightness profiles}
\label{sec:av-prof}

In this section we present average SB profiles for our sample. 
We have obtained three different stacked profiles by averaging with equal weights the profiles obtained from the NB images of (i) all the 61 quasars, 
(ii) the radio-quiet sample (39 quasars), and (iii)
the radio-loud sample (15 quasars)\footnote{We remind the reader that currently we do not have 
information on the radio emission for 6 quasars in our sample, which are thus excluded from
both the radio-quiet and radio-loud sample.}.  

\begin{figure*}
       \centering
       \includegraphics[width=0.7\textwidth, clip]{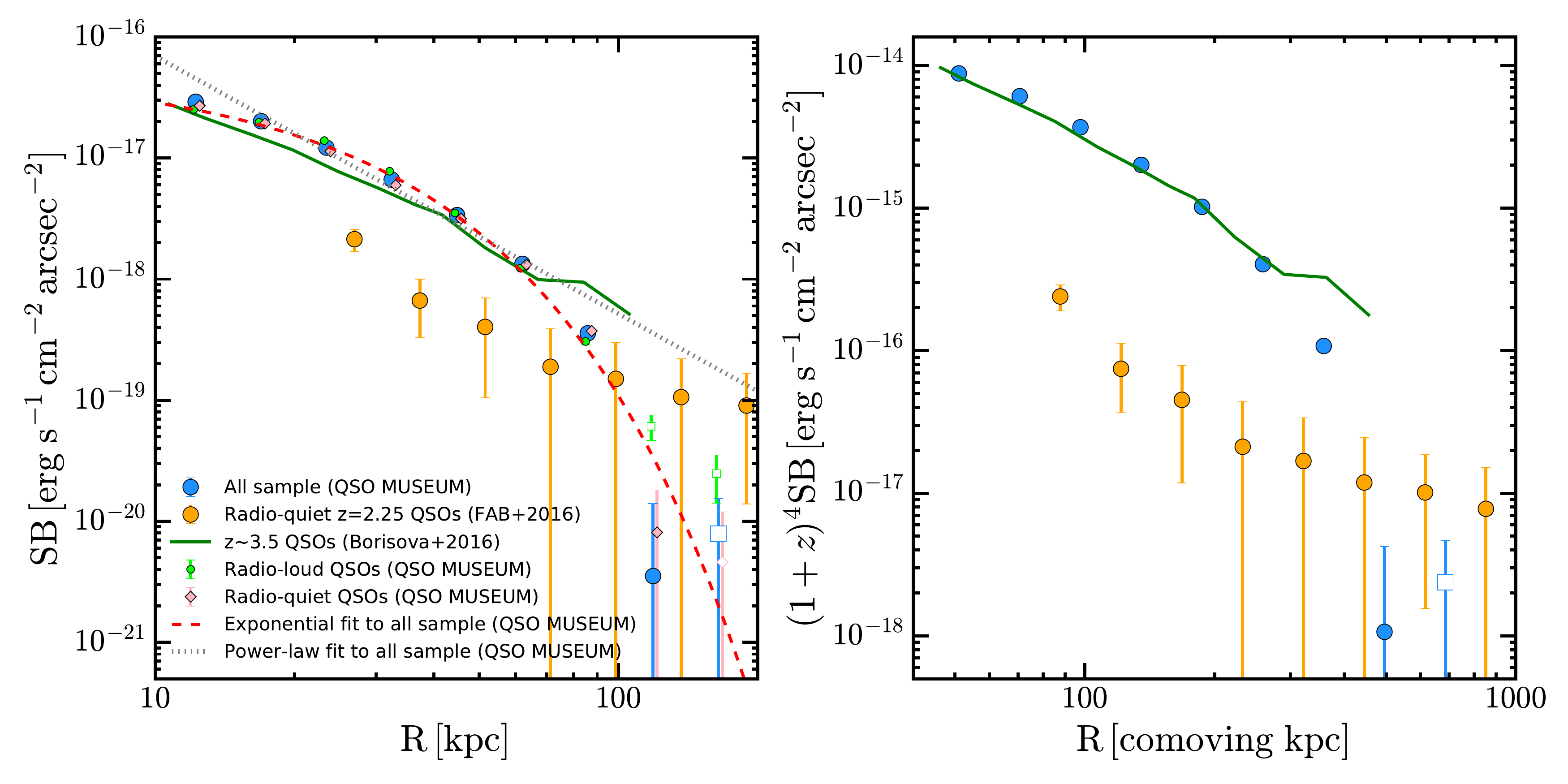}
    \caption{Average \lya\ SB profiles for the nebulosities discovered in our QSO MUSEUM survey. Left panel: we show the average profile for the whole 
    sample (large blue; 61 targets), the radio-quiet subsample (pink diamond; 39 quasars), and radio-loud subsample (small green; 15 quasars)
    in observed SB units versus the physical radius. For the radio-quiet and radio-loud subsamples we slightly shifted the datapoints along the $x$-axis to allow a better visualization.
    For each profile, open squares indicate negative datapoints.
    For the whole sample we show the best exponential fit (dashed red line) and the best power-law fit (dotted gray line). In addition, we compare our dataset with the average profile 
    for (i) $z\sim2.25$ quasars by \citet{fab+16}, and for (ii) $z\sim3.5$ quasars by \citet{Borisova2016}. Right panel: the average profile for the QSO MUSEUM sample is compared to 
    the average profiles by \citet{fab+16}, and \citet{Borisova2016} after correcting for the cosmological dimming and using comoving units for the radii.
    By comparing these datasets it appears that $z\sim3$ quasars are surrounded by a brighter glow in the \lya\ line (see Sections~\ref{sec:av-prof} and \ref{sec:redshiftEvol} for details and discussion).}
    \label{fig:averageProf}
\end{figure*}

The left panel in Figure~\ref{fig:averageProf} shows these three SB profiles versus the 
physical radius of the annuli used for the profile extraction. In particular, we indicate 
the whole sample in blue (large datapoints), while the sub-samples of the radio-quiet and radio-loud objects in pink and green (small datapoints),
respectively. We note that there is only a marginal difference between the radio-quiet and radio-loud subsamples at radii $20$~kpc~$<R<50$~kpc, with 
the radio-loud objects showing a slightly higher SB.
For comparison with \citet{Borisova2016}, we fit these stacked profiles with an exponential 
SB$(r)=C_e{\rm exp}^{-r/r_h}$ or a power law SB$(r)=C_p r^{\alpha}$. 
We found that the obtained stacked profiles are better fit by an exponential profile with scale length $r_h=15.7\pm0.5$~kpc, $r_h=15.5\pm0.5$~kpc, and 
$r_h=16.2\pm0.3$~kpc, for the whole sample, the radio-quiet and the radio-loud objects, respectively. 
The aforementioned excess in SB at intermediate scales for the radio-loud sample 
thus determine the small tentative difference in scale length between the two subsamples.
Further, our quasars are surrounded by 
\lya\ nebulosities characterized by scale lengths roughly $3\times$ larger than those for
LAEs at similar redshifts (e.g., \citealt{Leclercq2017}), probably reflecting the 
different halo masses, and also the ability of the QSOs to power observable \lya\ emission further out.

We note that our best fit power laws have slopes of $\alpha=-1.96$, $\alpha=-1.93$, and $\alpha=-1.81$, for the whole sample, the radio-quiet 
and the radio-loud objects, respectively.  These values are consistent with the power law slope reported by \citet{Borisova2016} 
for their average profile, i.e. $\alpha=-1.8$.
However in our case the best fit profile appears to be the exponential.
For comparison purposes, the average profile from \citet{Borisova2016} is shown in Figure~\ref{fig:averageProf} (solid green line), 
together with our best exponential fit (dashed red), and best power law fit (dotted gray)
to our whole sample. 

For completeness, we have then computed the average SB profile of the whole sample and of the radio-quiet
sub-sample both excluding the quasar with ID~13 (or PKS~1017+109), which shows a rare ELAN 
with multiple AGN companions (see Section~\ref{sec:ELAN}; \citealt{fab+2018}).
The scale-lenghts for the profiles slightly shrinks $r_h=15.4\pm0.7$~kpc and $r_h=15.2\pm0.7$~kpc, respectively for the whole sample and for the radio-quiet sample.
However the uncertainties are larger than the variations.
The brightness and extent of an ELAN thus seems to affect the average SB profiles even with a large sample of 
objects. Table~\ref{Tab:profilesfit} lists all the details of the profile fitting.

We then compare our stacked SB profile with the work by \citet{fab+16}. 
This work produced a stacked \lya\ SB profile for the $z\sim2$ quasar's CGM 
by targeting 15 $z\sim2.253$ radio-quiet quasars with the NB technique, and achieving similar depths to our extracted NB data. Indeed, those NB data are a factor of $\sim2$
less deep than our MUSE NB images, but the cosmological SB dimming of the form $(1+z)^{-4}$ should compensate this difference. 
Those quasars are characterized by an average $M_i(z=2) = -27.21$, and are thus slightly fainter than the quasar in this study (average $M_i (z=2)=-28.05$; see Table\ref{tab:sample}).
However 8 out of 15 of those quasars have luminosities similar to quasars in the current sample. In particular two have magnitudes 
$M_i(z=2)=-28.75$ (SDSS~J093849.67+090509.7), and 
$M_i(z=2)=-28.06$ (SDSS~J084117.87+093245.3), which are comparable also to quasars in the sample of \citet{Borisova2016}.
Notwithstanding such differences in luminosity, \citet{fab+16} did not report any dependence of the individual profiles of the extended \lya\ emission 
with respect to the luminosity of the quasar themselves.

As discussed in \citet{fab+16}, the NB imaging technique could be affected by flux losses if the \lya\ line is in reality at a different wavelength (thus redshift) than expected, 
and thus falls outside or at the edge of the used filter. \citet{fab+16} have shown that only error distributions on the redshift with width comparable
to the width of the NB filter (in this case about 2500~km~s$^{-1}$) can result in significant flux losses (see their appendix B), and concluded that flux losses should be 
of the order of $3$\% for their sample.
Before proceeding with the comparison, we discuss an additional piece of evidence suggesting that flux losses are indeed unlikely to dominate the observations in \citet{fab+16}.
In section~\ref{sec:morph} we have shown that the \lya\ extended emission around the targets of our QSO MUSEUM survey peaks at wavelengths close to 
the maximum of the \lya\ emission of the targeted quasars. We have thus checked the velocity separation between the quasar's systemic redshift\footnote{In \citet{fab+16} the quasar's
systemic redshfits are more robust than here (uncertainty of $\sim270$~km~s$^{-1}$).} 
and the peak of the \lya\ line for each of the quasars in \citet{fab+16}. We have found a median (mean) shift of $\Delta v=-92$~km~s$^{-1}$ ($-250$~km~s$^{-1}$), with 
only two quasars having a large displacement, $\Delta v=-1468$~km~s$^{-1}$ (SDSS~J131433.84+032322.0) and $\Delta v=-1789$~km~s$^{-1}$ (SDSS~J085233.00+082236.2). 
Since the quasar's systemic redshifts were chosen to conservatively
fit within the NB filter, this occurrence indicates that those observations should not be significantly affected by flux losses
at least for 13/15 of the quasars in their sample.

We thus show in Figure~\ref{fig:averageProf} the average profile for the \lya\ emission around $z\sim2.253$ 
quasars as determined by \citet{fab+16} (large orange datapoints).
It appears that $z\sim3$ quasars are surrounded by a brighter glow in the \lya\ line. This is even more evident if we correct the SB of the two datasets for 
the cosmological dimming, and use the comoving distances instead of the physical ones. We show this comparison in the right panel of Figure~\ref{fig:averageProf}.
This result has to be verified with more statistical studies of the $z\sim2$ population, e.g. with KCWI.
Nevertheless we discuss the implications of such a finding in Section~\ref{sec:redshiftEvol}.

\begin{table}
\scriptsize
\caption{Fit of the average Ly$\alpha$ surface brightness profile.}
\centering
\setlength\tabcolsep{3pt}
\begin{tabular}{lcccccccc}
\hline
\hline
Sample   &  \#     &  $\langle z_{\rm Ly\alpha} \rangle$	 &  $C_e^{\rm  \ \ a}$ & $r_h^{\rm \ \ a}$ & $\chi^{2 \ \ {\rm b}}_e$ & $C_p^{\rm  \ \ c}$ & $\alpha^{\rm  \ \ c}$ &  $\chi^{2 \ \ {\rm d}}_p$     \\
                       &        &       & ($10^{-18}$)   & (kpc) & & ($10^{-18}$) & (kpc) &  \\
\hline
All                    & 61     & 3.209 & 56.8 & 15.7 & 5.1 & 5124.3 & -1.96 & 40.1 \\
All but ID13$^{\rm e}$ & 60     & 3.210 & 56.6 & 15.4 & 4.8 & 4103.7 & -1.90 & 35.1 \\
\hline
RQ$^{\rm f}$ & 39 & 3.221 & 54.4 & 15.5 & 8.2 & 4250.6 & -1.93 & 38.7 \\
RQ but ID13$^{\rm e}$  & 38 & 3.223 & 55.0 & 15.2 & 7.5 & 3596.5 & -1.88 & 33.5 \\
RL$^{\rm f}$ & 15 & 3.174 & 56.2 & 16.2 & 0.6 & 3240.4 & -1.81 & 45.2 \\
\hline
\end{tabular}
\label{Tab:profilesfit}
\flushleft{$^{\rm a}$ Best-fit parameters assuming SB profile ${\rm SB}(r)=C_e {\rm exp}(-r/r_h)$. $C_e$ is in unit of $10^{-18}$~erg~s$^{-2}$~cm$^{-2}$~arcsec$^{-2}$.}
\flushleft{$^{\rm b}$ Reduced chi square for the exponential fit.}
\flushleft{$^{\rm c}$ Best-fit parameters assuming SB profile ${\rm SB}(r)=C_p r^\alpha$. $C_p$ is in unit of $10^{-18}$~erg~s$^{-2}$~cm$^{-2}$~arcsec$^{-2}$.}
\flushleft{$^{\rm d}$ Reduced chi square for the power-law fit.}
\flushleft{$^{\rm e}$ ID13 (PKS-1017+109) is not included as it is surrounded by an ELAN.}
\flushleft{$^{\rm f}$ RQ and RL stand for radio-quiet and radio-loud. Note that 6 targets currently do not have information in radio.}
\end{table}

\subsubsection{Averaging covering factor of detected extended Lyman-Alpha emission}

Given the diversity in morphology of the discovered nebulosities, it is difficult to visually assess which fraction of the area around the quasars 
contributes to the aforementioned profiles.
We have thus complemented the previous analysis by estimating the covering factor in logarithmic bins for the extended \lya\ emission 
above the ${\rm S/N} = 2$ threshold used to extract the 3D mask in Section~\ref{sec:cubex}.
Specifically, for each logarithmic bin used for the extraction of the radial profiles in Section~\ref{sec:ind_profiles}, 
we have simply defined the covering factor $f_{\rm C}$
as the ratio between the area spanned by the \lya\ emission above the ${\rm S/N}=2$ threshold and the total area of that radial bin.
We have thus obtained radial profiles for $f_{\rm C}$ for each discovered \lya\ nebulosity.

\begin{figure}
       \centering
       \includegraphics[width=0.98\columnwidth, clip]{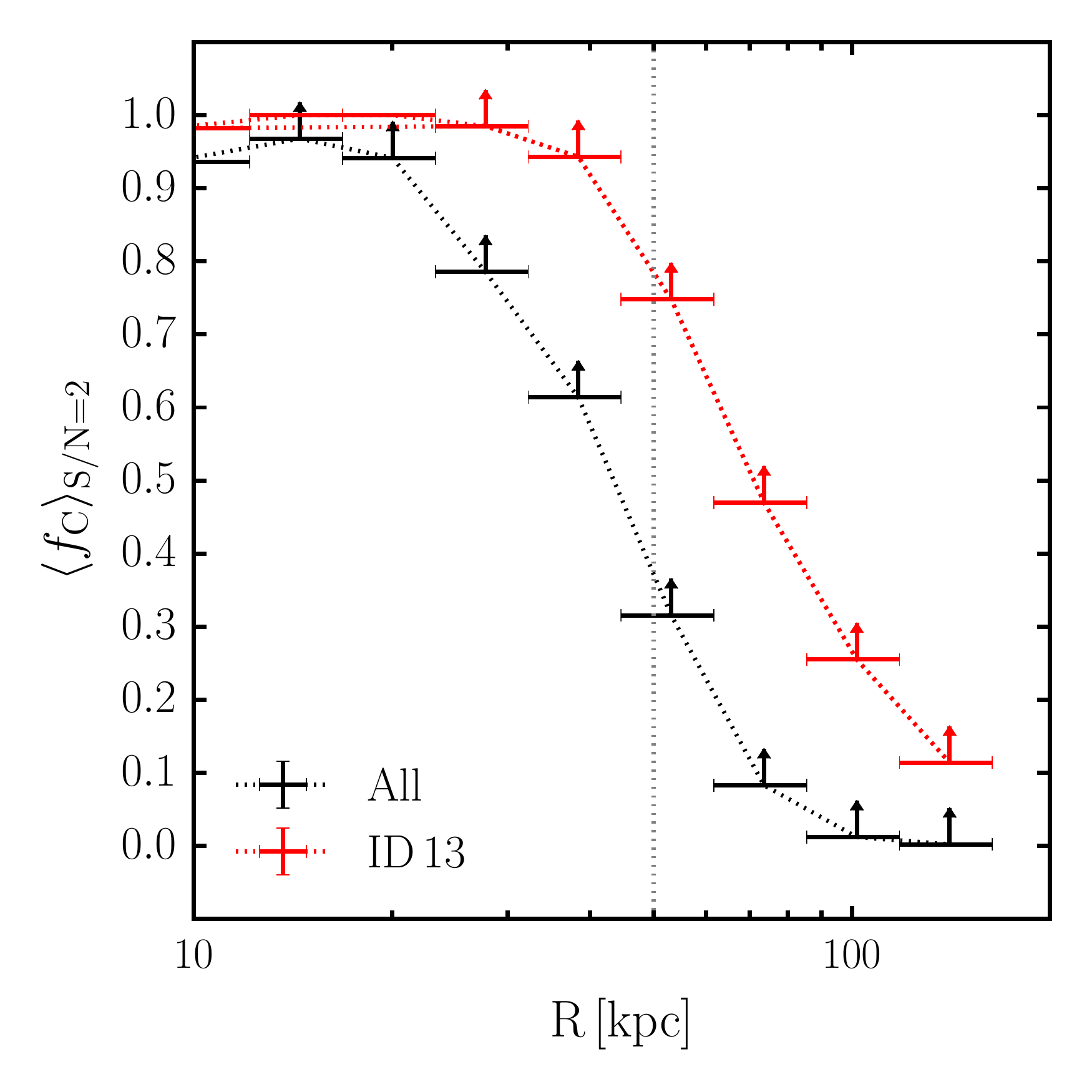}
    \caption{Covering factor of the \lya\ emission above ${\rm S/N}>2$ in logarithmic bins around the quasars. To obtain this plot, we 
    use the same circular apertures used for the profiles extraction (see section~\ref{sec:ind_profiles}). Given that we are not sensitive to 
    very faint \lya\ emission (SB$_{\rm Ly\alpha}\lesssim10^{-18}$~\unitcgssb), we show the data as lower limits: 
    for our whole sample (black) and for the target with ID 13 (or PKS~1017+109) (red).
    For reference, the gray vertical line indicates $R=50$~kpc.}
    \label{fig:coveringFactor}
\end{figure}

The average covering factor for our whole sample $\langle f_{\rm C} \rangle_{{\rm S/N}=2}$ is then estimated by averaging with equal weights the information from each individual target.
In Figure~\ref{fig:coveringFactor} we thus show $\langle f_{\rm C} \rangle_{{\rm S/N}=2}$ in logarithmic bins of radius $R$ from the quasar. 
This covering factor has to be treated with caution. Indeed, by construction, it depends on the depth of the dataset and on the ${\rm S/N}$ threshold used for the analysis.
In addition, close to the quasar position, the morphology of the nebulosities, and thus the area covered by the \lya\ emission, are uncertain due to the PSF subtraction.
For these reasons, in Figure~\ref{fig:coveringFactor} the estimated values are formally regarded as a lower limit for the area covered 
by \lya\ emitting gas around our sample of quasars. 

Figure~\ref{fig:coveringFactor} shows that the detected extended \lya\ emission covers a fraction $>30$\% for radii $R<50$~kpc (vertical gray line), 
while it quickly decrease to lower values at larger distances.
For comparison, we show in red the profile of the covering factor for the target with ID 13 (or PKS~1017+109), 
which is associated with an ELAN. This object is clearly characterized by a
larger covering factor than the average population, with $f_{\rm C}>0.8$ for $R<50$~kpc, and $f_{\rm C}>0.1$ at $R=140$~kpc (our last bin).

The Quasar Probing Quasar (QPQ) series of papers (e.g., \citealt{QPQ1,qpq4}) studied the $z\sim2$ quasar CGM by using bright background 
sightlines of higher-$z$
quasars as a probe of the foreground quasar halo. These studies have shown that the $z\sim2$ quasar CGM is permeated by optically 
thick gas with high covering factor >0.4 out to 200 projected kpc in a $\pm 1500$~km~s$^{-1}$ velocity interval (\citealt{QPQ5,qpq6}). 
A direct comparison between these absorption studies and our current work is complicated by the differences inherent to the two techniques. 
Indeed the absorption experiment  can only rely on a one-dimensional probe percing through the studied foreground system.
The seeing, the uncertainty on cloud sizes, 
the fact that one probes the quasar ionizing luminosity or $n_{\rm H}^2$ in emission, 
but optical depth in absorption (\citealt{qpq4}), and (possibly) the different redshift targeted (see Section~\ref{sec:redshiftEvol}), all hamper a statistical one-to-one
comparison.
Notwithstanding these uncertainties, our data suggests that the optically thick gas -- if present also at $z\sim3$ in similar quantities -- is not responsible for 
the Lyman-alpha emission at $>50$~kpc distance from the central quasar as its covering factor is much higher than the one of the 
emitting gas (at least down to the current SB limits), and such optically thick phase should be bright enough (if illuminated by the quasar) 
to be detected in our observations as demonstrated in \citet{qpq4}. To get similar lower limits on the covering factor at 50~kpc or larger distances 
as in our data, one would have to consider only the lowest column density systems observed in the QPQ series.
We will focus on a detailed comparison between absorption and emission studies (also in individual systems at $z\sim3$) in future works.

\subsection{Compact line emitters associated with the extended Lyman-Alpha emission}
\label{sec:comp_sel}

The targets of QSO MUSEUM have been chosen without any prior constraint on their environment\footnote{We remind the reader that we 
selected our targets by avoiding very crowded fields in optical observations (see Section~\ref{sec:obs}). This should not biased our selection as most of the optical sources 
for which a field result to be crowded at the depth of SDSS-DR12, 
DSS2 and USNO-A2/B1 are low-$z$ objects.}.
However, the morphology, extent, and level of the \lya\ emission could be influenced by the presence of 
active companions, satellites, nearby systems and large scale structures (e.g., \citealt{Cai2016, fab+2018}). 
It has indeed been shown a tendency for ELANe to be associated with overdensities of galaxies and QSOs (\citealt{hennawi+15,Cai2016,fab+2018}, Arrigoni Battaia
et al. in prep.).
In this section we thus report on our attempt to find associated compact emitters.
Given the depth of our data and the FOV probed, this analysis is not conclusive but 
is intended as a first attempt to 
characterizing the small scale environment in which our objects reside. 

To identify galaxies at the same redshift of the targeted quasars, we rely on the identification of emission or absorption lines
at the redshift of interest. Specifically, we search for (i) compact continuum sources with clear line features and (ii)
compact \lya\ emitters (LAEs) whose continuum might have been too faint to be detected in the former case.
Thus, here, we do not attempt to obtain the redshift of continuum objects with no clear emissions or absorptions as 
it would be highly uncertain given the depth of our data. To avoid contamination from the noisy edges of the datacubes, we
restrict this analysis to the central $55\arcsec \times 55\arcsec$ (or $418$~kpc~$\times418$~kpc).

More specifically, we proceed as follows. First, we obtained a catalogue of continuum source candidates by running 
\textsc{CubExtractor} on the white-light images shown in Figure~\ref{fig:Prof_one} with a threshold of $2\sigma$ above the background root mean square 
and with a minimum detection area of 
5 pixels. Using the segmentation maps produced by \textsc{CubExtractor}, we have then extracted the 1D spectra for each 
identified source from the MUSE datacubes, resulting in an average of 54 sources per field for a total of 3280 sources. 
The inspection of such spectra has been conducted to identify emission or absorption within $\pm3000$~km~s$^{-1}$
from the systemic redshift of each targeted quasars.   
This large velocity window should allow us to find systems associated with most of the quasars and
nebulosities even though their current redshifts significantly differ (see Table~\ref{tab:sample} and Section~\ref{sec:morph}).
This search resulted in the discovery of 11 continuum selected sources characterized by some line emission close to the location of 
expected transitions for the redshift of interest. Apart from the contamination of six foreground objects with detected [\ion{O}{ii}], or H$\delta$ line emission,
we found (i) four active companions with detected \lya, and tentative \civ, \heii, and (ii) one object with only one line emission that 
we interpreted as \lya. 
In addition, for each field, we look for signatures of AGN activity for all the continuum-detected sources, identifying two higher z quasars and one lower z quasar.
For all these sources\footnote{Two companions (AGN1 and QSO2) and the continuum object with \lya\ emission (LAE1) have been discovered around the 
quasar with ID 13, and already reported in \citet{fab+2018}. We do not show them again here (the same applies for LAE2 associated with the same system). }, 
we show their spectra in Appendix~\ref{app:individual_fields}, while we indicate their positions with the corresponding names 
on the NB images in Figure~\ref{fig:Prof_one}, and listed them in Table~\ref{tab:ContinuumSources}. 
In particular, we note that the quasar with the larger number (2, of which 1 is a type-I) of active companions detected (ID 13 or PKS~1017+109) also presents the more extended 
and brightest of the \lya\ nebulosities (see \citealt{fab+2018}).

\begin{figure}
       \centering
       \includegraphics[width=0.98\columnwidth, clip]{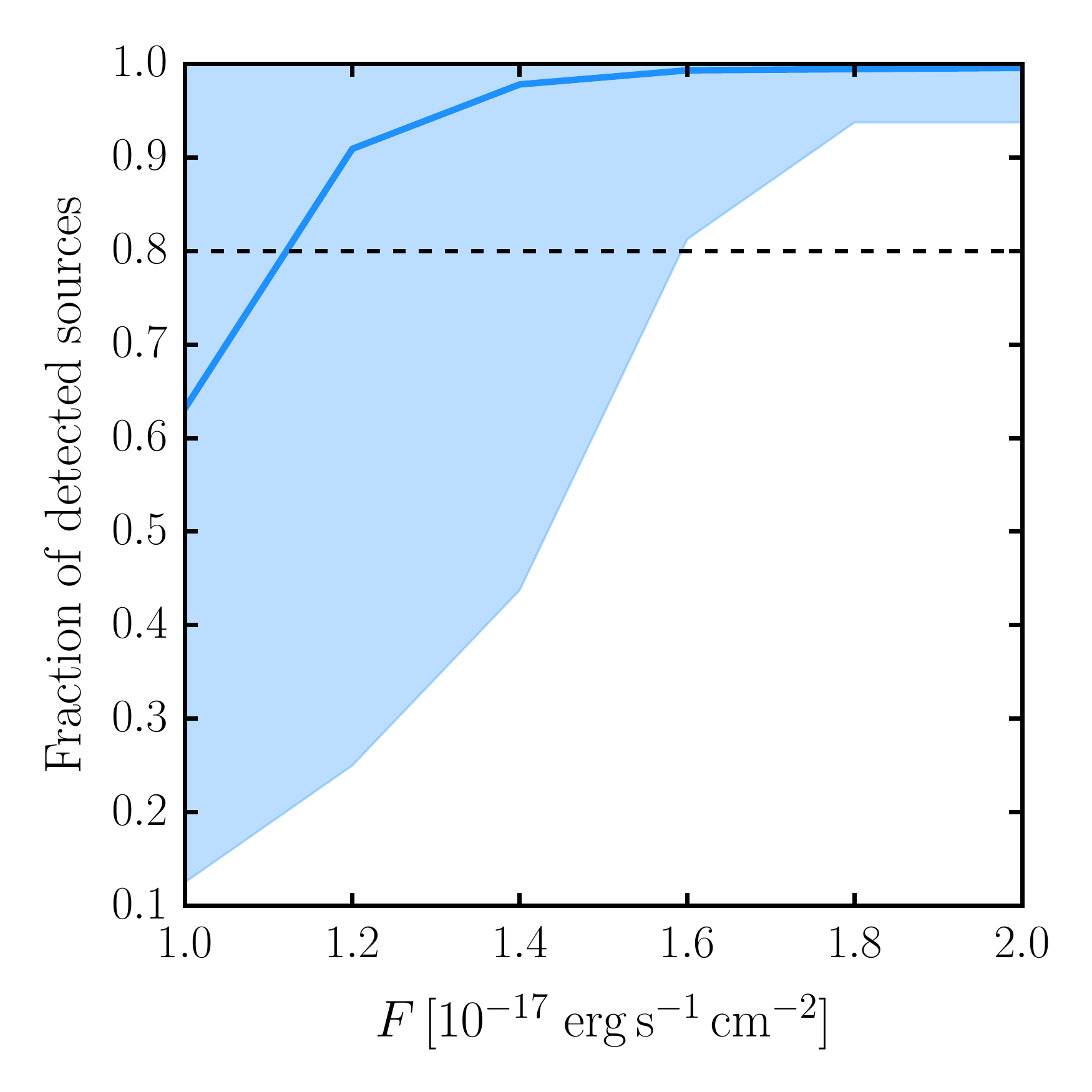}
    \caption{Results of a completeness test for the detection of \lya\ emitters associated with the targeted systems. The blue solid line indicates the 
    average recovered fraction, while the blue shaded region indicates the spread between the maximum and minimum of the recovered fraction for the individual fields.
    With our selection criteria, we are thus $\geq 80$~\% complete for fluxes $> 1.1 \times 10^{-17}$~erg~s$^{-1}$~cm$^{-2}$.}
\label{fig:completeness}
\end{figure}

The aforementioned analysis however misses all the line emitters whose faint continua were not detected in the white-light image. 
To search for such objects, we run \textsc{CubExtractor} to select groups of contiguous voxels within the PSF and continuum subtracted MUSE datacubes.
Specifically, we search for line emitters within a window of $\pm3000$~km~s$^{-1}$. To avoid 
contamination from spuriuous sources, 
we conservatively extract candidate sources with a minimum volume of 50 voxels at ${\rm S/N}\geq5$, and masking the location of residuals from stars that would otherwise 
contaminate our catalogue.
Note that the selected threshold would allow us to surely detect bright ($F>1.1 \times 10^{-17}$~erg~s$^{-1}$~cm$^{-2}$) 
line emitters associated with the quasars in most of our fields, if present.
We have indeed tested our selection criteria by introducing 100 mock sources with fluxes in the range $(1-2)\times10^{-17}$~erg~s$^{-1}$~cm$^{-2}$ 
into the PSF and continuum subtracted datacube for each targeted field. Each of the mock sources has a profile 
defined by a two-dimensional Gaussian in the spatial direction with FWHM equal to the estimated
seeing (on average FWHM~$=1.07$ arcsec), and a Gaussian with FWHM=$2.5$~\AA\ in the spectral direction. 
As it is clear from Figure~\ref{fig:completeness}, 
the aforementioned selection criteria allow us to be, on average, $\geq 80\%$ 
complete for line fluxes $F>1.1 \times 10^{-17}$~erg~s$^{-1}$~cm$^{-2}$.
Note however that for ten of our fields, the completeness is $\geq 80\%$ only at higher fluxes 
$F\gtrsim1.3 \times 10^{-17}$~erg~s$^{-1}$~cm$^{-2}$.

Applying this method to the data, we identify an average of 1 source per field. To avoid the selection of spurious sources, 
we have then visually inspected the ``optimally extracted'' line flux images, 
and extracted a 1D spectrum for each of the selected sources. We classify as \lya\ emitters only those compact sources which appear to 
not have other emission lines identifiable as e.g. H$\alpha$, [\ion{O}{iii}], H$\beta$, [OII]. 
In addition, for some fields, our selection criteria identify small portions of
the \lya\ nebulosities. We classify such portions as additional LAEs only
when they appear to be compact, centrally concentrated sources. This last choice, which happens 
in 9 cases out of 20 detected portions of nebulae\footnote{LAE2 in ID~8 (or UM683); LAE1, LAE2, AGN1 in ID~13 or (PKS~1017+109); 
LAE1 and LAE2 in ID~46 (or Q2355+0108); LAE1 in ID~47 (or 6dF~J0032-0414); LAE1 and LAE3 in
ID~56 (or TEX1033+137).}, 
has to be confirmed with additional data targeting the continuum emission of such
candidate companions. 

Following this procedure, we identify a total of 22 LAEs when considering all our sample. We list their coordinates, their redshifts and information on the \lya\
line emission in Appendix~\ref{app:individual_fields}, while we indicate them with their respective numbers on the NB images shown in Figure~\ref{fig:Prof_one}.

Overall, following our selection criteria, most of our fields (48) do not show the presence of LAEs down to our sensitivity limits 
and apertures used\footnote{For the quasar with ID~58 (Q1205-30) we were able to compare our analysis with NB data reaching the same depth (\citealt{Fynbo2000}; Appendix~\ref{app:individual_fields}), 
confirming that there are no LAEs in this particular field down to out limits.}. 
If we use 
the luminosity function for field LAEs by \citealt{Cassata2011} (bin $3<z<4.55$)
and assume a comoving volume of $231$~Mpc$^3$ defined by the MUSE FoV and the window of $\pm 3000$~km~s$^{-1}$,
we find that we would expect $0.6\pm0.3$ LAEs per field\footnote{The errors listed for the expected number of LAEs in this paragraph encompass the
different estimates due to the sensitivity of each individual field, log$(L_{\rm Ly\alpha}/[{\rm erg~s^{-1}}])=42-42.1$.}. 
A similar value, $0.7\pm0.4$, is obtained by adopting the more recent luminosity function of \citet{Drake2017}. 
If we thus take into account the estimates of the expected number of LAEs per field (ranging between about 0.3 and 1) and the 
fact that we are 80\% complete, we obtain a total expected number of LAEs in the range $24-30$, not far from the total of 22 found\footnote{Note that also 
the 5 objects with continuum detection reported in this section (4 companions + 1 \lya\ emitter) could be classified as LAE from their \lya\ ${\rm EW}_{\rm rest}$, summing up to 27 LAEs in total.}.
Therefore, at these luminosities, our quasar environments seem to be similar to the field, suggesting 
that on average quasars do not inhabit the most dense environments at their redshift (\citealt{ts12,Fanidakis13}).
Nevertheless, the fields with more than one detection are clear outliers, and probably reflect the presence 
of underlying overdensities and/or effects of powering mechanisms, e.g., 
chance alignment between the position of the LAE and the ionization cones of the quasars (e.g., \citealt{hr01}).

Some of the detected LAEs, expecially the one 
with high equivalent widths (${\rm EW}_{\rm rest}>240$~\AA; \citealt{cantalupo12}) 
might indeed represent candidates for the so-called  ``dark galaxies'', i.e. dense pockets of gas 
fluorescently illuminated by the targeted quasars (e.g. \citealt{hr01}).
Using the luminosity function from \citet{cantalupo12}, who reported candidate 
LAEs\footnote{The LAEs in \citet{cantalupo12} are not spectroscopically confirmed.} around one $z\sim2.4$ hyper-luminous quasar 
(bolometric luminosity $L_{\rm bol}>10^{47}$~erg~s$^{-1}$), 
we would expect to see 1 LAE per field above our luminosity limit\footnote{Note however that not all our quasars are hyper-luminous.}. 
\citet{cantalupo12} found indeed only 1 LAE with luminosity high enough to be detected by our observations, and at a 
'MUSE distance' from the quasar. This LAE has an exceptional ${\rm EW}_{\rm rest}=327\pm28$\AA.
Recently, \citet{Marino2018} also searched for ``dark galaxy'' candidates with deep MUSE observations around a sample of 5 QSOs 
(4 at $z\gtrsim 3.7$ and 1 at $z\approx 3.2$) and 1 type-II AGN.
Unfortunately, from the unprecise redshifts reported for their LAEs in their Table 5, we are not able to verify how many LAEs they detected down to our luminosity limit and
within our velocity window for the whole sample. However, we can see that they detected only 2 LAEs 
for their 4 QSOs at $z\gtrsim3.7$ matching our criteria (see their Fig.~7, right panel). Both of these sources are classified as 
candidate ``dark-galaxies'' (see their Table 3). On the contrary, no bright ``dark-galaxy'' candidates are reported within $\pm 3000$~km~s$^{-1}$ for their $z\approx3.2$ 
QSO (see their Table 3 and Fig.~4).
In our sample, we only have 1 LAE (LAE2 in the SDSS~J1209+1138 field; see Table~\ref{MEGAtabLAEs}) with an exceptional equivalent width of $EW_{\rm rest}>420$~\AA\ that could be 
considered as a ``dark-galaxy'' candidate as usually done in the literature.
We conclude that object-to-object variations can play an important role at these 
high luminosities, and within our small field-of-view, explaining the difference with \citet{cantalupo12}.

Finally, we stress that the analysis presented in this section is inevitably limited in identifying companions 
within the \lya\ nebulosities themselves, unless
they are characterized by a brighter emission than the extended \lya. We thus do not exclude the presence of 
faint galaxies embedded within the extended \lya\ emission
surrounding each quasar in our sample.

\subsection{Kinematic information from the extended Lyman-Alpha emission}
\label{sec:kinematics}

In this section we present the maps for the flux-weighted velocity centroid and the flux-weigthed velocity dispersion of the extended \lya\ emission calculated 
as explained in section~\ref{sec:cubex} (following the same analysis of \citealt{Borisova2016}). 
We reiterate that we do {\it not} fit any function to the line profile to
avoid neglecting fundamental information on the gas properties imprinted by the \lya\ resonant scattering, which is likely in play here.

\begin{figure*}
       \includegraphics[width=0.85\textwidth]{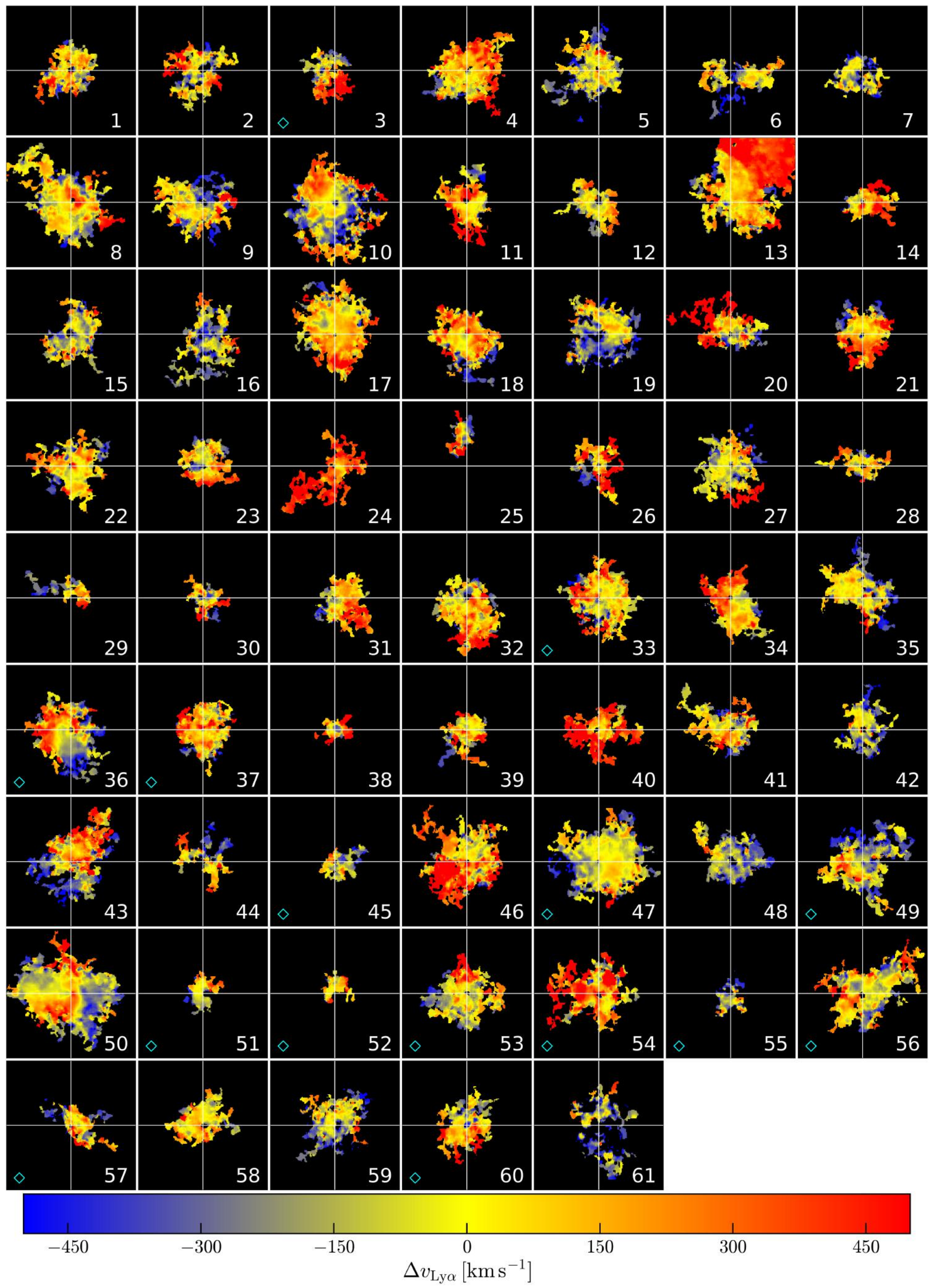}
    \caption{Atlas of the 61 maps for the flux-weigthed velocity cendroid of the \lya\ emission around the quasars in the QSO MUSEUM sample.
    The flux-weighted centroid is calculated as first moment of the flux distribution with respect to the peak of the 
    \lya\ emission of each nebulosity $z_{\rm peak\, Ly\alpha}$ (Table~\ref{tab:sample}). Each image is presented on the same scale of Figure~\ref{fig:SBmaps}, $30\arcsec\times30\arcsec$ 
    (or about $230$~kpc~$\times 230$~kpc), even though we use only the information within the extracted 3D-masks to compute these ``velocity maps'' (see Section~\ref{sec:cubex}).
    In each image the white crosshair indicates the position of the quasar prior to PSF subtraction. 
    A cyan diamond in the bottom-left corner indicates that the quasar is radio-loud.
    The detected extended \lya\ emission shows complex and noisy 
    maps on tens of kiloparsecs, while more coherent patterns on larger scales (see ID 13).}
    \label{fig:Velmaps}
\end{figure*}

Figure~\ref{fig:Velmaps} shows the atlas of the 61 maps for the flux-weighted velocity centroid with respect to $z_{\rm peak\, Ly\alpha}$ (Table~\ref{tab:sample})\footnote{Following 
the analysis in section~\ref{sec:morph}, all the discovered \lya\ nebulosities would appear mainly as red in Figure~\ref{fig:Velmaps}, i.e. as largely redshifted, 
if the flux-weighted velocity centroid would have been calculated with respect to the quasars sytemic redshifts (see Figure~\ref{fig:histoDeltaV} for the velocity shifts).}.
In agreement with \citet{Borisova2016}, the majority of the maps looks noisy and difficult to interpret as ordered motions of any kind (rotation, infall, or outflow), though 
each nebulosity shows peculiar structures. 
On the other hand, the largest and brightest nebula (ID 13) show a remarkable velocity pattern on hundreds of kiloparsec, with the furthest portion clearly redshifted. 
This coherent velocity pattern has been interpreted as signature of inspiraling motions within the quasar's halo (\citealt{fab+2018}). 

The difference in complexity between the maps of very extended neulosities and smaller systems 
could be due to the resonant nature of the \lya\ line which sample more turbulent regions on scales of tens of kpc. Such small scales around the quasars are likely more
affected by both turbulent motions due to 
galaxy formation processes (feedback, interactions) and the resonant scattering of \lya\ photons coming from 
the quasar itself (see Section~\ref{sec:powering} for further discussion). 

\begin{figure*}
       \includegraphics[width=0.85\textwidth]{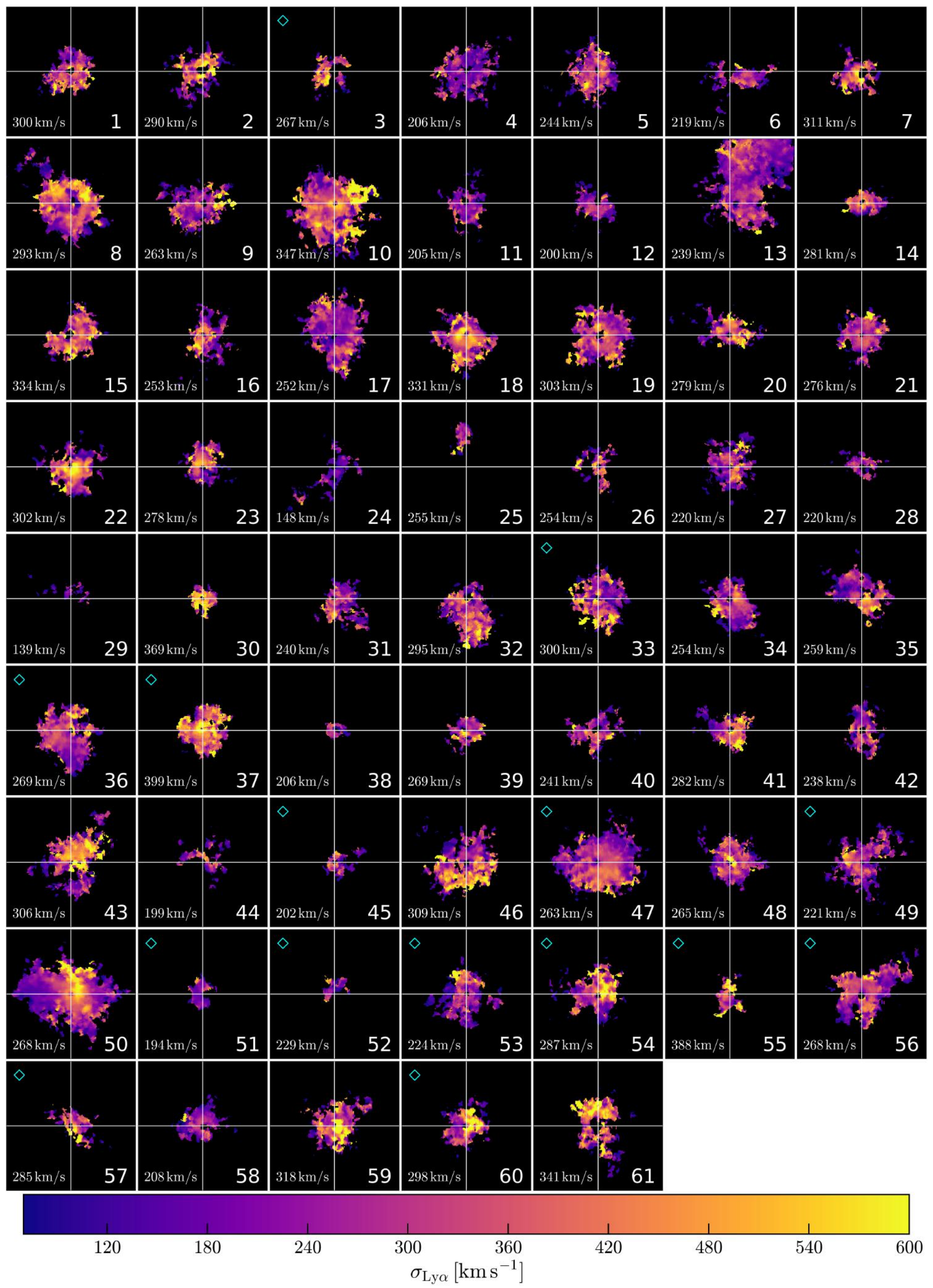}
    \caption{Atlas of the 61 maps for the flux-weigthed velocity dispersion of the \lya\ emission around the quasars in the QSO MUSEUM sample.
    The velocity dispersion is calculated as second moment of the flux distribution. Each image is presented on the same scale of Figure~\ref{fig:SBmaps}, $30\arcsec\times30\arcsec$ 
    (or about $230$~kpc~$\times 230$~kpc), even though we use only the information within the extracted 3D-masks to compute these ``velocity-dispersion maps'' (see Section~\ref{sec:cubex}).
    In each image the white crosshair indicates the position of the quasar prior to PSF subtraction. 
    A cyan diamond in the top-left corner indicates that the quasar is radio-loud.
    We indicate in the bottom-left corner of each image the average velocity dispersion 
    $\langle \sigma_{\rm Ly\alpha} \rangle$ for each map. The extended \lya\ nebulae show an average velocity dispersion 
    $\langle \sigma_{\rm Ly\alpha} \rangle < 400$~km~s$^{-1}$ (or FWHM$<940$~km~s$^{-1}$), 
    irrespective of the radio-loudness of the targeted quasars.}
    \label{fig:Sigmamaps}
\end{figure*}

Figure~\ref{fig:Sigmamaps} shows the atlas of the 61 maps for the flux-weighted velocity dispersion. In the bottom left corner 
of each image we indicate the average velocity dispersion $\langle \sigma_{\rm Ly\alpha} \rangle$ within each nebulosity.
We find that the extended \lya\ emission has  
$\langle \sigma_{\rm Ly\alpha} \rangle < 400$~km~s$^{-1}$ (or FWHM$<940$~km~s$^{-1}$) in all the
cases, irrespective of the radio-loudness of the targeted quasars. 
These estimates are similar to the velocity widths observed in absorption in the CGM surrounding $z\sim2$ quasars ($\Delta v>300$~km~s$^{-1}$; \citealt{qpq3,qpq9}).

As done in \citet{qpq9}, we can compare these values with the expected velocity dispersion for a dark matter halo hosting quasars.
In particular, it has been found that the maximum of the average one-dimensional root-mean-square (rms) velocity $\sigma_{\rm rms-1D}$ is related to the maximum circular velocity 
within a dark matter halo by $\sigma_{\rm 1D}=V_{\rm circ}^{\rm max}/\sqrt{2}$ (\citealt{Tormen1997}). If we assume 
dark matter halos of M$_{\rm DM}\sim10^{12.5}$~M$_{\odot}$ (\citealt{white12}) at $z\sim3$, 
an NFW profile (\citealt{Navarro1997}) with a concentration parameter of $c=3.7$ (\citealt{Dutton2014}), 
the maximum circular velocity is $V_{\rm circ}^{\rm max}=360$~km~s$^{-1}$.
Hence $\sigma_{\rm rms-1D}=250$~km~s$^{-1}$.
The motions within all these \lya\ nebulae 
have thus amplitudes consistent with gravitational motions expected in dark matter halos hosting quasars, 
with part of the velocity dispersion likely due to \lya\ resonant scattering and instrument resolution (FWHM$\approx2.83$\AA\ or 170 km~s$^{-1}$ at $5000$\AA). 

\begin{figure}
       \centering
       \includegraphics[width=0.95\columnwidth]{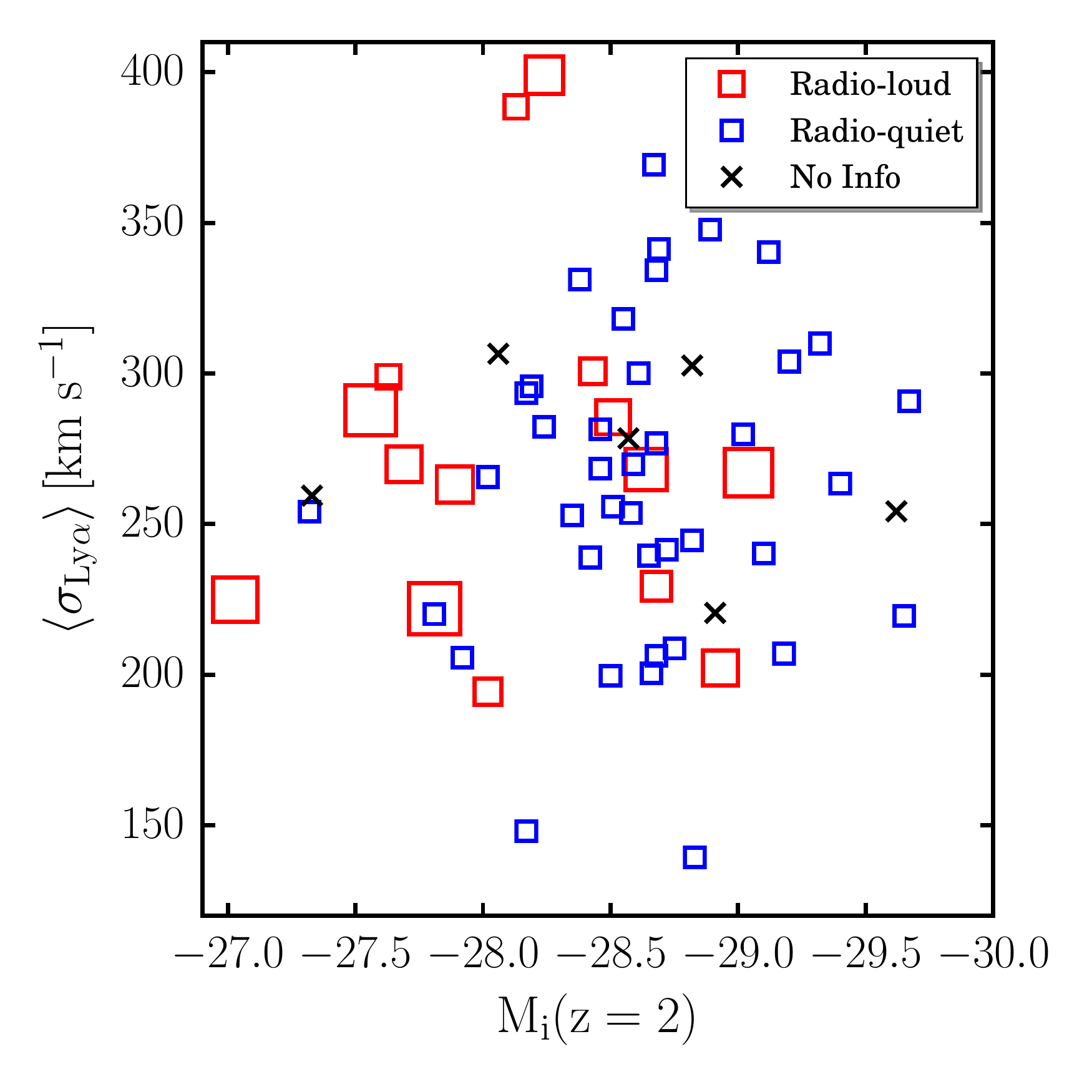}
    \caption{Plot of the average flux weighted velocity dispersion $\langle \sigma_{\rm Ly\alpha} \rangle$ of each \lya\ nebulosity versus the absolute 
    $i$-band magnitude normalized at $z=2$, $M_i(z=2)$, using 
    \citet{Ross2013}. We indicate in blue the radio-quiet systems, while in red the radio-loud objects. The size of the symbol for the radio-loud objects symbolizes the
    flux listed in Table~\ref{tab:sample}, i.e. larger symbols mean larger fluxes following the relation ${\rm size}_{\rm radio} = {\rm size}_{\rm quiet} + 5\times{\rm log} (F_{\rm Radio})$.
    The crosses indicates the objects for which we do not have information on their radio flux. The average flux-weighted velocity dipersion do not depend 
    on the current quasar luminosity or strength of radio emission.} 
    \label{fig:sigmavsMi}
\end{figure}

To check for any dependence on the quasar luminosity and on the strength of the radio emission, in Figure~\ref{fig:sigmavsMi} 
we plot the average flux-weighted velocity
dispersion for each nebulosity as a function of $M_i(z=2)$ for each targeted quasar. 
On the same plot, we differentiate between objects with measured radio fluxes and radio-quiet object (red and blue), and use 
larger symbols for quasars characterized by stronger radio emission.
It is clear that there are no trends in both luminosity or radio strengths, confirming that the mechanisms responsible for 
the broadening of the \lya\ line on 50 kpc scales do not depend on these properties for at least our observed ranges ($F_{\rm Radio}<900$~mJy; 
$M_i(z=2)\gtrsim-29$ for radio-loud objects, and $M_i(z=2)>-29.7$ for radio-quiet).

We note however that the relatively quiescent \lya\ nebulosities around all the 15 radio-loud quasars in our sample run counter to the two 
radio-loud systems with $\sigma>425$~km~s$^{-1}$ (or FWHM$>1000$~km~s$^{-1}$) showed by \citet{Borisova2016}. 
Specifically, the two radio-loud objects in \citet{Borisova2016}, PKS~1937-101 and QB~2000-330, 
have a peak flux at $1.4$~Ghz of $838.8\pm25.2$~mJy and $446.8\pm15.7$~mJy (\citealt{Condon1998}), and a 
magnitude of $M_{i}(z=2)=-30.35$ and $-29.77$, respectively.
Even though we have in our sample quasars as strong in radio emission as these two, i.e. ID 49 and 54 ($>700$~mJy), our quasars
are 2.4 magnitudes fainter and show $\langle \sigma_{\rm Ly\alpha} \rangle = 221$, and $287$~km~s$^{-1}$ (or FWHM~$=519$, and $647$~km~s$^{-1}$). 
This difference thus suggests that -- if an outflow is present in the radio-loud systems --  (i) the winds are driven by the 
radiative output of the AGN
rather than by a jet, and/or (ii) the coupling between the \lya\ emitting gas on large scales (tens of kpc) and an outflow 
is more effective around more luminous AGN, and/or (iii) more luminous radio-loud AGN are able to sustain coherent winds (or jets) 
for longer timescales. 

It is interesting to note that also in the radio-quiet population there can be extended \lya\ nebulosities showing broad emission. 
In particular, \citet{Borisova2016} showed one radio-quiet quasar in their sample, J0124+0044 (z=3.783), with broad extended \lya\ emission.
The statistics from 
our work together with \citet{Borisova2016} suggests a very low rate, 1 out of 56 (39+17), which seems to not depend on the current 
quasar luminosity\footnote{The object in \citet{Borisova2016} with
broad \lya\ emission has $M_i(z=2)=-28.98$}. 
We further discuss these rare broad extended \lya\ nebulosities (for both radio-quiet and radio-loud systems) 
in Section~\ref{sec:timescale}.

\subsection{Spectral shape and asymmetry of the extended Lyman-Alpha emission}

As the \lya\ photons are expected to experience frequency excursions in the scattering process (\citealt{Dijkstra2017} and references therein),
and the resulting change in frequency depends both on local and unrelated structures, 
the shape of the \lya\ line can be fundamental in understanding 
the astrophyiscs of the studied system,
e.g. presence of inflows/outflows (e.g., \citealt{Laursen2009}), absorption from the foreground IGM.
For this reason, it is particular important to quantify the asymmetries of the \lya\ line profile, and/or the presence of different peaks (e.g., \citealt{Gronke2016}).  

Unfortunately, our data lack the necessary sensitivity to clearly assess the presence of asymmetric wings in the line profile.
This has been indeed tested by calculating the third moment of the flux distribution for the extended \lya\ emission as done for the 
smaller moments, leading to poorly constrained and inconclusive values.
By averaging over the whole nebulae we could achieve higher sensitivity, but we would complicatedly mix, and washed out the low S/N features we are looking for.

In addition, we do not detect clear double-peaks profile -- expected in the case of optically thick gas 
(e.g., \citealt{Neufeld_1990}) -- in any of the discovered \lya\ nebulosities.
However, this result may be driven by the spectral resolution of MUSE at these wavelengths (FWHM$\approx2.83$\AA\ or 170 km~s$^{-1}$ at $5000\AA$).

Therefore, for this kind of detailed analysis, we would need a dataset with a much higher S/N for the wings of the \lya\ emission, and with higher spectral resolution.
Now that the level of the \lya\ emission is known, we can plan such experiments.

\section{Discussion}
\label{sec:disc}

From the analysis presented in the previous sections, it is clear that we have detected extended \lya\ emission 
on tens of kpc scales around each of the targeted quasars. 
In this section we discuss some of the implications of the observations presented in this work, 
relying only on the information coming from the \lya\ emission.
In future studies, we will revisit these points with the use of additional diagnostics.

\subsection{Small detection rate for Enormous Lyman-Alpha Nebulae}
\label{sec:ELAN}

Recent observations have shown the existence of Enormous Lyman-Alpha Nebulae (ELAN; \citealt{cantalupo14,hennawi+15,Cai2016})
extending for hundreds of kiloparsecs around $z\sim2$ quasars, with SB$_{\rm Ly\alpha}\sim10^{-17}$~\unitcgssb.
Current statistics suggest that the ELAN are very rare structures which are found around a few percent of the 
targeted radio-quiet quasars at $z\sim2$ (\citealt{qpq4,hennawi+15}). One of the aim of the QSO MUSEUM survey, 
together with on-going efforts (e.g., \citealt{Cai2018}), is to better characterize the frequency of this phenomenon, and
understand why these nebulae are so uncommon (e.g., presence of companions, halo mass, illumination).

Thanks to our QSO MUSEUM survey, and the previous work of \citet{Borisova2016}, we have now amassed a sample of 80 ($61+19$) 
quasars at $z\sim3$ down to SB$_{\rm limit}\sim {\rm few}\times 10^{-18}$\unitcgssb. 
Of all these quasars, only one object shows the extreme observed surface brightnesses and extents
found in the ELAN at $z\sim2$, i.e. the ID 13 in this work or PKS~1017+109. All the other nebulosities show lower
levels of \lya\ emission or extends only to much smaller distances from the central quasar down to the 
same SB limits (see Figure~\ref{fig:asymmetry}).
The ELAN around the quasar PKS~1017+109 has been studied in detail in \citet{fab+2018}, where we have indicated the 
common peculiarities with the other ELAN discovered at $z\sim2$, with particular emphasis on the identification 
of active companions in their surroundings.
 
It has been proposed that the higher SB$_{\rm Ly\alpha}$ in these ELAN traces patches with higher volume densities 
than the average population at both redshifts (\citealt{hennawi+15,fab+16}) and/or gas reprocessing the additional
contribution to the ionizing radiation from the active companions (\citealt{fab+2018}) in the environment of ELAN.
The higher density patches could be due to substructures interacting with the main halo hosting the targeted quasars
(\citealt{fab+2018}).

\begin{figure*}
       \centering
       \includegraphics[width=0.75\textwidth, clip]{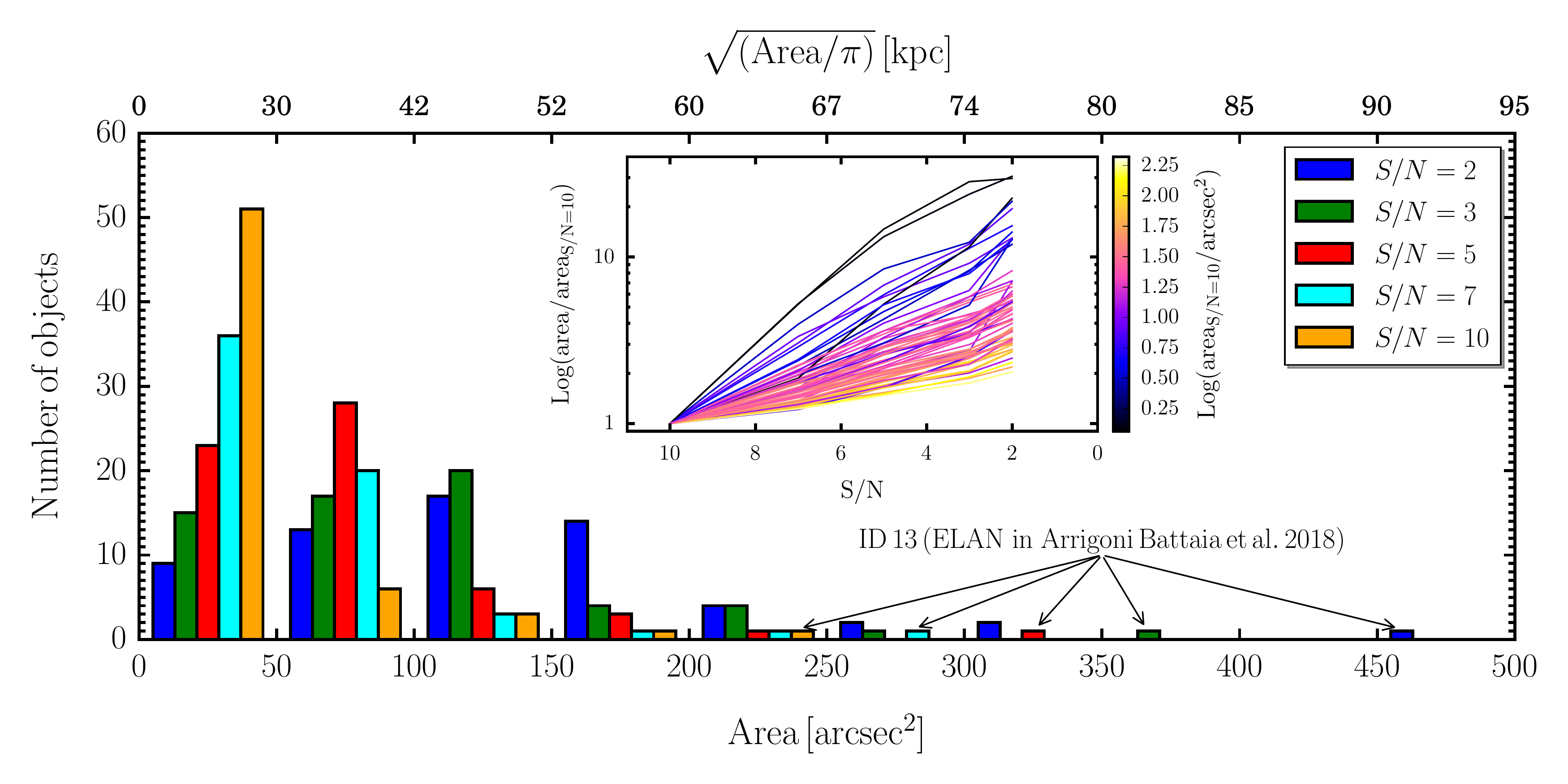}
    \caption{Histogram of the variation of the area for each \lya\ nebulosity for different S/N thresholds (different colors as indicated in the legend). 
    The ELAN discovered in QSO MUSEUM is indicated with
    arrows. The plot in the inset shows the logarithmic variation of the area (normalized to the area for ${\rm S/N}=10$) 
    of the \lya\ nebulae with respect to the S/N threshold. 
    In this plot, we color-coded the lines for each nebula following the area for ${\rm S/N}=2$ as listed in Table~\ref{Tab:LyaNeb}.
    This plot clearly state that ELAN can be detected with very fast integrations with current facilities. 
    Smaller \lya\ nebulae have the potentiality of ``growing'' faster than ELAN 
    when the exposure time is increased, probably reflecting the different scales sampled (CGM vs CGM+IGM).}
    \label{fig:AreaVSSN}
\end{figure*}

Further, to clarify how easy it is to detect ELAN with current facilities and 
how a detected \lya\ nebulosity around a quasar could ``grow'' with longer integration times, 
in Figure~\ref{fig:AreaVSSN} we show the histogram of the area spanned by each of the \lya\ nebulosities in our sample for different
S/N detection threshold (the different colors). For each histogram, we indicate the position of the 
nebula around the quasar with ID 13, or PKS~1017+109 (\citealt{fab+2018}). 
It is clear that ELAN are clearly visible as extended objects even with shorter integration times than used here, 
as the area for ${\rm S/N}=10$ is already $>200$~arcsec$^2$ for ID 13.
In the same figure, an additional x-axis indicates the corresponding radius for a circular nebula with the same area, i.e. $R=\sqrt{(Area/\pi)}$.
Even with the threshold of ${\rm S/N=10}$ this ELAN has $R>60$~kpc. 

The plot in the inset of Figure~\ref{fig:AreaVSSN} shows how the area grows for each nebulosity as a function of 
S/N with respect to the area for ${\rm S/N}=10$. In the same plot, we color coded the line for each \lya\ nebula 
depending on the final observed area for ${\rm S/N}=2$. This plot shows that the nebulae with the larger sizes (lines with lighter colors) do not increment
their extent as fast as smaller nebulae if deeper observations are conducted. This occurrence could reflect the different scales probed by 
the different nebulae. Indeed, while small \lya\ nebulae most probably trace the gas within the central portion of the dark matter halo hosting a quasar, 
nebulae with $100$~kpc sizes start to probe IGM regions where the densities are expected to be lower.
This has been also the case for the nebula around UM~287 (\citealt{cantalupo14}), which was detected after only 20 minutes of NB imaging with the 
Low Resolution Spectrograph (LRIS; \citealt{Oke1995})
instrument on the Keck telescope. The size of the nebula did {\it not} increase significantly with the total observing time acquired in those observations of 
$\sim10$~hours.

If the relation for the large nebulosities shown in the inset of Figure~\ref{fig:AreaVSSN} holds for even larger areas, 
one would need to increase the S/N by a factor of 8 to roughly double the detected area.
This would then require additional $\sim64$~hours on a previously detected 200~arcsec$^2$ \lya\ nebula in data as deep as our survey. 
However, the characteristics of the powering mechanisms and the volume density of the gas can 
both change on larger scales, most probably requiring longer integration times to detect the IGM.

\subsection{Is there a redshift evolution of extended Lyman-Alpha emission around quasars?}
\label{sec:redshiftEvol}

As anticipated in Section~\ref{sec:av-prof} and shown in Figure~\ref{fig:averageProf}, the current
observations of \lya\ halos around radio-quiet quasars hint to a clear difference between the 
$z\sim3$ and $z\sim2$ population. Indeed, the \lya\ emission seems to be much stronger at $z\sim3$
than at $z\sim2$, maybe suggesting a larger mass in the cool ($T\sim10^4$~K) phase of the CGM at $z\sim3$ than at 
$z\sim2$ ($M_{\rm cool}>10^{10}$~M$_{\odot}$; \citealt{QPQ5,QPQ7}). Here we investigate this hypothesis.

To further check the presence of a redshift evolution, we split our large sample of radio-quiet quasars
into two redshift bins, and look for any trend that would confirm the strong difference between $z\sim2$ and $z\sim3$.
Given the large uncertainties on the systemic redshift of our sample (see Section~\ref{sec:systemic_redshift}), we 
built the two redshift bins by maximizing the number of objects in each bin, while excluding 
quasars whose redshift is so uncertain that they could sit in the other bin both because of the uncertainty
on the systemic redshift $z_{\rm systemic}$, and/or the difference in redshift between the \lya\ nebula $z_{\rm peak\, Ly\alpha}$ 
and $z_{\rm systemic}$. We find that a cut $z_{\rm peak\, Ly\alpha}<3.191$ and $z_{\rm peak\, Ly\alpha}>3.229$ satisfies our
conservative criteria, resulting in two almost equally populated subsamples of 17 and 18 objects each, respectively. The selected quasars cannot change their bin even with a $3\sigma$ 
deviation from their current redshift.
The median redshift for the two samples is $z_{\rm low}=3.114$, and $z_{\rm high}=3.336$.
Such a redshift difference corresponds to about 0.15 Gyr in the standard cosmology assumed in this work. 
For each bin, we then computed the average profile as performed in Section~\ref{sec:av-prof}.

\begin{figure*}
       \centering
       \includegraphics[width=0.9\textwidth, clip]{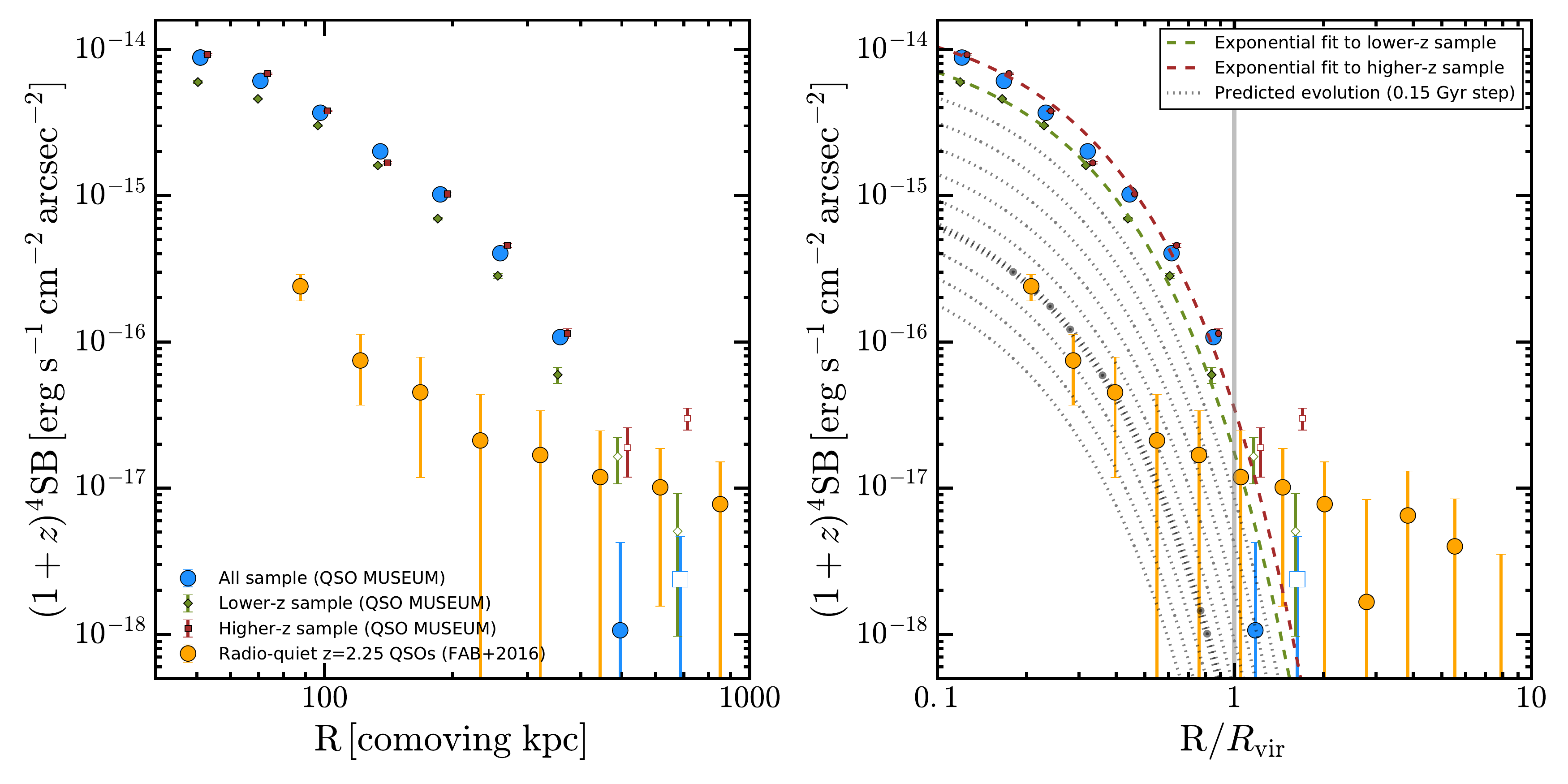}
    \caption{Redshift evolution of \lya\ emission around quasars. Left panel: we show the average profile for the whole 
    sample (large blue; 61 targets), the radio-quiet lower-$z$ subsample (small dark-green diamond; 17 quasars), and radio-quiet higher-$z$ subsample (small brown squares; 18 quasars)
    after correcting the SB for the cosmological dimming and using comoving units for the radii.
    For each profile, open symbols indicate negative datapoints.
    We compare our dataset with the average profile 
    for $z\simeq2.25$ quasars by \citet{fab+16}. Right panel: the average SB profiles presented in the left panel are now plotted with respect to the proper radius normalized to the virial radius 
    $R_{\rm vir}=[3M_{\rm halo}/(800\pi\rho_{\rm crit}(z))]^{1/3}$ expected for halos hosting quasars at these redshifts (see Section~\ref{sec:redshiftEvol} for details).
    We show the best exponential fit to the two subsamples (colored dashed lines). Using the ratio between the parameters of these two fits we predict the evolution for lower-redshifts in steps of about 0.15 Gyr (dotted lines).
    The thicker dotted line corresponds to about 1.05 Gyr, the difference between the higher-$z$ subsample and $z\simeq2.25$.  For reference, the gray vertical line indicates 1 Rvir.
    Our observations thus suggest a clear evolutionary trend (see Section~\ref{sec:redshiftEvol} for discussion).}
    \label{fig:EvolProf}
\end{figure*}

The left panel in Figure~\ref{fig:EvolProf} shows the comparison of the average surface brightness profiles for the 
two redshift bins with the profile for the whole sample, and the profile at $z\simeq2.25$ (\citealt{fab+16}) after correcting all for the SB dimming.
A difference between the higher and the lower redshift bin is evident, with the lower-$z$ profile being always
below the data-points for higher-$z$.
The average ratio between the data-points for the two subsamples is $1.5\pm0.1$, while the profile for $z\simeq2.25$
is a factor of about $17\pm8$ lower (two central points) with respect to the lower-$z$ subsample.   
These differences are thus significant given the estimated uncertainties.

What is the cause of such strong variations? We first explore the possibility that such differences are due to
the cosmological evolution of densities and scales. \citet{MasRibas2018} investigated the redshift scaling of a low SB
signal in the case of electron-scattering around quasars. They show that the cosmological evolution of densities and 
scales cancel out the cosmological effect of surface brightness dimming $(1+z)^{-4}$ in their particular case.
Here, we perform a similar calculation, but in the case of \lya\ emission from gas clouds that are optically thin to 
Lyman continuum photons (neutral Hydrogen column density $N_{\rm HI}\ll10^{17.2}$~cm$^{-2}$). 
This emission scenario is indeed the most favored scenario for extended \lya\ emission around AGN (e.g., \citealt{heckman91a,fab+15b,fab+16,Cai2016}; see also discussions in \citealt{qpq4,cantalupo14}).

Following the formalism in \citet{qpq4}, the observed \lya\ surface brightness in an optically thin scenario scales as SB$_{\rm Ly\alpha}\propto (1+z)^{-4}n_{\rm H}N_{\rm H}$, 
where $n_{\rm H}$ is the Hydrogen volume density, and $N_{\rm H}$ is the Hydrogen total column density.
It can be shown that $N_{\rm H}\propto n_{\rm H}R_{\rm vir}$ (see e.g. equation 2 in \citealt{qpq4}), where  $R_{\rm vir}$ is the virial radius of the quasars' halo,
$R_{vir}\equiv R_{200}=[3M_{\rm halo}/(800\pi\rho_{\rm crit}(z))]^{1/3}$ \footnote{Here we use as virial radius the radius $R_{200}$ at which the average density of a matter overdensity is 200 times the cosmic mean density.}.
To compute the dependence with redshift we thus need an estimate of the quasars' halo mass at the redshifts of interest.
The current consensus in the literature indicates that quasars at $z\sim2$ and $z\sim3$ inhabits dark-matter halos with similar masses ($M_{\rm halo}=10^{12.5}$~M$_{\odot}$) independent of their luminosities, 
though with large uncertainties (e.g., \citealt{daAngela2008,Kim&Croft2008,white12,Trainor2012} ).
If we then assume the same halo mass for the different redshifts, $R_{\rm vir}$ changes only because of the evolution of the critical density $\rho_{\rm crit}(z)$, thus increasing by 
a factor (1+z) towards lower redshifts.
Therefore, because the cosmological evolution sets $n_{\rm H}\propto (1+z)^3$, and $N_{\rm H}\propto n_{\rm H}R_{\rm vir}\propto(1+z)^2$, we find an overall scaling
of SB$_{\rm Ly\alpha} \propto (1+z)$ or SB$_{\rm Ly\alpha}^{\rm no\, dimming} \propto (1+z)^5$ when taking out the factor for the SB dimming as in Figure~\ref{fig:EvolProf}.
If the changes in the \lya\ profiles were caused simply by the cosmological evolution of density and scales, we would thus expect a ratio
of 1.3 and 3, respectively between the two subsamples at $z\sim3$, and between $z\simeq2.25$ and the lower-$z$ subsample.
These values differ with the aforementioned observed ratios ($1.5\pm0.1$ and $17\pm8$), suggesting that the profile evolution is much stronger than expected from
the cosmological density evolution.
This result matches well with the high densities ($n_{\rm H}> 1$~cm$^{-3}$) invoked to explain the levels of the observed \lya\ emission (\citealt{heckman91a, cantalupo14,fab+15b, hennawi+15}).
Such a high density gas most likely does not trace the evolution of the cosmic mean density.

We then focus on physical processes that could act on CGM scales.   
To have a better handle on the profiles evolution on such scales, we plot the SB values with respect to the
physical distance normalized to the virial radius of the quasars' halo at each redshift, $R_{\rm vir}$.
As previously done, we assumed the same halo mass for the redshift subsamples, finding $R_{\rm vir}^{z\sim2}\simeq 130$~kpc and $R_{\rm vir}^{z\sim3}\simeq 100$~kpc, 
respectively for $z\simeq2.25$ and $z\sim3$.
The right panel of Figure~\ref{fig:EvolProf} shows this test. 
At $z\sim3$ the exponential behavior of the SB profiles now seems to well describe the presence of emitting gas within a halo of radius $R_{\rm vir}$, with a consequent dimming of the \lya\ emission at larger distances.
At $z\sim2$ the \lya\ emission seems instead more concentrated in the central portion of the halos, with tentative indications of a fainter signal at larger separations.

\begin{table}
\scriptsize
\caption{Exponential fit of the Ly$\alpha$ surface brightness profiles of the two radio-quiet subsamples after correction for SB dimming and
normalization of physical distance with $R_{\rm vir}$ (right panel in Figure~\ref{fig:EvolProf}).}
\centering
\begin{tabular}{lccccc}
\hline
\hline
Sample   &  \#     &  $\langle z_{\rm Ly\alpha} \rangle$	 &  $C_e^{\rm  \ \ a}$ & $x_h^{\rm \ \ a}$ & $\chi^{2 \ \ {\rm b}}_e$  \\
     &    &    & ($10^{-14}$)   &  &   \\
\hline
Higher-$z$$^{\rm c}$ & 18  & 3.336 & 1.95 & 0.16 & 7.1 \\
Lower-$z$$^{\rm c}$ & 17 & 3.114 & 1.36 & 0.15 & 0.3  \\
\hline
\end{tabular}
\label{Tab:fit_subsamples}
\flushleft{$^{\rm a}$ Best-fit parameters assuming SB profile ${\rm SB}(x)=C_e {\rm exp}(-x/x_h)$, where $x=R/R_{\rm vir}$. 
$C_e$ is in unit of $10^{-14}$~erg~s$^{-2}$~cm$^{-2}$~arcsec$^{-2}$. Note that the SB has been multiplied by $(1+z)^4$ before the fit (Figure~\ref{fig:EvolProf}).}
\flushleft{$^{\rm b}$ Reduced chi square for the exponential fit.}
\flushleft{$^{\rm c}$ Both subsamples include only radio-quiet quasars (see Section~\ref{sec:redshiftEvol} for details on the selection).}
\end{table}

To quantitatively constrain the difference between profiles and check if there is any relation between their shapes, we proceed as follows.
First, we fit the two subsamples at $z\sim3$ with an exponential function ${\rm SB}(x)=C_e {\rm exp}(-x/x_h)$ (colored dashed lines in Figure~\ref{fig:EvolProf} and Table~\ref{Tab:fit_subsamples}).
Being  $\approx0.15$~Gyr one from the other, the ratio between the parameters of these profiles ($C_e$ and $x_h$) could define a trend for the evolution towards lower-$z$.
We then construct other exponential functions by using such ratios (thus in steps of 0.15 Gyr; dotted lines in the right panel of Figure~\ref{fig:EvolProf}).
A profile at $z=2.25$ should be at 1.05 Gyr (or about 7 steps; thicker dotted line) from the profile of the higher-$z$ sample. Surprisingly we find the datapoints of \citet{fab+16} at this location. 
Figure~\ref{fig:EvolProf} thus shows a striking agreement between the profile reported by \citet{fab+16} and the predictions based on $z\sim3$.
The difference we detect for the subsamples at $z\sim3$ is then possibly due to the same physical mechanism(s) that cause the difference between 
$z\sim3$ and $z\sim2$, and these most likely act on CGM scales.

Given the currently favored optically thin scenario for the \lya\ emission, the abrupt decrease in SB$_{\rm Ly\alpha}$ -- which follows SB$_{\rm Ly\alpha}\propto n_{\rm H}N_{\rm H}$ -- could be due to
a strong variation in $n_{\rm H}$ and/or $N_{\rm H}$. We have shown however that variations in densities have to be larger than the cosmological density evolution, making it unlikely that only variations in 
$n_{\rm H}$ are responsible for the differences between $z\sim2$ and $z\sim3$.
Variations in $N_{\rm H}$ -- or in other words in the mass of the cool optically thin gas phase within the CGM, $M_{\rm cool}\propto N_{\rm H}$ (e.g., \citealt{qpq4}) -- are thus a more plausible 
explanation. 

Intriguingly, theoretical works have predicted just such a scenario for halos with masses above the critical halo mass for shock stability $M_{\rm shock}\simeq 10^{12}$~M$_{\odot}$ (roughly constant with redshift; \citealt{Dekel&Birnboim2006,Dekel2009}).
As mentioned above, the quasars here studied should be indeed in this regime.
In this theoretical framework, if $M_{\rm halo}>M_{\rm shock}$ there are two possible scenarios for the CGM: i) the presence of cool streams embedded in a hot phase, and ii) the prevalence of the hot phase.
The smooth transition between these two scenarios, or the maximum halo mass for cold streams, is a function of redshift, and is set by the shock stability criterion for the cool streams to balance the 
virial shock-heating (see Section~4.3 in \citealt{Dekel&Birnboim2006}).  
At the quasar's halo mass at $z\sim2$ and $z\sim3$, $M_{\rm halo}=10^{12.5}$~M$_{\odot}$, the threshold between the two scenarios is at $z=2$ (see Figure 7 in \citealt{Dekel&Birnboim2006}).
As one expects a smooth transition from the presence of cool streams to their absence, we conclude that the observed decrease in \lya\ emission 
around quasars from $z\sim3$ to $z\sim2$ could be due to this phenomenon, namely the inability of cool dense gas to survive the shock-heating process in similarly massive halos at lower $z$. 
It is important to stress that this scenario would hold for any powering mechanism for the \lya\ emission (see Section~\ref{sec:powering}). In other words, a decrease with redshift in the mass 
of the cool gas component within the CGM would result in a decrease of the level of \lya\ emission.

Finally, our result could be linked to the evolution of the space density of bright quasars as 
a function of redshift (e.g., \citealt{Richards2006}). It has been shown indeed that the space density of quasars steadily increases from 
$z=5$ to lower redshifts, peaking between redshift $z=2$ and $z=3$. After this peak, the space density drops significantly, reaching at $z=1$ roughly the same value 
of $z=5$ (e.g., Figure 20 in \citealt{Richards2006}). This trend, together with the level of star-formation of the quasars' hosts,
could be related to the drastic changes within the quasars CGM which we start to witness.
If the cool phase within the CGM cannot be replenished by infalling gas from the IGM or from the transformation of hot gas into cool gas within the CGM, the mass of cool CGM gas would thus drastically decrease, 
probably preventing future episodes of quasar activity as this gas is also the fuel for the high star-formation rates estimated for quasars' hosts (e.g., \citealt{rtl+13}).
This scenario finds additional confirmation in the fact that the IGM becomes more tenuous at lower redshift (e.g., \citealt{Dave2010,McQuinn2016}), 
and the halo gas is expected to warm up while structure evolution progresses (e.g., \citealt{Dekel&Birnboim2006,vandeVoort2011}).

We stress again that our result is based on heterogeneous techniques 
and thus requires a confirmation with current IFU instruments at $z\sim2$ (e.g. KCWI) and higher redshifts.
Sensitive current facilities, such as MUSE and KCWI, will allow us to tackle these hypothesis
with statistical and homogeneous samples, painting a clear picture for the evolution of the cool gas phase around quasars 
across cosmic time.

\subsection{What is the combination of mechanisms powering the Lyman-alpha emission?}
\label{sec:powering}

To better understand the astrophysics of the \lya\ emitting gas, it is key to determine the
primary mechanism (or constrain the combination of mechanisms) which power the observed
\lya\ luminosities.
While studying LABs, i.e. \lya\ nebulae with area$>16$~arcsec$^2$ and 
$L_{\rm Ly\alpha}\sim 10^{43}-10^{44}$~erg~s$^{-1}$ at $z\sim2-6$, several studies have
usually invoked four mechanisms to power the extended \lya\ emission: 
 (i) shocks powered by outflows (e.g.,
\citealt{Mori2004}), (ii) \lya\ collisional excitation or so called gravitational cooling radiation
(e.g., \citealt{Haiman2000,Rosdahl12}), (iii) photoionization
by a central AGN or star-formation (e.g., \citealt{Overzier2013,Prescott2015b}), (iv) resonant scattering of \lya\ photons from embedded sources 
(e.g., \citealt{Dijkstra2008}). 
In addition, when the \lya\ emission encompasses very large scales and radio-loud objects are within the sample, 
additional processes could
inevitably affect the observed levels, e.g., presence of multiple active sources, shocks from
radio jets. All of these processes may contribute together to the observed \lya\
emission, but due to the complexity to simulate them simultaneously, they are usually assessed 
individually.
Given the similar levels of \lya\ emission of LABs and the nebulosities around quasars, 
we here consider in turn all the aforementioned processes.

As anticipated in Section~\ref{sec:kinematics}, the relatively quiescent ($\langle\sigma_{\rm Ly\alpha}\rangle<400$~km~s$^{-1}$) 
\lya\ line of the discovered nebulosities seems to be 
consistent with gravitational motions within the massive halos currently expected to host
quasars. These large-scale nebulosities are thus unlikely powered by very fast outflows driven by the quasar
($v_{\rm s}>1000$~km~s$^{-1}$). Due to our spatial resolution and to our 
analysis technique, we cannot however rule out the presence of winds on smaller scales (few kpc). 
Such small-scale winds could act as additional sources of ionizing photons ($F_{\rm UV}\propto v_{\rm s}^3$; e.g., \citealt{Allen2008}), and thus contribute to power the 
\lya\ emission on large scales through the recombination channel.
In addition, especially in the case of radio-loud objects, a collimated jet may be responsible for 
part of the emission. As stated in Section~\ref{sec:kinematics}, we do not find evidences for 
violent kinematics in radio-loud objects. However, in Appendix~\ref{app:FIRST} we show that the extended \lya\ emission
for the radio-loud objects with detections from the FIRST survey (spatial resolution of 5 arcsec) matches well
the extent of the detected radio emission. This occurrence probably reveals a link between the \lya\
emission and the synchrotron radiation usually invoked to explain the radio data (see e.g. \citealt{vanOjik1996,mileyd08} for the alignment seen in HzRGs).
 
Further, it has been routinely shown that collisional excitation of the \lya\ line would require a fine
tuning between the density and the temperature of the gas to reproduce the observations (e.g., \citealt{Dijkstra2009_hd}). 
This is due to the 
exponential dependence on temperature of the collisional excitation coefficient and, being a
collisional process, to the dependence on density squared in the ionized case.
Thus, as it has been stressed previously in the case of \lya\ nebulae
(e.g., \citealt{Borisova2016}), collisional excitation seems to be unlikely the dominant mechanism
when strong ionizing sources are present.

Indeed, the only requirement for a photoionization scenario (by the quasar or other ionizing sources)
is at least a partial illumination of the cool emitting gas by the targeted AGN. 
As explained in \citet{qpq4}, we can solve for two limiting regimes in the case of photoionization from a source with ionizing luminosity $L_{\nu_{\rm LL}}$.
In the first case, the source or the level of radiation is not strong enough to highly ionize the gas, allowing a large fraction of neutral Hydrogen to be present, i.e. the gas is 
optically thick to Lyman continuum photons ($N_{\rm HI}\gg 10^{17.2}$~cm$^{-2}$). In this approximation the cool gas will behave like a `mirror',
converting a portion of the impinging ionizing radiation into \lya\ emission. The SB thus follows the relation SB$_{\rm Ly\alpha}\propto L_{\nu_{\rm LL}}$.
In the second case instead, the source of radiation is bright enough to keep the gas highly ionized, i.e. the gas is optically thin to 
Lyman continuum photons ($N_{\rm HI}\ll 10^{17.2}$~cm$^{-2}$). In this regime, as anticipate in Section~\ref{sec:redshiftEvol}, it can be shown that SB$_{\rm Ly\alpha}\propto n_{\rm H} N_{\rm H}$. 
Therefore, in the optically thin regime, the observed surface brightness should not depend 
on the luminosity of the targeted quasars, but only on the physical properties of the gas, i.e., the volume density
$n_{\rm H}$, and the column density $N_{\rm H}$.

In addition, if there is enough neutral Hydrogen (i.e. the gas is optically thick in the \lya\ transition, $N_{\rm HI}\gtrsim10^{14}$~cm$^{-2}$), the resonant scattering of
\lya\ photons emitted by the central AGN or embedded sources could be a viable process to produce extended \lya\ emission (e.g., \citealt{Dijkstra2017}, and references therein).
In this case, the \lya\ emission on large scales is due to a `random walk' in frequency and in space of centrally produced \lya\ photons. This  
process is caused by the absorption and consequent re-emission of photons by neutral hydrogen atoms in the quasar CGM.
If scattering of \lya\ photons from the quasar is the main powering mechanism, we would thus expect 
the observed SB$_{\rm Lya}$ to be proportional to the \lya\ emitted by the quasar, SB$_{\rm Ly\alpha}\propto L_{\rm Ly\alpha}^{\rm QSO}$.

\begin{figure}
       \centering
       \includegraphics[width=0.9\columnwidth, clip]{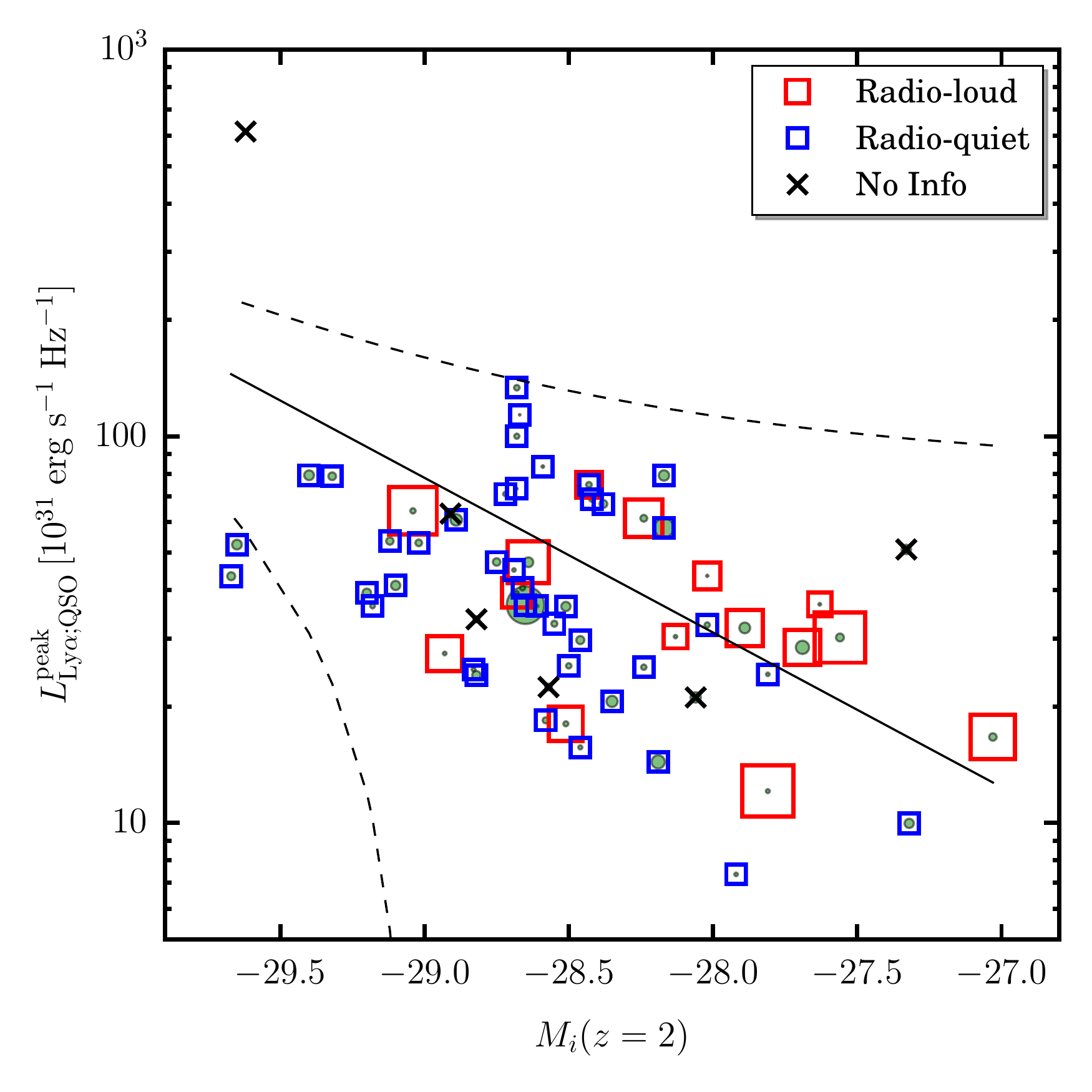}
    \caption{Plot of the peak \lya\ luminosity of the targeted quasars $L_{\rm Ly\alpha;QSO}^{\rm peak}$ versus their absolute $i$-mag normalized at $z=2$, $M_i(z=2)$.
    As done in Figure~\ref{fig:sigmavsMi}, we indicate in blue the radio-quiet systems, while in red the radio-loud objects (larger sizes = larger $F_{\rm Radio}$). The crosses indicate the objects for
    which we do not have information on their radio flux. The green circles are drawn for all the objects and their sizes indicates a product of the area of the nebulosities and of the
    distance between the center of the nebulosities and the targeted quasars (see Figure~\ref{fig:asymmetry}). The solid and dashed black lines 
    show a linear fit to the data and the 1 rms distance from the fit, respectively (see Section~\ref{sec:powering} for details).}
    \label{fig:LLyaMi}
\end{figure}

On top of this, $L_{\rm Ly\alpha}^{\rm QSO}$ is expected to increase with increasing quasar's $L_{\nu_{\rm LL}}$. 
We show this trend in Figure~\ref{fig:LLyaMi} where we plot the peak \lya\ luminosity of each quasar $L_{\rm Ly\alpha;QSO}^{\rm peak}$ against their absolute 
$i$-mag normalized at $z=2$. The first quantity has been calculated by considering the maximum at the \lya\ transition of the quasar's spectra 
shown in Figure~\ref{fig:1DspecComparison}.
In this plot we indicate the radio-loudness of our targets (red and blue) and 
we associate a green circle to each data-point. This green circle indicates the product of the area of each nebulosity and the distance between the
center of the nebulosity and the targeted quasars as defined in section~\ref{sec:asymm}.
The data in Figure~\ref{fig:LLyaMi} appear to be slightly correlated but a linear regression through the datapoints resulted in a 
Pearson correlation coefficient of $r_{\rm PCC}=0.4$, which is not statistically meaningful.
We however overlay in the figure the obtained fit to the data, 
$L_{\rm Ly\alpha;QSO}^{\rm peak}/(10^{31}\, {\rm erg~s~Hz^{-1}})=12.376\times10^{-(M_i(z=2)+27)/2.5}$ (black solid line), together with the 
1 rms distance from the fit (dahsed lines).

The aforementioned last three cases, namely photoionization of optically thick or thin gas, and resonant scattering are thus closely related, difficult to disentangle, and possibly present simultaneously.
However, if any of these mechanisms plays a dominant role in the production of extended \lya\ around quasars with no effect from the other
mechanisms, it should naively imprint a well defined dependence in the observed \lya\ levels as explained previously.
We have thus explored our data to find any of the aforementioned relations for the SB$_{\rm Ly\alpha}$.
In particular, we checked for variation of the average SB$_{\rm Ly\alpha}$ with current ionizing luminosity, and with the peak 
\lya\ luminosity of the targeted quasars $L_{\rm Ly\alpha;QSO}^{\rm peak}$. 
In addition, we have also looked for any dependence for the peak of the \lya\ emission from the nebulosities $L_{\rm Ly\alpha;Neb}^{\rm peak}$ 
against the same quantities. Consistently with $L_{\rm Ly\alpha;QSO}^{\rm peak}$, we have computed the peak of the \lya\ emission of the nebulosities using the spectra 
shown in Figure~\ref{fig:1DspecComparison}.

\begin{figure*}
       \centering
       \includegraphics[width=0.75\textwidth, clip]{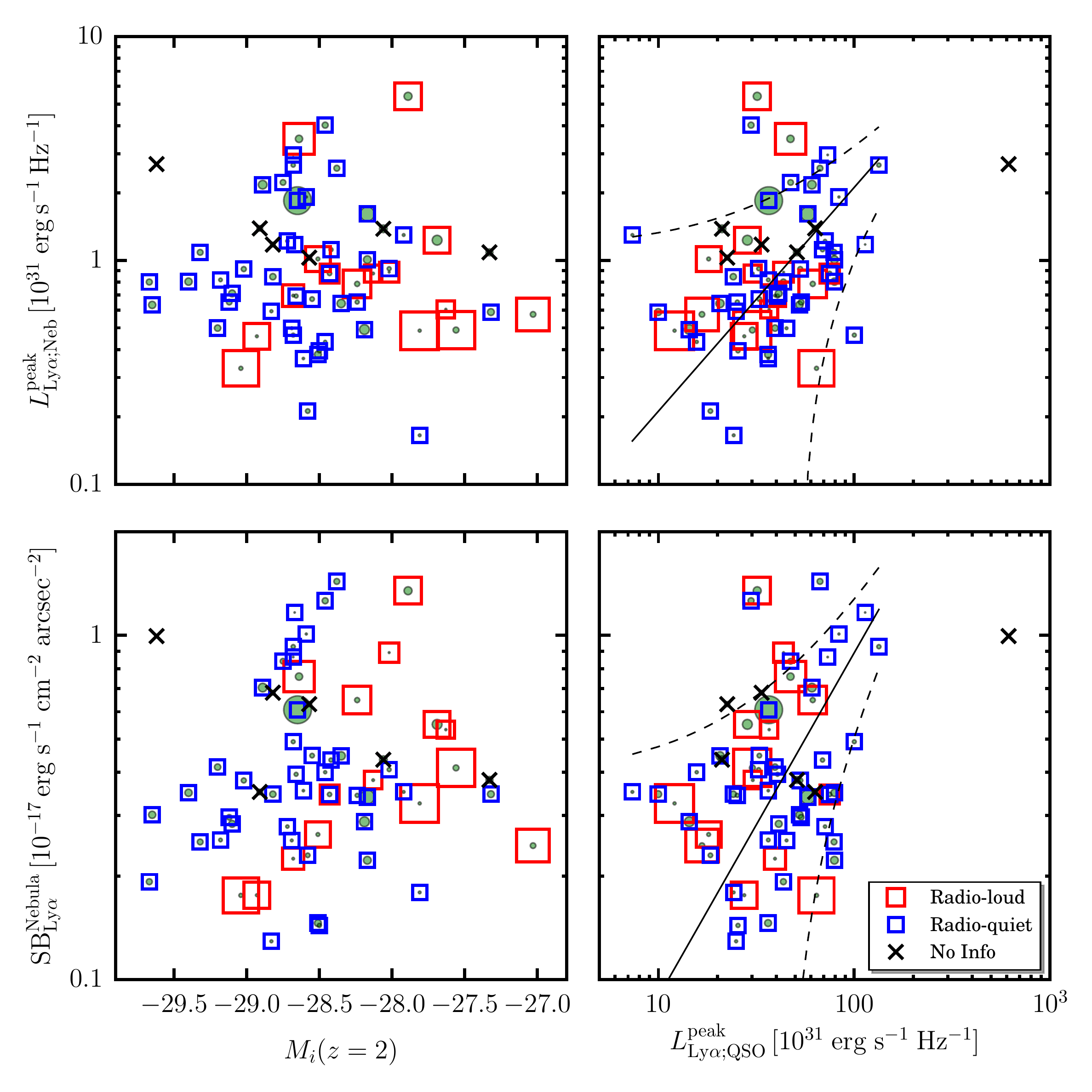}
    \caption{Lower left: plot of the average \lya\ surface brightness (SB$_{\rm Ly\alpha}$) of all the discovered nebulosities versus the absolute $i$-mag normalized at $z=2$ for the 
    targeted quasars. Lower right: plot of the SB$_{\rm Ly\alpha}$ versus the peak \lya\ luminosity for the targeted quasars (as derived from our MUSE data).
    Upper left: plot of the peak \lya\ luminosity for the discovered nebulosities versus $M_i(z=2)$ for the 
    targeted quasars. Upper right: plot of the peak \lya\ luminosity for the discovered nebulosities 
    versus the peak \lya\ luminosity for the targeted quasars (as derived from our MUSE data).
    As done in Figure~\ref{fig:sigmavsMi}, we indicate in blue the radio-quiet systems, while 
    in red the radio-loud objects (larger sizes = larger $F_{\rm Radio}$). The crosses indicate the objects for
    which we do not have information on their radio flux. The green circles are drawn for all 
    the objects and their sizes indicates a product of the area of the nebulosities and of the
    distance between the center of the nebulosities and the targeted quasars (see Figure~\ref{fig:asymmetry}). In the right panels, 
    the solid and dashed black lines show a linear fit to the data and the 1 rms distance from the fit, respectively (see Section~\ref{sec:powering} for details).}
    \label{fig:VariousPlots}
\end{figure*}

Figure~\ref{fig:VariousPlots} shows this test. The two left panels of this figure reveal that there is no clear dependence on 
the quasar luminosity (spanning roughly three magnitudes) for both the 
SB$_{\rm Ly\alpha}$ and for $L_{\rm Ly\alpha;Neb}^{\rm peak}$, with a consistent number of the nebulosities sitting at similar values around
$\sim0.4\times10^{-17}$~\unitcgssb\ and $\sim0.8\times10^{31}$~erg~s$^{-1}$~Hz$^{-1}$, respectively for SB$_{\rm Ly\alpha}$ and $L_{\rm Ly\alpha;Neb}^{\rm peak}$.
Given the absence of any dependence with quasar luminosity, this occurrence seems to rule out a fully optically thick regime for the \lya\ emitting gas, 
while hints to the possibility of this gas being optically thin to the ionizing radiation. This is usually found to be indeed 
the case if one assumes the quasar to shine on gas at distances comparable with halo scales (hundreds of kpc; \citealt{fab+15b,fab+2018}).
The nebulosities however are characterized by a wide scatter around the aforementioned values, especially for SB$_{\rm Ly\alpha}$, indicating that other
processes probably drive the level of the \lya\ emission.

The two right panels of Figure~\ref{fig:VariousPlots} show our test for the dependence of the \lya\ emission of the nebula (peak and average) with respect to the
\lya\ luminosity of the quasars, and thus probe a pure scattering scenario. In an attempt to verify the presence of a linear correlation within the dataset, we have performed a linear fit to the data and calculated the 
Pearson correlation coefficient $r_{\rm PCC}$. For both plots, we do not find any strong correlation, with $r_{\rm PCC}=0.2$ and $r_{\rm PCC}=0.3$ for the upper right panel and lower right
panel respectively. For completeness, we indicate in the plots our linear fits which resulted to be 
$L_{\rm Ly\alpha;Neb}^{\rm peak}=0.02121\times10^{31} (L_{\rm Ly\alpha;QSO}^{\rm peak}/10^{31}[{\rm erg~s^{-1}~Hz^{-1}}])$~erg~s$^{-1}$~Hz$^{-1}$, and
SB$_{\rm Ly\alpha}^{\rm Nebula}=0.00891\times10^{-17} (L_{\rm Ly\alpha;QSO}^{\rm peak}/10^{31}[{\rm erg~s^{-1}~Hz^{-1}}])$\unitcgssb. 
On the same plots, we also indicate the 1 rms distance from the fits with the dashed lines.
The absence of clear correlations could indicate that the resonant scattering of \lya\ photons is not a dominant mechanism in shaping the emission from the 
discovered nebulosities on the full scales encompassed by the 2$\sigma$ isophote used to calculate the SB$_{\rm Ly\alpha}$ for each nebula. 
This can be due by the fact that (i) additional powering mechanisms (i.e. recombination) and/or (ii) the variability of the quasar luminosity (e.g., \citealt{Peterson2004}) could have washed out a clearer dependance.
The two right panels of Figure~\ref{fig:VariousPlots} and the quality of the PSF subtraction on small scales  ($\sim10$~kpc) prevent us to probe the presence of a tighter relation at smaller scales.
However, a signature of the relation between the \lya\ emitted by the quasar 
and each \lya\ nebula could be represented by the small velocity difference between the quasar \lya\ and the nebulae themselves shown in section~\ref{sec:morph}. 
Indeed this small difference in velocity could reflect that 
the \lya\ emission from the quasar and the peak of the nebula (always at small projected separation from the quasar) follow the same path of least resistance to escape the system in the scattering process.
It is important to stress that -- if resonant scattering is the only mechanism in play -- only a very small fraction of the \lya\ emission
produced by the central quasar is needed to match the
total \lya\ emission observed in the detected extended nebulae, 
i.e., the \lya\ luminosity of a nebula is $\approx3\%$ of the peak of the \lya\ emission of the quasar (average of our whole sample).

Finally, as done in Figure~\ref{fig:LLyaMi}, in Figure~\ref{fig:VariousPlots} we associate a green circle to each data-point to indicate the extent and spatial asymmetry of each 
nebulosity. By looking at these datapoints, we do not find any correlation between the quasar ionizing luminosity and \lya\ emission and the
size/spatial asymmetry of the nebulae. 

Summarizing, the current data at the \lya\ line location suggest a scenario in which the \lya\ emission is powered by a combination of photoionization of relatively optically thin gas, and resonant scattering. 
In addition, we cannot rule out a contribution from small-scales winds or jets in ionizing the surrounding gas distribution. We will investigate the powering mechanisms
in a future work focused on additional diagnostics covered by our observations.

\subsection{Insights on the observability of fast AGN winds on large scales}
\label{sec:timescale}

Our study, together with \citet{Borisova2016}, has shown that we can routinely detect 
\lya\ nebulosities for tens of kpc with fast MUSE observations, and we have argued in Section~\ref{sec:powering}
that these nebulosities are likely powered by a combination of mechanisms dominated by the radiation 
of the central quasar.
As already noticed in Section~\ref{sec:kinematics}, the same samples result in only 1 broad \lya\ 
nebulosity around a radio-quiet quasar out of 56 ($17+39$),
and only 2 out of 17 ($15+2$) for radio-loud objects.
If the broad \lya\ emission is a signature of AGN winds on large scales (tens of kpc), these observations
suggest a low probability of detecting them out to 50 kpc, i.e. $2\%$ and $12\%$, respectively for the radio-quiet and radio-loud samples.

Such low probabilities could reflect i) a bad coupling between a relatively hot wind and the cool gas phase and/or 
ii) that winds do not emit \lya\ emission or emit \lya\ at much fainter levels than here probed, and/or 
iii) the difficulties to propagate coherent 
winds on large scales (tens of kpc) for long period of times\footnote{A wind with a velocity of $0.01c$ would require $\sim10^{7}$~yr to reach 50 kpc.} 
by the observed stochastic and episodic AGN winds
generated on much smaller scales (e.g., \citealt{KingPounds2015}). 

The first two options seem quite unlikely given that fast winds (with shifts and line widths $>1000$~km~s$^{-1}$) on scales of tens of kpc have been reported 
also in cold tracers around quasars (e.g., [\ion{C}{ii}]~158~$\mu$m; \citealt{Cicone2015}), and the shock front should behave like a quite hard ionizing source, thus powering strong \lya\ emission (e.g., \citealt{Allen2008}).
Our snapshot observations could however be missing a broad and diffuse \lya\ component.

Regarding the propagation of winds on large scales, in addition to the uncertainties in the triggering of a sustained wind, we would expect a lag between the light propagation
from the quasar and the wind propagation due to the different times at which they probably switch on, 
and at the different speeds (light vs shock). This lag could be the cause of the 
observed low probabilities. Indeed, young quasars or quasars with an inefficient propagation of the wind would be characterized by 
a quiescent \lya\ nebula (due to the quasar illumination)
more extended than a turbulent region affected by the wind (if any). This scenario has been illustrated around HzRGs, where 
large velocity dispersion in \lya\ are found within the smaller scales perturbed by the jet (FWHM$>1000$~km~s$^{-1}$), but 
smaller velocities dispersions on larger scales beyond the radio structures or in closer regions not affected by the jet (FWHM$\sim300-700$~km~s$^{-1}$; e.g., \citealt{VillarMartin2003, Vernet2017}).

If this is indeed the case, we would expect already in our data to see  
intermediate nebulae with large velocity dispersions on few kiloparsecs, and smaller (quiescent) velocity dispersions on larger scales. 
Given the current spatial resolution of our data, and the PSF subtraction algorithm used, we cannot firmly assess this scenario
below approx. $1-2$~arcsec from the quasar with the current data. 

We note however that several authors argued for ubiquitous extreme kinematics ($>1000$~km~s$^{-1}$) around quasars 
on tens of kpc scales (usually $<30$~kpc)
from low to very high redshift ($z\sim6$), and basically along the whole electromagnetic 
spectrum and thus gas phases (e.g., \citealt{Greene2011,Liu2014,Greene2014,Brusa2015,Perna2015,Cicone2015,Rupke2017}).
Such scales would corresponds to about 4 arcsec at $z\sim3$.
Nevertheless, both our flux-weigthed dispersion maps (Figure~\ref{fig:Sigmamaps}) and the maps presented in \citet{Borisova2016} 
(their Figure~7), do not show at face value
higher widths close to the central quasars.
This result would even be exacerbated if the \lya\ line that we observe is broaden by scattering, as is expected. 
The observed \lya\ line width could then be even smaller if measured through non-resonant tracers.
As noted in Section~\ref{sec:kinematics}, the motions within the \lya\ nebulae thus seems to be consistent with gravitational motions.

This evidence thus encourage a new effort in characterizing the small-scale gas phase around quasars, while linking it to the larger scales 
(100~kpc) targeted in this study. 
The new AO system GALACSI (\citealt{Stuik2006}) on MUSE will be fundamental to addressing the aforementioned points, while taking into account known 
issues dealing with
small-scale emission around bright quasars (e.g., PSF deblending; \citealt{Husemann2016, HusemannHarrison2018}). 
Indeed, \citet{Husemann2016} showed that previously reported 
spatially resolved [\ion{O}{iii}] emission-line widths on kpc scales around a sample 
of $z\sim0.5$ quasars (\citealt{Liu2014}) are significantly narrower than the one before PSF
deblending.

\section{Summary and conclusions}
\label{sec:summ}

To characterize the CGM of massive halos and to uncover additional ELAN, 
we exploit the unprecedented capabilities of the MUSE instrument on the ESO/VLT to build the 
``fast'' survey (45 minutes/source) QSO MUSEUM targeting 61 quasars at a median redshift of $z=3.17$ ($3.03<z<3.46$).
With quasars characterized by different radio-loudness (39 radio-quiet; 15 radio-loud; 6 with currently no radio information) 
and spanning roughly three magnitudes $-29.67\leq M_i(z=2)\leq-27.03$, 
this survey expands previous works targeting quasars at these cosmic epochs (e.g., \citealt{Borisova2016}).
In this work we present the results of our survey for the \lya\ emission only. We plan to present 
the results for additional diagnostics in future works.
By analyzing the datacubes at the location of the \lya\ transition, we thus find that

\begin{enumerate}

\item After PSF and continuum subtraction, each of the targeted quasars show a 
\lya\ nebulosity whose bulk emission is on radii $R<50$~kpc above our detection limits (SB$_{\rm Ly\alpha}=8.8\times10^{-19}$~\unitcgssb; 
$2\sigma$ in 1 arcsec$^2$ in a single channel of $1.25\AA$), with an average surface brightness of the order of 
SB$_{\rm Ly\alpha}\sim0.4\times10^{-17}$~\unitcgssb.
The average maximum projected distance covered by the \lya\ emission from each quasar is 80~kpc.
This extended \lya\ emission is characterized by
diverse morphologies. In particular, most of the discovered nebulae appear to be symmetric (median asymmetry $\alpha=0.71$), while few exceptions 
show more elongated morphologies.  

\item Given the diverse morphologies, the discovered \lya\ nebulosities show very diverse radial profiles, which however result in 
a stacked profile best described by an exponential fit, SB$(r)=C_e{\rm exp^{-r/r_h}}$ with scale-length of $r_h=15.7\pm0.5$~kpc (whole sample). 
This radial profile is consistent till $R\simeq60$~kpc with the profile shown in \citet{Borisova2016}, at larger radii our profile better indicates an exponential behavior.
The radio-loud sample has a slight excess in SB at $20$~kpc$<R<50$~kpc, resulting in a tentatively larger scale-length $r_h=16.2\pm0.3$~kpc, with respect
to the radio-quiet sample $r_h=15.5\pm0.5$~kpc.

\item  Down to our detection limits, the gas emitting \lya\ emission covers a fraction $>30\%$ of the area around the quasars for radii $R<50$~kpc.
At larger distances, the covering fraction rapidly decrease (Figure~\ref{fig:coveringFactor}).

\item The peak of the \lya\ emission of the nebulosities appear to be closer in redshift to the peak of the \lya\ emission of the quasars 
$z_{\rm peak\, QSO\, Ly\alpha}$ than to the 
current systemic redshift of the quasars $z_{\rm systemic}$. We find a median shift of the \lya\ emission of $\Delta v=782$~km~s$^{-1}$, and $\Delta=144$~km~s$^{-1}$ from 
$z_{\rm peak\, QSO\, Ly\alpha}$ and $z_{\rm systemic}$, respectively. 
More precise systemic redshift are needed to verify this occurrence.

\item We discovered 27 LAEs with
$F>1.1\times10^{-17}$~erg~s$^{-1}$~cm$^{-2}$ (our $\geq 80\%$ completeness level). 
This low detection rate agrees with the currently known luminosity functions for field LAEs (\citealt{Cassata2011, Drake2017}), 
which would predict $24-30$ LAEs to be found.
We find fields with 
more than one detection which appear to be clear outliers and probably reflecting the presence of underlying overdensities,
and/or effects of powering mechanisms, e.g., chance alignment between the position of the LAE and the ionization cones of the quasars.
In addition, only three fields show the presence of active companions.
We note that the quasar with the larger number (2, of which 1 type-I) of active companions detected (ID 13; \citealt{fab+2018})
also presents the more extended and brightests of the discovered \lya\ nebulosities (see point (ix)).

\item The velocity maps obtained from the first moment of the flux distribution of the discovered \lya\ nebulosities
appear to be noisy and difficult to interpret as ordered motions of any kind (rotation, infall, or outflow; see also \citet{Borisova2016}), though each
nebulosity show peculiar structures on velocities of the order of few hundreds km~s$^{-1}$ (Figure~\ref{fig:Velmaps}). Nebulae with 
larger extents than the average easily show coherent velocity structures (e.g., ID 13; \citealt{fab+2018}). The difference
in complexity between the maps of very extended nebulosities and smaller systems could be due to the resonant nature of the
\lya\ line which sample more turbulent regions on tens of kpc around the quasar.

\item All the discovered \lya\ nebulosities, irrespective of the radio-loudness, are characterized by relatively quiescent 
kinematics ($\langle \sigma_{\rm Ly\alpha} \rangle<400$~km~s$^{-1}$ (or FWHM$<940$~km~s$^{-1}$).
These estimates are similar to the velocity widths observed in absorption in the CGM surrounding $z\sim2$ quasars ($\Delta v>300$~km~s$^{-1}$; 
\citealt{qpq3,qpq9}). The motions within all these \lya\ nebulae have amplitudes consistent with gravitational motions expected in dark matter halos hosting 
quasars (M$_{\rm DM}\sim10^{12.5}$~M$_{\odot}$; \citealt{white12,Trainor2012}).
This point is even strengthen by the fact that
part of the velocity dispersion could be due to \lya\ resonant scattering and 
instrument resolution (FWHM$\approx2.83$\AA or $170$~km~s$^{-1}$ at 5000\AA).

\item The relatively quiescent \lya\ nebulosities around all the 15 radio-loud quasars in our sample run counter to 
the two radio-loud system with $\sigma_{\rm Ly\alpha}>425$~km~s$^{-1}$ showed by \citet{Borisova2016}.
This occurrence might be driven by the higher luminosity of the Borisova's quasars for equal radio emission (see Section~\ref{sec:kinematics}), 
and/or might reflect the probabilities of propagating a coherent AGN winds on halo scales (see Section~\ref{sec:timescale}).
Note that this would be also valid in the case of radio-quiet systems (Section~\ref{sec:timescale}).

\item Of all the discovered \lya\ nebulosities, only ID 13 (or PKS~1017+109) 
shows the extreme observed surface brightnesses and extent found in the case of 
$z\sim2$ ELAN (\citealt{cantalupo14,hennawi+15}). 
The ELAN around PKS~1017+109 has been studied in detail in \citet{fab+2018}. 
In this work we clearly show that ELAN are indeed easily
detected as extended sources even with faster observations than here used (Figure~\ref{fig:AreaVSSN}).

\item The current statistics of \lya\ nebulae around radio-quiet quasars at $z\sim2$ (\citealt{fab+16}) suggests a lower level of \lya\ emission with respect to 
the nebulae in the QSO MUSEUM sample and other surveys at $z\sim3$ (\citealt{Borisova2016}). Even though the current statistics at $z\sim2$ still 
rely on a relatively small sample of NB data and have thus to be confirmed with on-going campaigns
using IFU instruments, we here show that such a difference would fit well with 
a decrease in cool halo mass in the CGM of similarly massive halos from $z\sim3$ to $z\sim2$ (\citealt{Dekel&Birnboim2006}).
These changes in \lya\ profiles can thus reflect a different balance between cool and hot gas phase at different redshifts
(see discussion
in Section~\ref{sec:redshiftEvol}). Future statistical surveys at different redshift are key in testing this hypothesis.

\item There are hints suggesting that the discovered \lya\ nebulosities are powered by recombination 
and scattering of \lya\ photons in optically thin gas (Section~\ref{sec:powering}). 
The presence of fast outflows ($v_s>1000$~km~s$^{-1}$) on $R\sim50$~kpc seems to be disfavored fo the majority of the sample. 
However we cannot exclude the presence of winds or jets on scales of few kpc.
  
\end{enumerate}

As shown by the aforementioned list of results, the large dataset of QSO MUSEUM allowed us to reveal the common features characterizing 
the extended \lya\ emission around quasars, but also to find extreme and rare outliers, i.e., the brightest and more extended nebulosities or ELAN.
Given their high surface brightnesses, these objects are the ideal laboratory for the study of the astrophysics in play on halo-scales.

Overall, QSO MUSEUM unveils the complexity inherent in the study of large-scale (hundreds of kpc) systems. Each of the discovered \lya\ 
nebulosities would indeed require a detailed individual study using multiwavelength and deeper observations to firmly characterize them 
(e.g., geometry, environment, powering mechanisms), and understand the particular moment at which these systems are observed (e.g., ongoing
AGN outflow, merger(s), quasar lifetime).
Notwithstanding this complexity, our survey open the path to a better comprehension on how massive systems form and evolve along the cosmic history, 
and would be the benchmark for future statistical and homogeneous investigations.

\section*{Acknowledgements}

We thank Max Gronke and Aura Obreja for fruitful discussions.
Based on observations collected at the European Organisation for Astronomical Research in the Southern Hemisphere under ESO programmes 094.A-0585(A), 095.A-0615(A), 095.A-0615(B) and 096.A-0937(B).
We acknowledge the role of the ESO staff in providing high-quality service-mode observations, which are making this project feasible in a shorter time-scale.
JXP acknowledges support from the National Science Foundation (NSF) grant AST-1412981.
SC gratefully acknowledges support from Swiss National Science Foundation grant PP00P2\_163824.
EL is supported by a European Union COFUND/Durham Junior Research Fellowship (under EU grant agreement no. 609412).

%%%%%%%%%%%%%%%%%%%%%%%%%%%%%%%%%%%%%%%%%%%%%%%%%%



\bibliographystyle{mnras}
\bibliography{allrefs} % 

%%%%%%%%%%%%%%%%%%%%%%%%%%%%%%%%%%%%%%%%%%%%%%%%%%

%%%%%%%%%%%%%%%%% APPENDICES %%%%%%%%%%%%%%%%%%%%%

\appendix

\section{Quasar PSF subtraction away from the Lyman-alpha line}
\label{app:PSFsubtraction}

\begin{figure}
       \centering
       \includegraphics[width=1.0\columnwidth, clip]{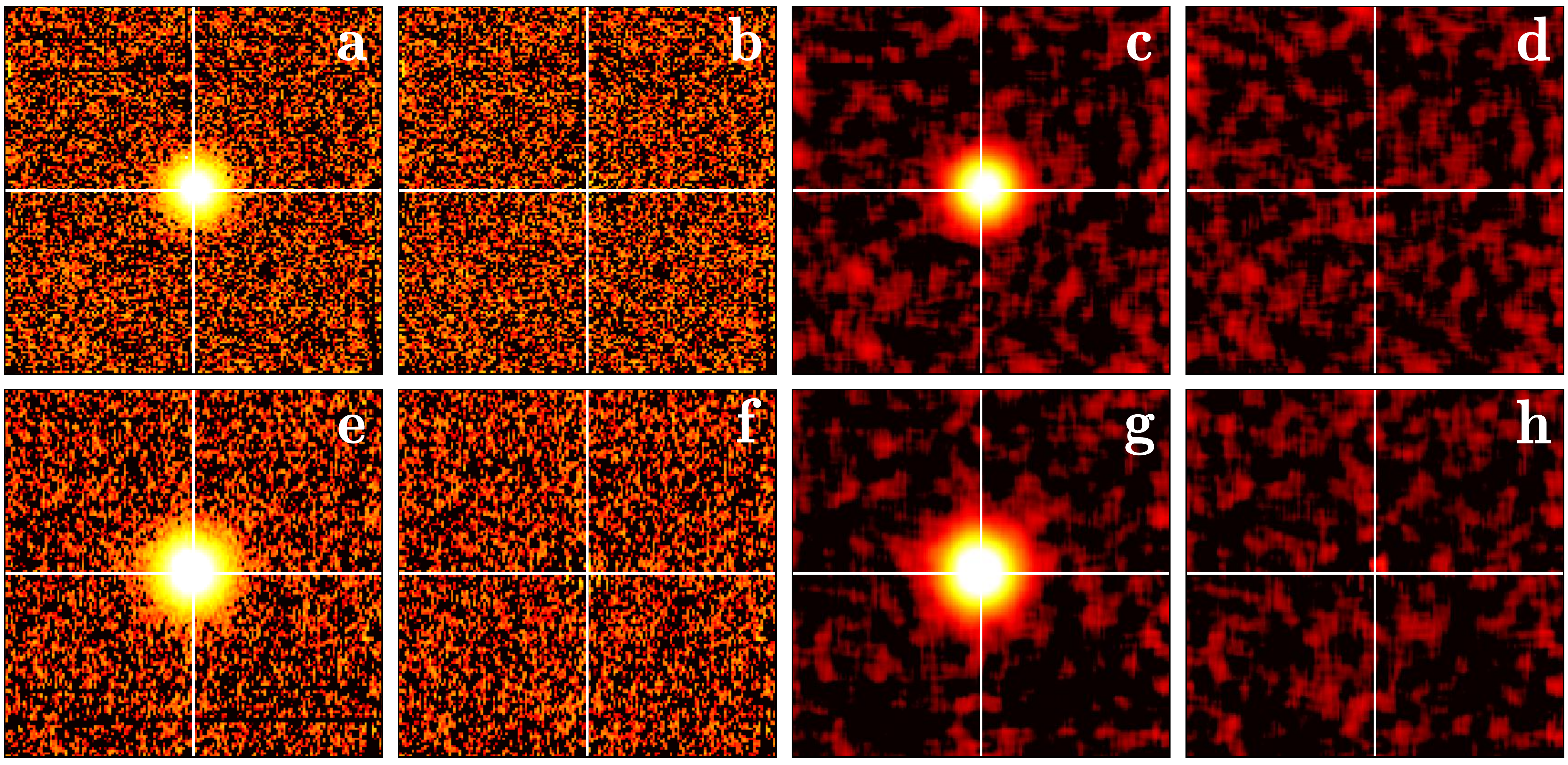}
    \caption{Quasar PSF subtraction off the \lya\ line. For the first two quasars in our sample (SDSSJ~2319-1040, top; UM~24, bottom) we show
    30~\AA\ NB images centered at the quasar position (white crosshair), and at rest-frame 1260~\AA\ for (i) the final datacube (`a' and `e'), 
    (ii) the final datacube after continuum and PSF subtraction (`b' and `f'), (iii) the final datacube smoothed with a boxcar kernel of 3 spaxel (`c' and `g'), 
    and (iv) the the final datacube after continuum and PSF subtraction smoothed with a boxcar kernel of 3 spaxel (`d' and `h'). 
    Each image is 30\arcsec~$\times$~30\arcsec. Clearly there are no significant residuals that could contaminate our analysis.}
    \label{fig:OutLya}
\end{figure}

In this appendix, we show how our procedure for the subtraction of the continuum and the PSF of the quasar
works in a wavelength range where no line emission is expected at the quasar redshift. At such a location 
we thus expect no residual emission at the levels probed by our observations. We proceed as follows.
We focus on the portion of the final datacube at rest-frame 1260~\AA,
and we construct NB images by collapsing the layers within 30~\AA. We then construct NB images of the same wavelength range
for the continuum and PSF subtracted datacube. 
In Figure~\ref{fig:OutLya} we show the results of this test for the first two quasars of our sample (SDSSJ~2319-1040, top; UM~24, bottom).
Panel `a' and `e' of Figure~\ref{fig:OutLya} present the central 30\arcsec$\times$~30\arcsec of the 
final datacube with the clear PSF of the quasar at the center of the image.
Panel `b' and `f' show the same FOV, but after the continuum and PSF subtraction.
The other panels show the same images but smoothed with a boxcar kernel of 3 spaxels.
From these images, it is clear that there is no significant systematic underestimation of the 
PSF of the quasar, which could lead to residuals. We are thus confident that the extended \lya\ emission detected around our sample is 
real. 

\section{Comparison between radio emission and Ly$\alpha$ emission}
\label{app:FIRST}

In this appendix we focus on the radio-loud objects in our sample, and compare the extended \lya\ emission detected in this work with the radio emission.
Of the 15 radio-loud quasars, only 10 have radio data taken with a spatial resolution good enough to be compared with our observations.
Indeed, five sources have been detected in the NVSS (\citealt{Condon1998}) which is characterized by a resolution of 45\arcsec, while 10 in the 
FIRST survey (\citealt{Becker1994}) at a resolution of 5\arcsec.
The resolution of FIRST thus approaches the resolution of our observations, allowing us to qualitatively compare the two types of emission.
Further, quasar with ID 17 has a FIRST detection at close separation. Here, we will investigate the nature of such source to understand if it is physical associated to the quasar.

\begin{figure*}
       \centering
       \includegraphics[width=0.95\textwidth, clip]{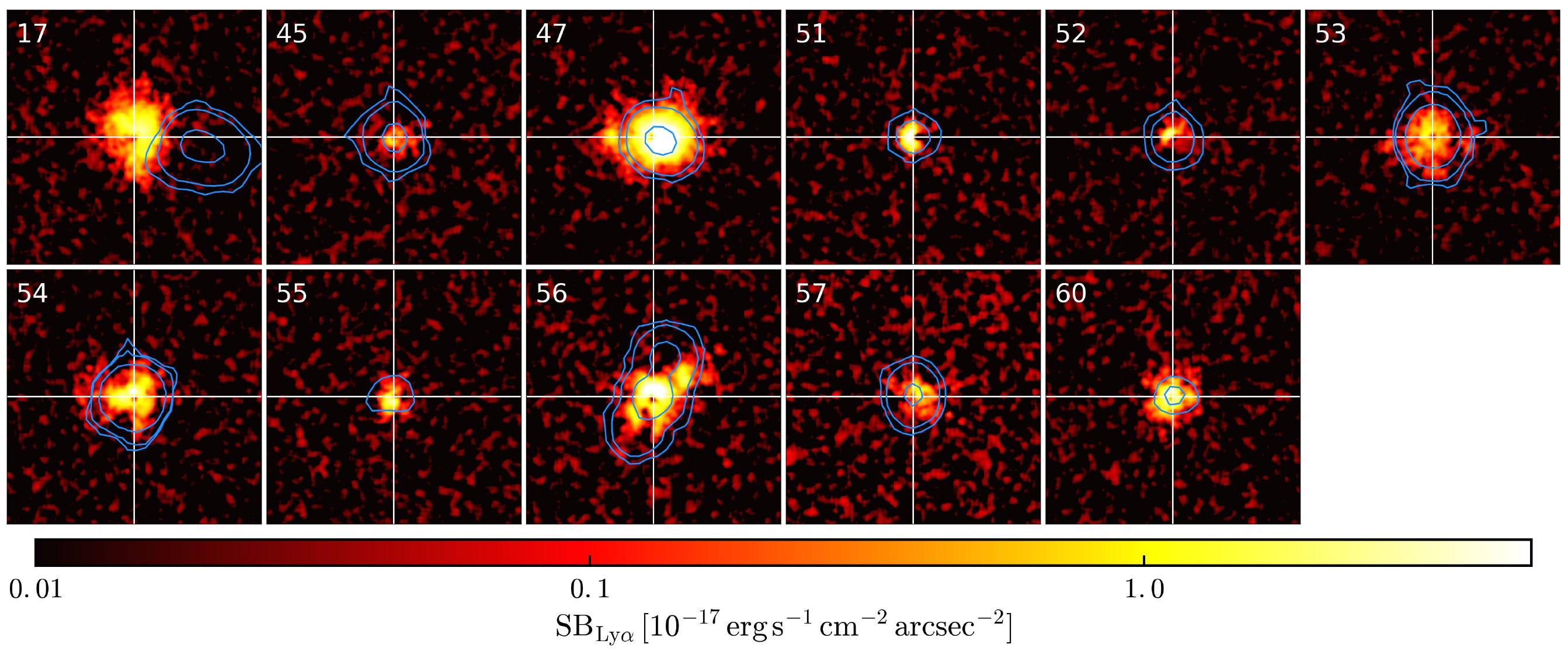}
    \caption{Comparison between the detections within the FIRST survey (\citealt{Becker1994}) and the discovered \lya\ nebulae in our sample.
    We show the $50\arcsec \times 50\arcsec$ (or approx. 380~kpc~$\times$~380~kpc) ``optimally extracted'' NB maps for the targets detected in the FIRST survey with overlayed the contours for S/N$=3,10,100$ for the radio emission. 
    Down to the current spatial resolutions and sensitivity limits, the \lya\ emission and the radio emission are well matched for objects whose radio emission 
    is centered at the quasar location. North is up, east is to the left.}
    \label{fig:FIRST_comp}
\end{figure*}

\begin{figure}
       \centering
       \includegraphics[width=0.95\columnwidth, clip]{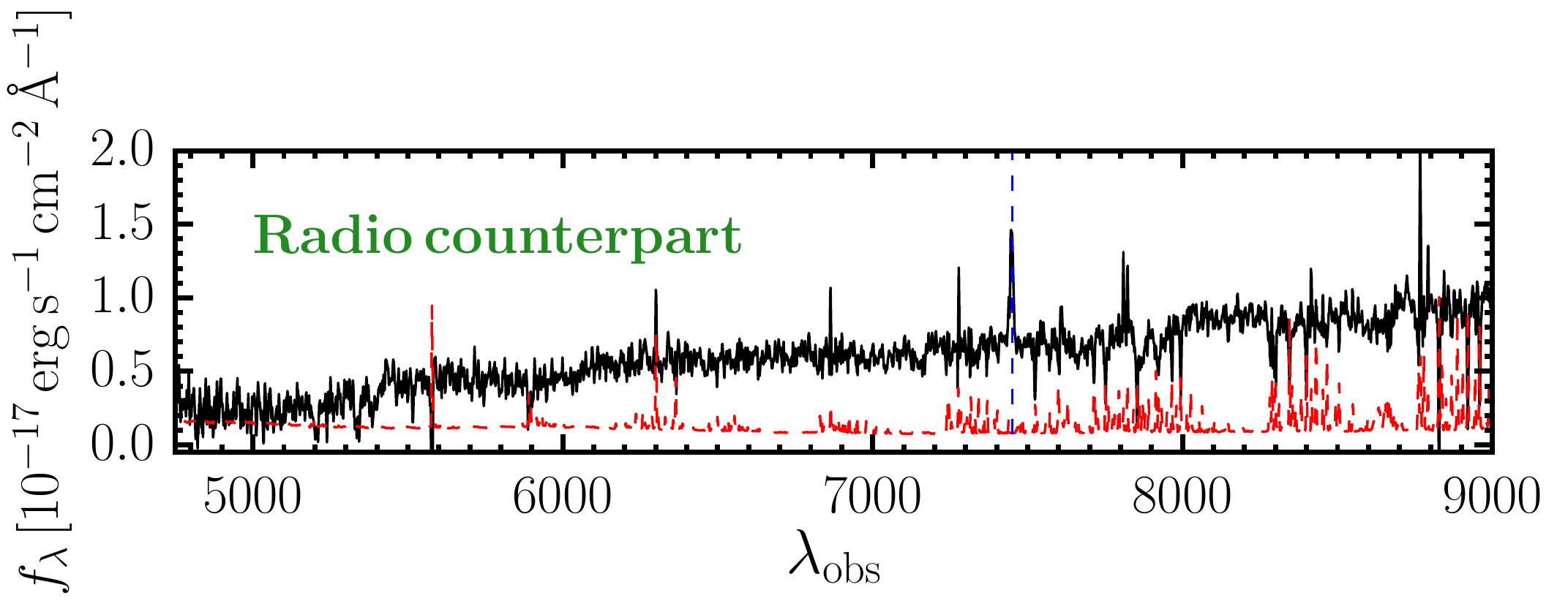}
    \caption{Spectrum of the counterpart to the radio emission at 13.5\arcsec\ (or $100$~kpc) from the quasar with ID 17. 
    The spectrum has been extracted from a circular region with a 2\arcsec\ radius at the peak of the radio emission. The red dashed line indicates the error spectrum within the same aperture.
    At this location some sky lines are still visible as residuals.
    The source is characterized by a continuum with a red upturn, and an emission line at about 7450\AA (blue dashed). Some absorption lines are tentative at the current depth (see e.g. at about 5200\AA\ and 5340\AA).}
    \label{fig:radio_counterpart}
\end{figure}

In Figure~\ref{fig:FIRST_comp} we show the comparison between the optimally extracted images of these 11 \lya\ nebulae and the S/N=3, 10, 100 isophotes for the 1.4~GHz emission
detected with FIRST (blue). Overall, the extended \lya\ emission matches well the location and extent of the radio emission, with the bulk of the \lya\ emission concentrated within the radio region.
Exceptions to this are ID 56 and ID 17. ID 56 shows an elongated morphology towards the North in the FIRST data, suggesting the likely presence of a radio-jet. The \lya\ emission however presents a different orientation, 
possibly reflecting the position of a close LAE or substructure (see Figure~\ref{fig:Prof_one}) rather than following the radio emission.
ID 17 instead shows radio emission displaced by $\approx13.5\arcsec$ (or $100$~kpc) from the quasar position. We have checked the MUSE datacube at that location and found a morphologically disturbed continuum source 
(see Figure~\ref{fig:Prof_one}) with a red continuum slope, and only a strong line emission at about 7450\AA (see Figure~\ref{fig:radio_counterpart}). 
Given the presence of only one line emission and no clear absorption at the current depth, we are not able to firmly classify this object. However, it is likely that such object is not at the same redshift of the quasar with ID 17. 
The radio emission is thus not physically associated to this system. We thus include ID 17 in our radio-quiet sample.

All the firmly selected radio-loud objects thus show a similarity between the radio contours and the \lya\ emission at the current resolutions. 
Future works are needed to investigate the relation between the \lya\
emission, the synchrotron radiation usually invoked to explain the radio data, and other mechanisms responsible for hard photons over large scales around radio sources (e.g. extended X-rays from inverse Compton; \citealt{Erlund2006}). 
Future efforts will have to take into account the similarities in velocity dispersion and velocity shifts for the extended \lya\ emission between the radio-loud and radio-quiet samples (Sections~\ref{sec:kinematics} and \ref{sec:timescale}).

\section{Additional notes on some individual fields}
\label{app:individual_fields}

In this appendix we report (i) the information on the LAEs discovered, (ii) the spectra and coordinates for the companion AGN, for the line emitters discovered 
during our search for LAEs, and for the foreground or background quasars discovered.
In addition, we discuss the field of Q1205-30 (or ID 58) that had been previously observed (\citealt{Fynbo2000,Weidinger04,Weidinger05}), and that thus 
provide a great sanity check of our analysis.

\subsection{The LAEs sample}

In Table~\ref{MEGAtabLAEs} we list the coordinates, and several properties for the discovered LAEs in the 61 fields (see Section~\ref{sec:comp_sel} for the details on the LAEs selection).
Note that we don't list here the LAEs around the quasar with ID 13 as those are already listed in \citet{fab+2018}.

\begin{table*}
\caption{Information on the Ly$\alpha$ emission for the LAEs discovered around the targeted quasars.}
\centering
\resizebox{\textwidth}{!}{
\begin{tabular}{llccccccccccc}
\hline
\hline
Object & QSO-field &  RA  	    &  Dec	       & $d$	& Line Center	 & Redshift	    & $F_{\rm Ly\alpha}$			  & $f_{\lambda}$				    & EW$_{\rm rest}$ & $\sigma_v$ &   $\Delta v_{\rm Neb}$   & $\Delta v_{\rm QSO}$   \\
       & ID (name) & (J2000)	    &  (J2000)         & (\arcsec)    & (\AA)		 &		    & (10$^{-17}$ cgs)  		  & (10$^{-20}$ cgs)				    & (\AA)	    & (km s$^{-1}$)&   (km s$^{-1}$)	      & (km s$^{-1}$)	       \\
\hline 																									    
LAE1   & 7 (SDSSJ1209+1138)&  12:09:17.742    &  11:38:44.56     & 13.7	     & 5004.91$\pm$0.03 & 3.117$\pm$0.001  & 1.88$\pm$0.02		     &  5.6$\pm$12.8				       & $>$81         & 137$\pm$3    &   -654$\pm$73		 &    0$\pm$421 	  \\
LAE2   & 7 (SDSSJ1209+1138)&  12:09:17.298    &  11:38:48.77     & 20.2	     & 5004.91$\pm$0.03 & 3.117$\pm$0.001  & 2.76$\pm$0.03		     &  1.6$\pm$18.2				       & $>$420        & 172$\pm$2    &   -654$\pm$73		 &   -5$\pm$421 	  \\
LAE3   & 7 (SDSSJ1209+1138)&  12:09:17.623    &  11:38:54.89     & 24.2	     & 5023.15$\pm$0.03 & 3.132$\pm$0.001  & 3.02$\pm$0.03		     &  3.9$\pm$14.2				       & $>$183        & 176$\pm$2    &    436$\pm$73		 & 1092$\pm$421 	  \\
LAE4   & 7 (SDSSJ1209+1138)&  12:09:16.759    &  11:38:47.85     & 24.6	     & 5029.23$\pm$0.03 & 3.137$\pm$0.001  & 3.13$\pm$0.03		     & -9.8$\pm$10.2				       & $>$74         & 152$\pm$2    &    799$\pm$73		 & 1456$\pm$421 	  \\
LAE5   & 7 (SDSSJ1209+1138)&  12:09:19.090    &  11:38:17.78     & 21.0	     & 5012.21$\pm$0.03 & 3.123$\pm$0.001  & 2.53$\pm$0.03		     & -8.6$\pm$11.4				       & $>$54         & 184$\pm$3    &   -218$\pm$73		 &  437$\pm$421 	  \\
\hline
LAE1   & 8 (UM683)	  & 03:36:28.701     &  -20:19:24.97	& 27.0         & 5023.15$\pm$0.03 & 3.132$\pm$0.001  & 1.10$\pm$0.02	      &  -7.9$\pm$4.8	  & $>$56	    & 112$\pm$3    &	    0$\pm$73	       &   0$\pm$421	      \\
LAE2   & 8 (UM683)	  & 03:36:27.484     &  -20:19:32.29	&  9.2         & 5023.15$\pm$0.03 & 3.132$\pm$0.001  & 1.35$\pm$0.02	      &  -4.4$\pm$5.1	  & $>$64	    & 128$\pm$2    &	    0$\pm$73	       &   0$\pm$421	      \\
\hline
LAE1   & 11 (Q-N1097.1)	 &  02:46:35.498    &  -30:04:35.22    & 26.9	      & 4981.82$\pm$0.03 & 3.098$\pm$0.001  & 2.92$\pm$0.03  	     &  -3.8$\pm$10.9	 & $>$65	   & 157$\pm$2    &	   -73$\pm$73	      &   1470$\pm$421	     \\
\hline
LAE1   & 20 (SDSSJ1429-0145) &  14:29:02.011    & -01:45:14.14     & 16.6	      & 5386.63$\pm$0.04 & 3.431$\pm$0.001  & 3.12$\pm$0.03 	     &  -5.0$\pm$8.3     & $>$84	   & 259$\pm$3	  &	  407$\pm$68	      &  2456$\pm$420	 \\
\hline	 	 	 															    					   
LAE1   & 21 (CT-669)	 &  20:34:25.781    &  -35:37:36.46    & 11.6	      & 5134.99$\pm$0.03 & 3.224$\pm$0.001  & 2.26$\pm$0.04 	     &  -4.6$\pm$10.8    & $>$50	   & 314$\pm$6	  &	  426$\pm$71	      &  355$\pm$421	 \\
\hline	 	 	 															    					   
LAE1   & 26 (Q-2204-408) &  22:07:33.216    &  -40:37:00.16    & 12.8	      & 5077.85$\pm$0.03 & 3.177$\pm$0.001  & 31.40$\pm$0.07 	     &  -119$\pm$150    & $>$50	   & 98$\pm$4	  &	  -143$\pm$72	      &  -287$\pm$421	 \\
\hline	 	 	 															    					   
LAE1   & 35 (CTS-C22.31) &  02:04:37.863    &  -45:59:16.16    & 25.7	      & 5162.95$\pm$0.03 & 3.247$\pm$0.001  & 3.66$\pm$0.03 	     &  -1.8$\pm$8.2    & $>$106	   & 221$\pm$2	  &	  71$\pm$71	      &  0$\pm$421	 \\
\hline	 	 	 															    					   
LAE1   & 39 (SDSSJ0100+2105) &  01:00:27.486    &  21:05:24.91     & 18.5 	      & 4964.80$\pm$0.03 & 3.084$\pm$0.001  & 6.37$\pm$0.03 	     &  -20.6$\pm$11.1    & $>$141	   & 116$\pm$1	  &	  -951$\pm$73	      &  -1170$\pm$421	 \\
\hline	 	 	 															    					   
LAE1   & 42 (SDSSJ0219-0215) &  02:19:39.060    &  -02:15:59.59    & 18.5	      & 4907.66$\pm$0.03 & 3.037$\pm$0.001  & 3.63$\pm$0.03 	     &  6.6$\pm$17.8    & $>$136	   & 202$\pm$2	  &	  74$\pm$74	      &  -371$\pm$422	 \\
\hline	 	 	 															    					   
LAE1   & 46 (Q2355+0108) &  23:58:08.881    &  01:25:12.86	& 5.3	       & 5346.52$\pm$0.04 & 3.398$\pm$0.001  & 3.19$\pm$0.02	      &  -1.9$\pm$11.0    & $>$66	    & 129$\pm$1    &	    205$\pm$68       &  889$\pm$421	\\
LAE2   & 46 (Q2355+0108) &  23:58:08.913    &  01:25:08.05	& 5.1	       & 5357.46$\pm$0.04 & 3.407$\pm$0.001  & 2.73$\pm$0.03	      &  4.1$\pm$18.6	  & $>$151	    & 197$\pm$3    &	    819$\pm$68       &  1504$\pm$421	 \\
\hline	 	 	 															    					   
LAE1   & 47 (6dFJ0032-0414) &  00:32:05.856    &  -04:14:15.02    & 8.5	      & 5048.68$\pm$0.03 & 3.153$\pm$0.001  & 2.92$\pm$0.03 	     &  -1.0$\pm$14.4    & $>$49	   & 167$\pm$2	  &	  -648$\pm$72	      &  -216$\pm$421	 \\
\hline	 	 	 															    					   
LAE1   & 56 (TEX1033+137) &  10:36:26.554    &   13:26:56.87	&  6.7         & 4978.17$\pm$0.03 & 3.095$\pm$0.001  & 8.33$\pm$0.03	      &  -12.7$\pm$13.6   & $>$149	    & 280$\pm$1   &   -146$\pm$73	       &  440$\pm$421	  \\
LAE2   & 56 (TEX1033+137) &  10:36:27.405    &   13:26:42.61	&  12.7        & 4976.95$\pm$0.03 & 3.094$\pm$0.001  & 3.70$\pm$0.03	      &  -6.7$\pm$ 30.3   & $>$30	    & 200$\pm$2   &   -220$\pm$73	       &  367$\pm$421	  \\
LAE3   & 56 (TEX1033+137) &  10:36:27.219    &   13:26:49.75	&  5.4         & 4980.60$\pm$0.03 & 3.097$\pm$0.001  & 4.14$\pm$0.03	      &  -0.8$\pm$27.7    & $>$36	    & 240$\pm$2   &   0$\pm$73  	       &  587$\pm$421	  \\
\hline
\end{tabular}}
\flushleft{We list the coordinates and the distance $d$ (arcsec) from the quasar for each LAE discovered. For the Ly$\alpha$ emission-line in the spectrum of each LAE, we report 
the line center, the redshift, the line flux (in unit of 10$^{-17}$ erg s$^{-1}$ cm$^{-2}$), the underlying continuum flux $f_{\lambda}$ (in unit of 10$^{-20}$ erg s$^{-1}$ cm$^{-2}$ \AA$^{-1}$), the rest-frame equivalent width, 
and the line width as $\sigma_v$ of a Gaussian fit to the line. In addition, $\Delta v_{\rm Neb}$ and $\Delta v_{\rm QSO}$
are the velocity shift from the peak of the Ly$\alpha$ emission of the nebula and the systemic of the quasar targeted, respectively.
The uncertainty on the latter is dominated by the uncertainty on the systemic redshift of the quasar.
}
\label{MEGAtabLAEs}
\end{table*}

\subsection{The continuum detected sources and the low-redshift line-emitters contaminants}

\begin{figure}
       \centering
       \includegraphics[width=0.95\columnwidth, clip]{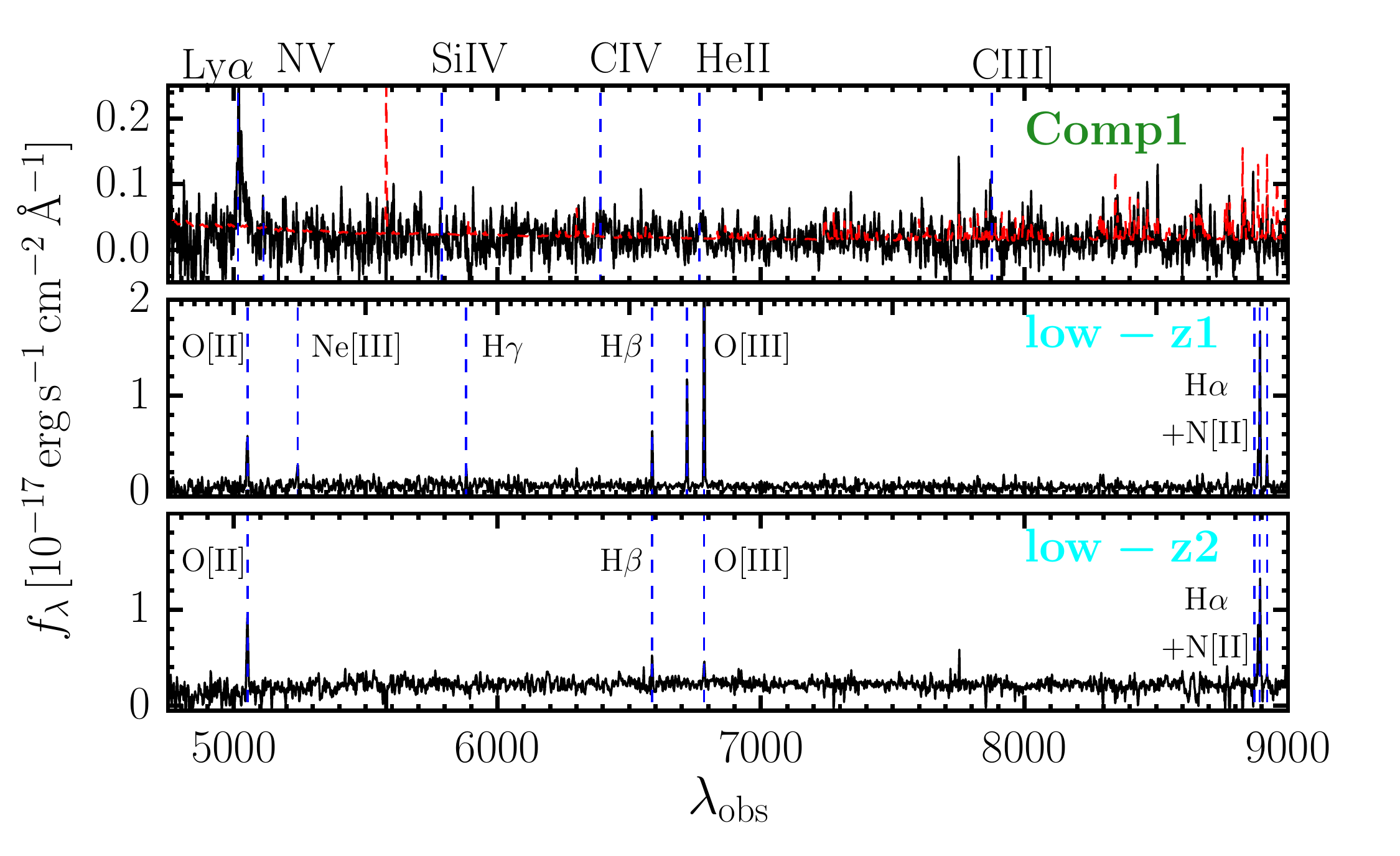}
    \caption{One dimensional spectrum of the companion AGN of ID 14 (or SDSSJ2100-0641), together with the two spectra
    of two lower redshift [\ion{O}{ii}] emitters discovered during our LAEs search. The spectra are obtained from circular apertures
    with 1.5\arcsec\ radius at the coordinates indicated in Table~\ref{tab:ContinuumSources}. As the companion AGN is quite faint, 
    we plot the noise spectrum within the same aperture (red). \civ and \heii emission
    are present at a very faint level. The location of important line emissions is indicated by blue vertical dashed lines.}
    \label{fig:compID14}
\end{figure}

\begin{figure}
       \centering
       \includegraphics[width=0.95\columnwidth, clip]{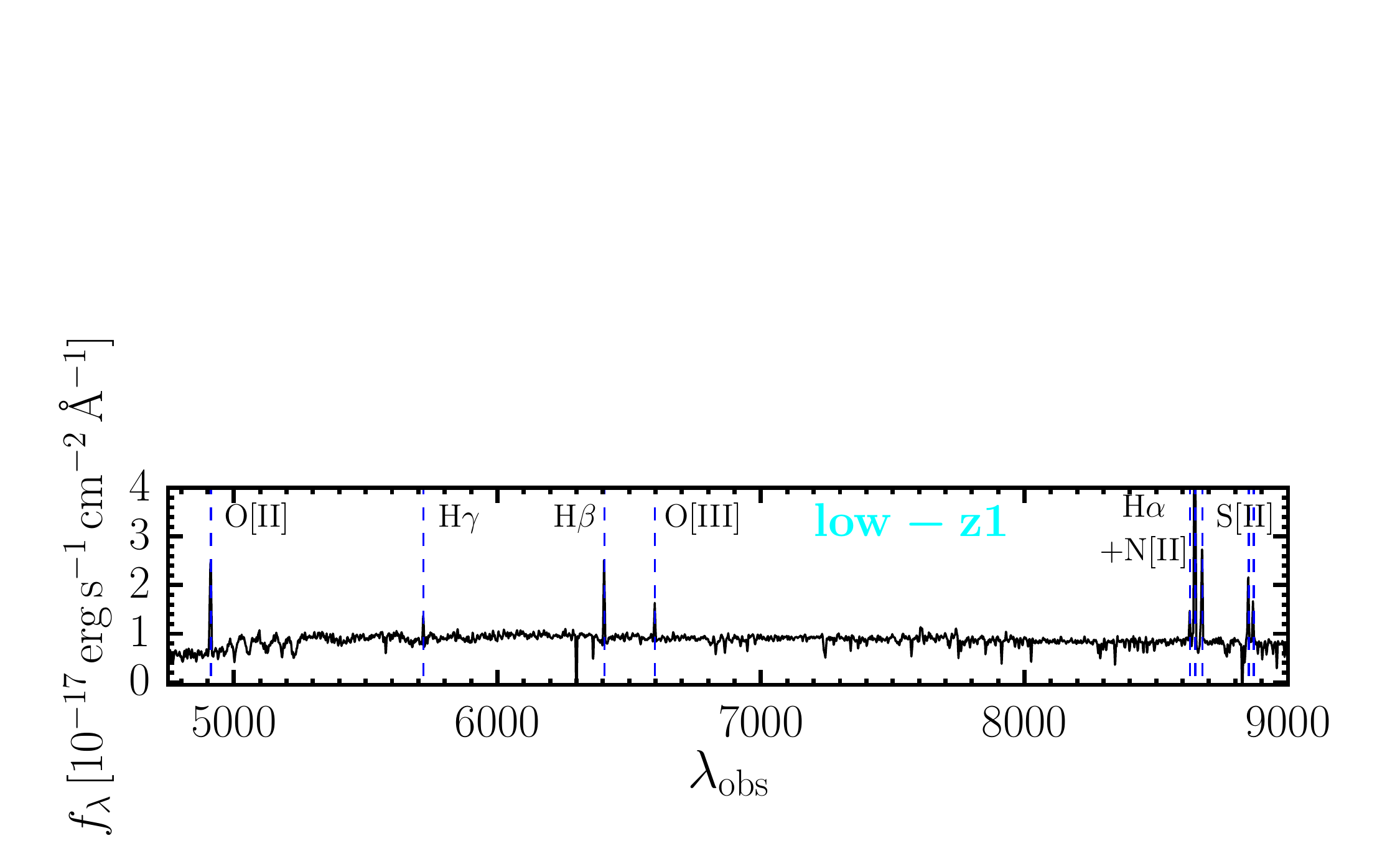}
    \caption{One dimensional spectrum of a lower redshift [\ion{O}{ii}] emitter discovered during our LAEs search around ID 15 (or SDSSJ1550+0537). 
    The spectrum is obtained from a circular aperture with 1.5\arcsec\ radius at the coordinates indicated in Table~\ref{tab:ContinuumSources}. 
    The location of important line emissions is indicated by blue vertical dashed lines.}
    \label{fig:compID15}
\end{figure}

\begin{figure}
       \centering
       \includegraphics[width=0.95\columnwidth, clip]{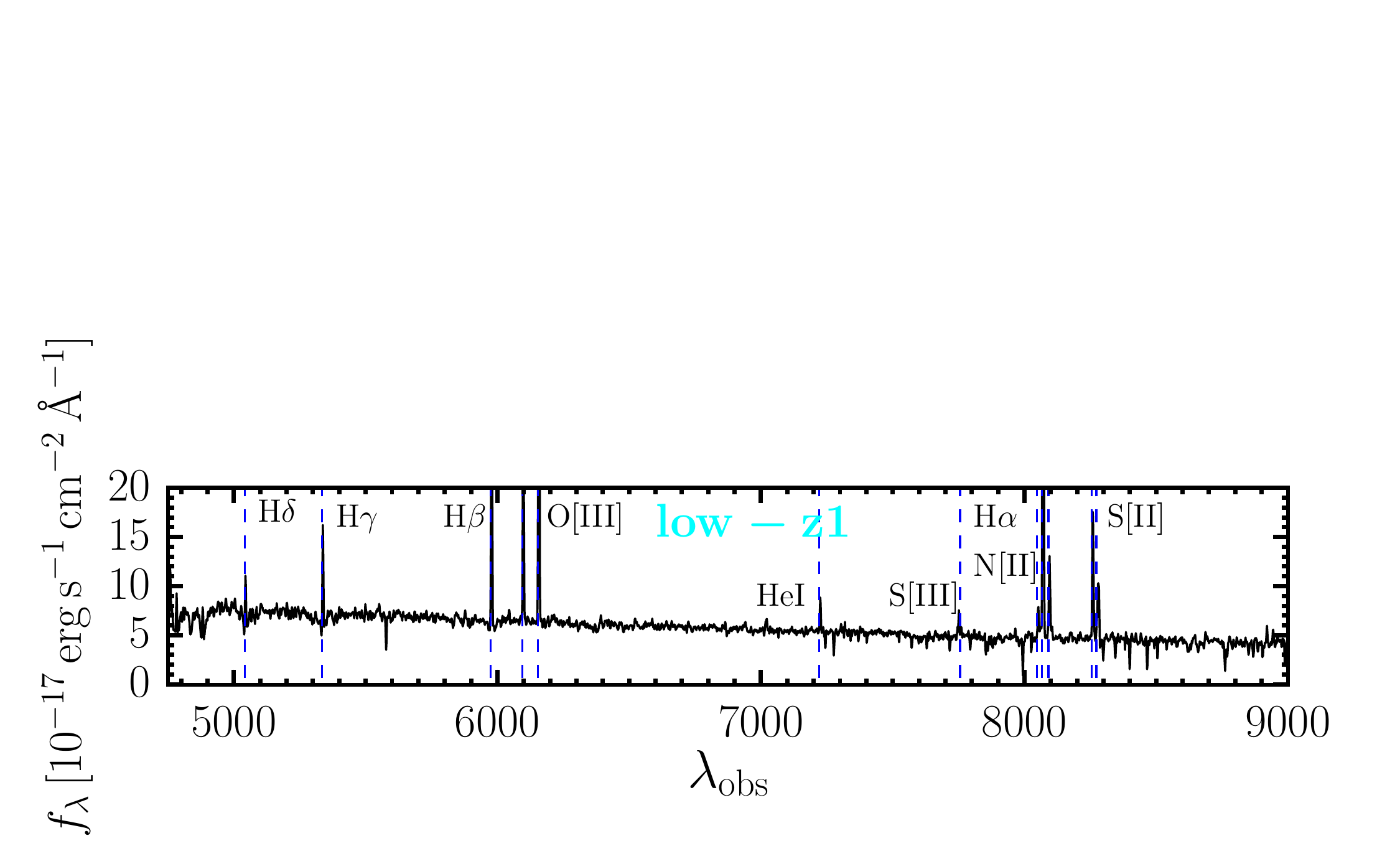}
    \caption{One dimensional spectrum of a lower redshift H$\delta$ emitter discovered during our LAEs search around ID 26 (or Q-2204-408). 
    The spectrum is obtained from a circular aperture with 1.5\arcsec\ radius at the coordinates indicated in Table~\ref{tab:ContinuumSources}. 
    The location of important line emissions is indicated by blue vertical dashed lines.}
    \label{fig:compID26}
\end{figure}

\begin{figure}
       \centering
       \includegraphics[width=0.95\columnwidth, clip]{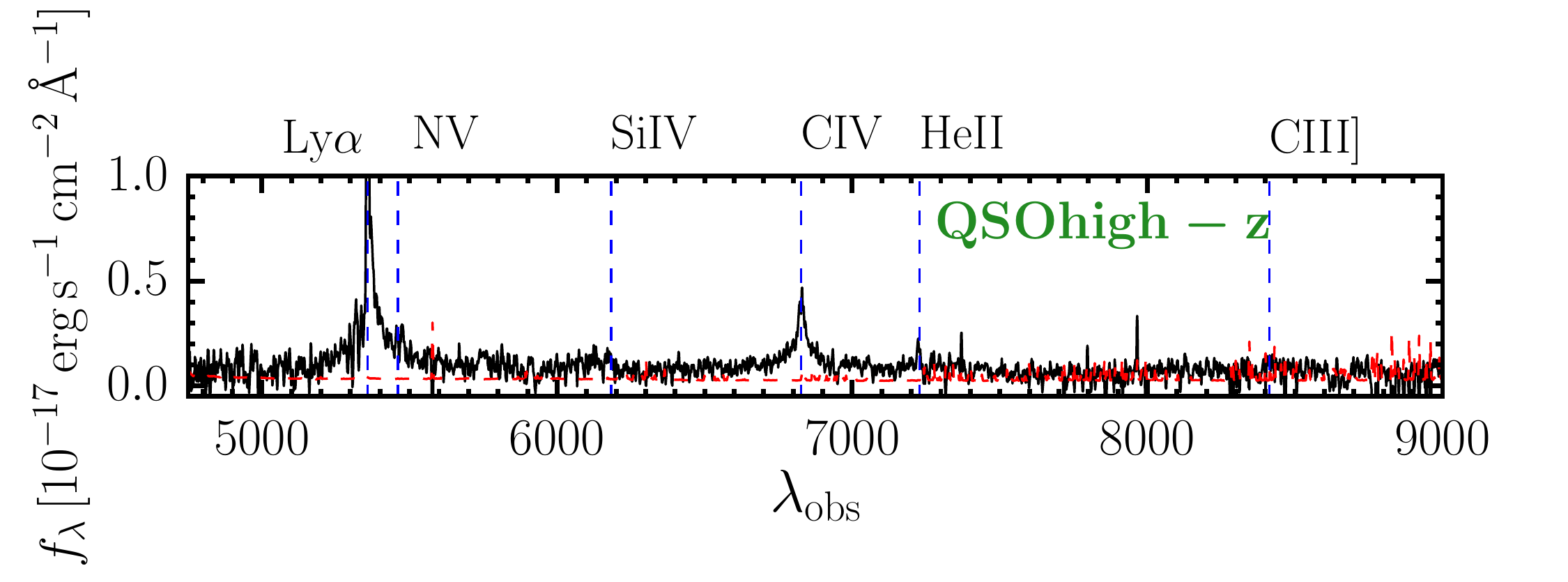}
    \caption{One dimensional spectrum of a higher redshift QSO ($z=3.407\pm0.006$) discovered around ID 31 (or UM670). 
    The spectrum is obtained from a circular aperture with 1.5\arcsec\ radius at the coordinates indicated in Table~\ref{tab:ContinuumSources}. 
    The location of important line emissions is indicated by blue vertical dashed lines. The noise spectrum within the same aperture is shown in red.}
    \label{fig:compID31}
\end{figure}

\begin{figure}
       \centering
       \includegraphics[width=0.95\columnwidth, clip]{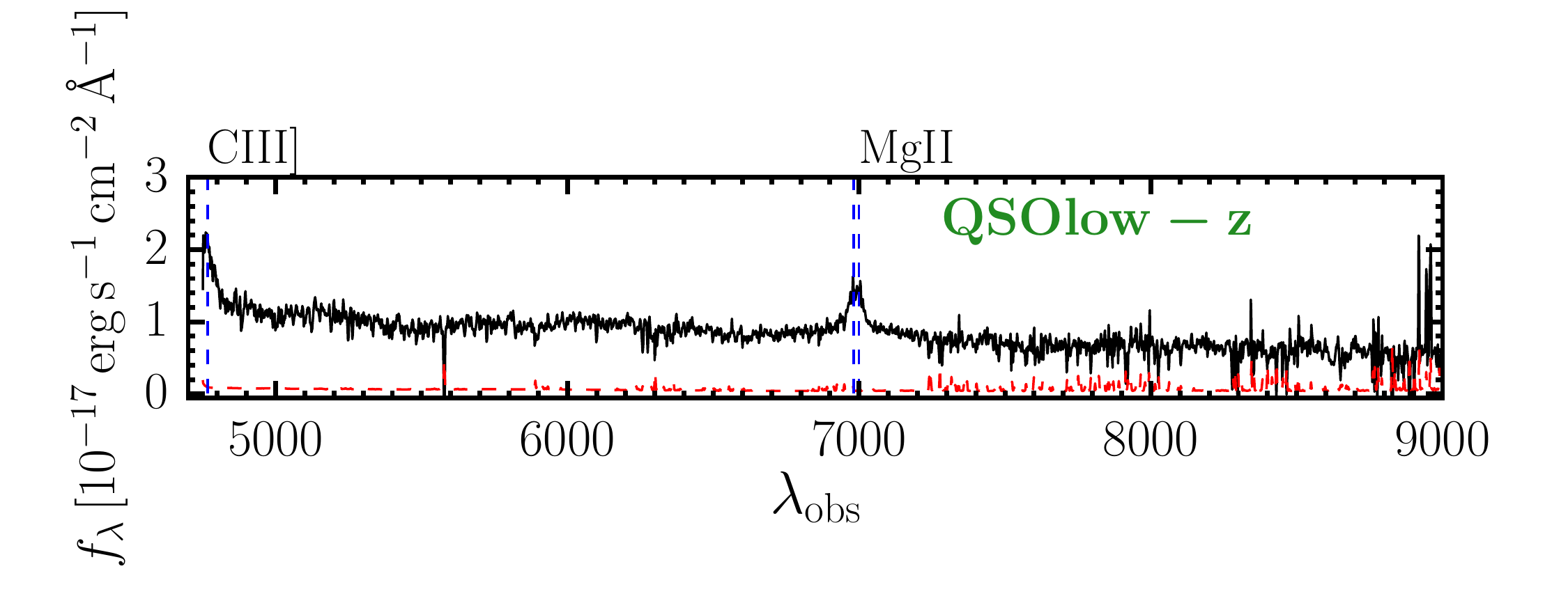}
    \caption{One dimensional spectrum of a lower redshift QSO ($z=1.497\pm0.003$) discovered around ID 32 (or Q-0058-292). 
    The spectrum is obtained from a circular aperture with 1.5\arcsec\ radius at the coordinates indicated in Table~\ref{tab:ContinuumSources}. 
    The location of important line emissions is indicated by blue vertical dashed lines. The noise spectrum within the same aperture is shown in red.}
    \label{fig:compID32}
\end{figure}

\begin{figure}
       \centering
       \includegraphics[width=0.95\columnwidth, clip]{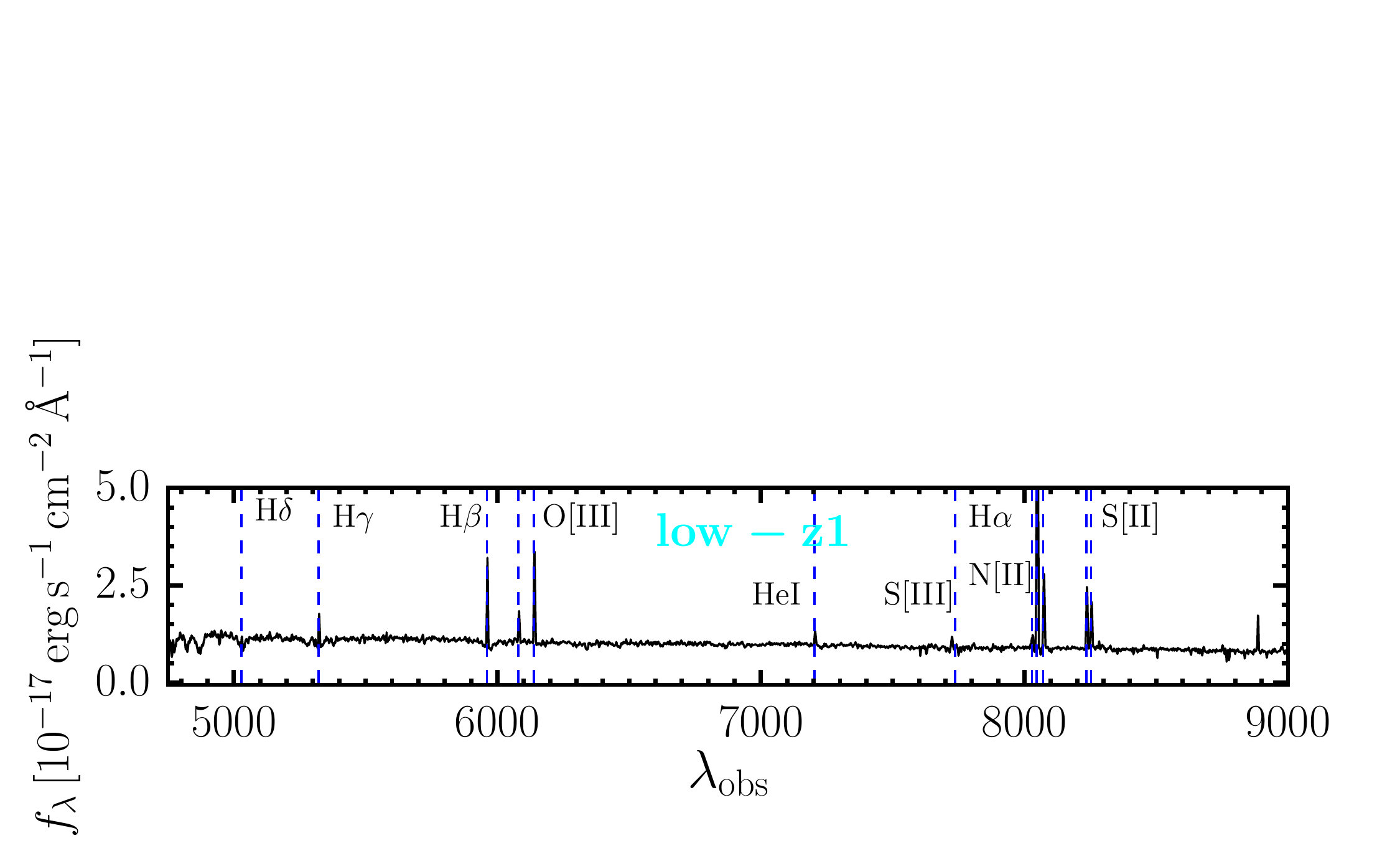}
    \caption{One dimensional spectrum of a lower redshift H$\gamma$ emitter discovered during our LAEs search around ID 46 (or Q2355+0108). 
    The spectrum is obtained from a circular aperture with 1.5\arcsec\ radius at the coordinates indicated in Table~\ref{tab:ContinuumSources}. 
    The location of important line emissions is indicated by blue vertical dashed lines.}
    \label{fig:compID46}
\end{figure}

\begin{figure}
       \centering
       \includegraphics[width=0.95\columnwidth, clip]{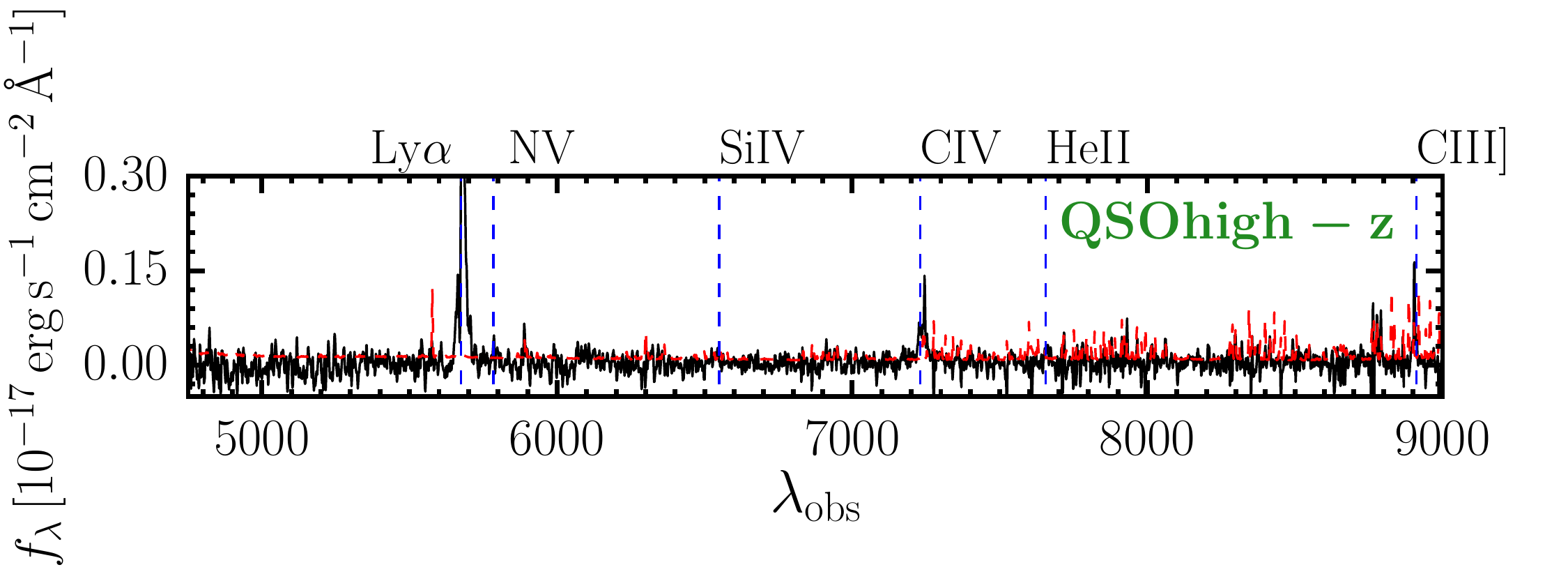}
    \caption{One dimensional spectrum of a higher redshift QSO ($z=3.67\pm0.01$) discovered around ID 48 (or UM679). 
    The spectrum is obtained from a circular aperture with 1.5\arcsec\ radius at the coordinates indicated in Table~\ref{tab:ContinuumSources}. 
    The location of important line emissions is indicated by blue vertical dashed lines. The noise spectrum within the same aperture is shown in red.
    Note that strong sky lines are present at the location of the \civ and \ion{C}{iii}] lines, making their characterisation more uncertain.}
    \label{fig:compID48}
\end{figure}

\begin{figure}
       \centering
       \includegraphics[width=0.95\columnwidth, clip]{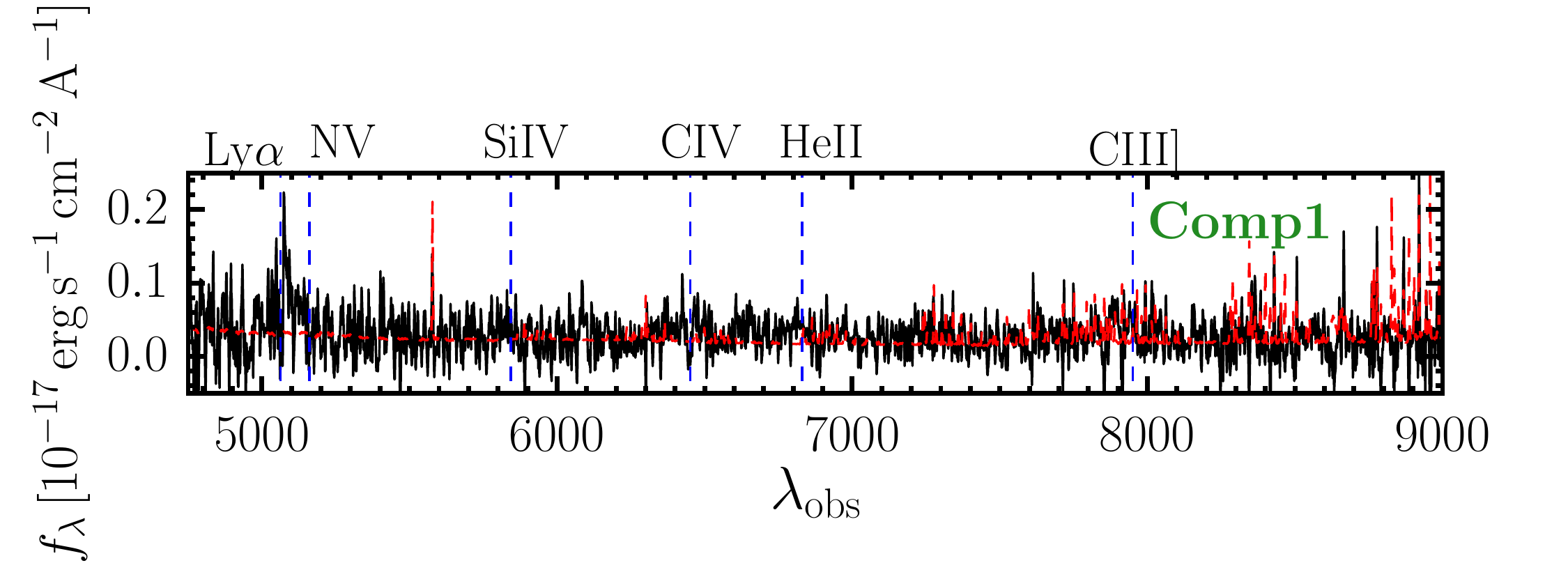}
    \caption{One dimensional spectrum of the very faint companion AGN of ID 53 (or SDSSJ0905+0410). The spectrum is obtained from a circular aperture
    with 1.5\arcsec\ radius at the coordinates indicated in Table~\ref{tab:ContinuumSources}. We plot the noise spectrum within the same aperture (red). 
    The location of important line emissions is indicated by blue vertical dashed lines.
    There are tentative evidences for \civ and \heii emissions at a very faint level, while \lya\ appear to be broad. }
    \label{fig:compID53}
\end{figure}

\begin{figure}
       \centering
       \includegraphics[width=0.95\columnwidth, clip]{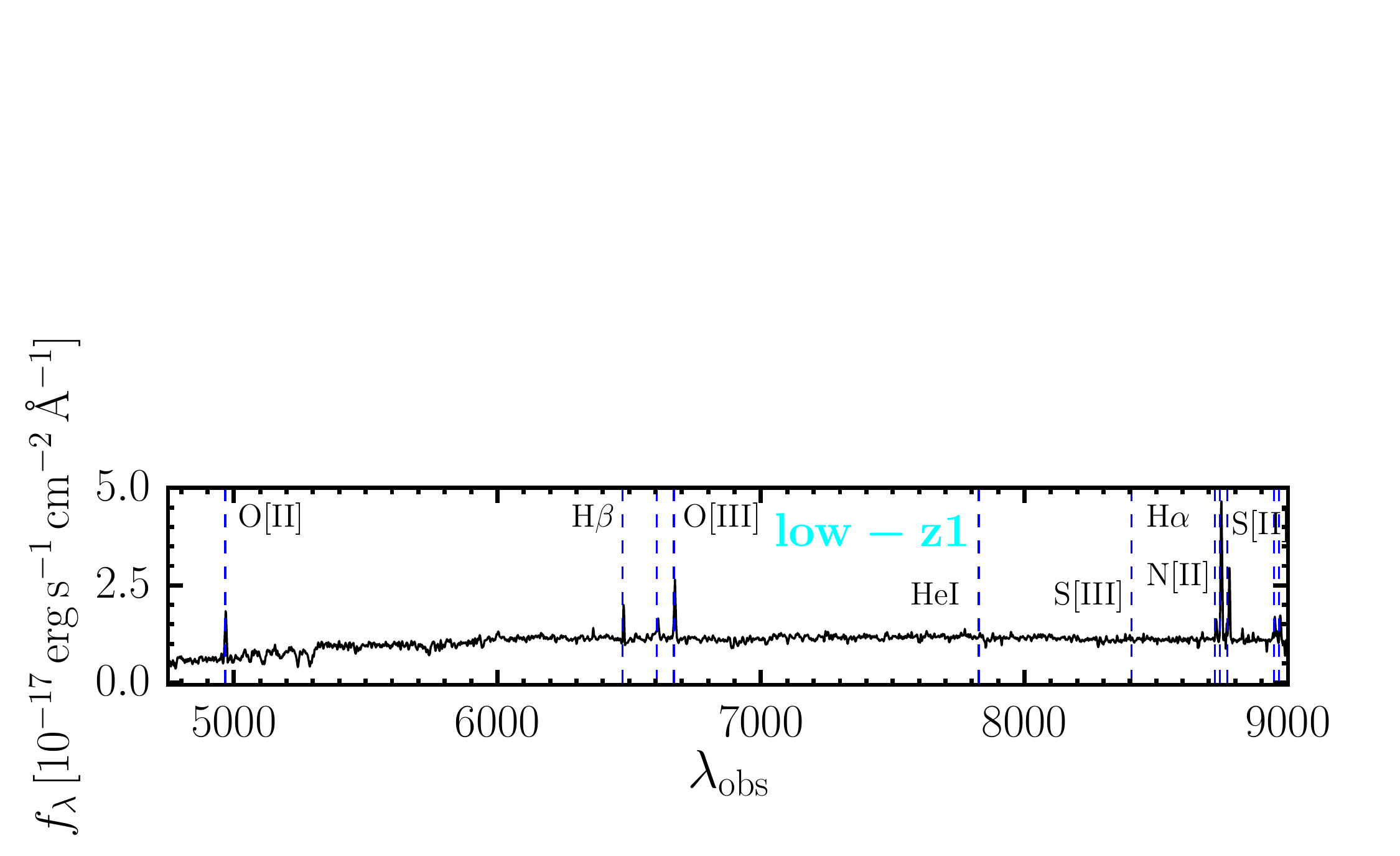}
    \caption{One dimensional spectrum of a lower redshift [\ion{O}{ii}] emitter discovered during our LAEs search around ID 61 (or LBQS1209+1524). 
    The spectrum is obtained from a circular aperture with 1.5\arcsec\ radius at the coordinates indicated in Table~\ref{tab:ContinuumSources}. 
    The location of important line emissions is indicated by blue vertical dashed lines.}
    \label{fig:compID61}
\end{figure}

In Table~\ref{tab:ContinuumSources} we list the coordinates of the continuum detected sources (companions or background and foreground quasars), and of the 
low-$z$ line emitters that were found in our search for LAEs. 
The companions of ID~13 are not listed here, but they could be found in \citet{fab+2018}.
We show the spectra of these sources 
in Figures~\ref{fig:compID14}, \ref{fig:compID15}, \ref{fig:compID26}, \ref{fig:compID31} , \ref{fig:compID32}, \ref{fig:compID46}, 
\ref{fig:compID48}, \ref{fig:compID53}, and \ref{fig:compID61}.

\begin{table}[h]
\caption{Coordinates of the continuum detected sources reported in Section~\ref{sec:comp_sel} and shown in Figure~\ref{fig:Prof_one}.}
\centering
\resizebox{\columnwidth}{!}{
\begin{tabular}{llcc}
\hline
\hline
Object	 & QSO-field        & RA	      &  Dec		\\
	 & ID (name)        &(J2000)	      &  (J2000)	\\
\hline	  						
Comp1	  &  14 (SDSSJ2100-0641)  & 21:00:25.459    & -06:42:05.97	\\
low-z1	  &  14 (SDSSJ2100-0641)  & 21:00:24.327    & -06:41:20.15	\\
low-z2	  &  14 (SDSSJ2100-0641)  & 21:00:26.160    & -06:41:18.82	\\
\hline
low-z1	  &  15 (SDSSJ1550+0537)  & 15:50:36.950    &  05:37:54.38	\\
\hline
low-z1	  &  26 (Q-2204-408)      & 22:07:34.069    & -40:36:33.66	\\
\hline
QSOhigh-z &  31 (UM670)           & 01:17:22.359    & -08:41:43.43	\\
\hline
QSOlow-z  &  32 (Q-0058-292)      & 01:01:05.618    & -28:57:41.14	\\
\hline
low-z1    &  46 (Q2355+0108)      & 23:58:10.253    &  01:24:53.61	\\ 
\hline
QSOhigh-z &  48 (UM679)           & 02:51:46.350    & -18:14:25.10	\\
\hline
Comp1     &  53 (SDSSJ0905+0410)  & 09:05:48.546    &  04:09:54.17	\\
\hline
low-z1    &  61 (LBQS1209+1524)   & 12:12:33.919    &  15:07:16.45 	\\
\hline
\end{tabular}}
\label{tab:ContinuumSources}
\end{table}

\subsection{The field of the quasar Q1205-30 (or ID 58)}

The extended \lya\ emission around the quasar Q1205-30 (or ID 58) has been discovered and analised in detail by 
\citet{Fynbo2000,Weidinger04,Weidinger05}. These works relied on NB imaging and spectroscopic observations with two position angles
down to surface brightness limits of about $4.4\times10^{-18}$~erg~s$^{-1}$~cm$^{-2}$~arcsec$^{-2}$ ($2\sigma$; \citealt{Fynbo2000}), and $6\times10^{-18}$~erg~s$^{-1}$~cm$^{-2}$~arcsec$^{-2}$ ($2\sigma$; at the \lya\ line location, see e.g. Figure 4 in \citealt{Weidinger05}), respectively for the NB and spectroscopic data.
The \lya\ emission has been reported to extend out to about 30~kpc from the quasar using the spectroscopic data (\citealt{Weidinger04}), and has been
interpreted as emitted by highly ionized hydrogen falling into the quasar dark-matter halo.
We can now compare our MUSE observations of the same system and assess if we recover the same values. 
This comparison can be regarded as a sanity check for our overall analysis technique.

First, we look at the quasar systemic redshift $z_{\rm sys}$ assumed in those works. \citet{Weidinger05} reported $z_{\rm sys}=3.041\pm0.001$ based on 
blueshift-adjusted wavelengths of the five lines \ion{N}{v}, \ion{Si}{ii} / \ion{O}{i}, \ion{Si}{iv} / \ion{O}{iv}], \ion{C}{iv}, and \ion{C}{iii}] (also used in \citealt{Tytler1992}). 
In particular, they quoted in their Table~3 a redshift from \civ of $3.038\pm0.003$.
Both systemic redshifts well agree with $z_{\rm sys}$ assumed in our work and calculated from the \civ line following \citet{Shen2016}, $z_{\rm sys}=3.037\pm0.007$.

In addition, our observations are deeper than their spectroscopic data and NB imaging if individual layers are considered. 
Nevertheless, the MUSE observations reach similar depths to their NB if we use a 30\AA\ range ($2\sigma$ SB limit of $4.3\times10^{-18}$~erg~s$^{-1}$~cm$^{-2}$~arcsec$^{-2}$; Table~\ref{tab:ObsLog}). 
Accordingly, at our depths the \lya\ emission extend out to a larger maximum projected distance from the quasar: 68~kpc (Table~\ref{Tab:LyaNeb}). However, 
the profiles in the inner portion ($< 30$~kpc) covered by both works well agree, with the profile approaching SB$_{\rm Ly\alpha}\approx 10^{-17}$~erg~s$^{-1}$~cm$^{-2}$~arcsec$^{-2}$ at about 30~kpc (compare our Figure~\ref{fig:Prof_one} 
with Figure~4in \citealt{Weidinger05}). Also, we find agreement on the redshift calculated from the peak of the \lya\ extended emission in close proximity of the quasar.
\citet{Weidinger05} quoted a redshift of 3.049, which is very similar to our $z_{\rm peak\, Ly\alpha}=3.047$ (Table~\ref{tab:sample}).
Finally, \citet{Fynbo2000} used their NB imaging data (centered at $z=3.036$) to search for LAEs in a $315\arcsec \times315\arcsec $ field-of-view. Down to basically our same limit ($F=1.1\times10^{-17}$~erg~s$^{-1}$~cm$^{-2}$, $5\sigma$), 
they did {\it not} discover any LAE within the MUSE field-of-view in agreement with our observations. They however discover candidate LAEs at larger projected distances (see e.g. their Figure~1).

Summarizing, our MUSE observations cover in greater details the morphology and kinematics of the system of Q1205-30 (or ID 58) with respect to previous studies. Notwithstanding the slightly different depths, 
we find very good agreement between the works in the literature and our overall analysis. This test thus confirm the reliability of our analysis tools.

%%%%%%%%%%%%%%%%%%%%%%%%%%%%%%%%%%%%%%%%%%%%%%%%%%

% Don't change these lines
\bsp	% typesetting comment
\label{lastpage}
\end{document}